\documentclass[conference]{IEEEtran}
\IEEEoverridecommandlockouts
\usepackage{cite}
\usepackage{amsmath,amssymb,amsfonts}
\usepackage{graphicx}
\usepackage{textcomp}
\usepackage{xcolor}
\usepackage{booktabs}
\usepackage{url}
\usepackage{hyperref}
\def\BibTeX{{\rm B\kern-.05em{\sc i\kern-.025em b}\kern-.08em
    T\kern-.1667em\lower.7ex\hbox{E}\kern-.125emX}}

\makeatletter
\g@addto@macro\UrlBreaks{\do\-}
\makeatother

\usepackage{hyperref}
\usepackage{cleveref}
\usepackage{tablefootnote}
\usepackage[caption=false,font=footnotesize]{subfig}
\usepackage{enumitem}
\usepackage[dvipsnames]{xcolor}
\usepackage{tikz}
\usepackage{soul}
\usepackage{multirow}
\usepackage{booktabs}

\usepackage{listings}

\usepackage{balance}

\usepackage{def}

\begin{document}
\bstctlcite{bstctl:nodash}

\title{Auto-Scaling Heterogeneous Neural Processing Units for Energy and Cost-Efficient LLM Serving}

\author{
\IEEEauthorblockN{Yuqi Xue}
\IEEEauthorblockA{\textit{University of Illinois Urbana-Champaign}\\
Urbana, IL, USA \\
yuqixue2@illinois.edu}
\and
\IEEEauthorblockN{Jichuan Chang}
\IEEEauthorblockA{\textit{Google DeepMind}\\
Mountain View, CA, USA \\
jichuan@google.com}
\and
\IEEEauthorblockN{Jian Huang}
\IEEEauthorblockA{\textit{University of Illinois Urbana-Champaign}\\
Urbana, IL, USA \\
jianh@illinois.edu}
}

\maketitle

\begin{abstract}
% Modern cloud platforms have widely deployed neural processing units (NPUs) to accelerate AI workloads. To meet ever-increasing computing demands, NPU chips have been developed and evolved at an incredibly fast pace. This inevitably produces heterogeneous compute pools backed by different versions of NPU chips.
% However, due to the lack of system and architecture support for managing NPU heterogeneity in modern cloud platforms, it is unclear how to best utilize heterogeneous NPUs with minimal deployment overhead.
% However, due to the lack of system support for NPU heterogeneity in modern cloud platforms, it is not easy to enable efficient reuse of older-generation NPU chips for achieving sustainable AI while ensuring service-level quality guarantees. % at the same time. 

%Large language model (LLM) services are pervasive in the cloud today. 
To meet the ever-increasing computing demands of large language model (LLM) services, modern cloud platforms have widely deployed neural processing units (NPUs). These NPU chips have been developed and evolved at an incredibly fast pace. This inevitably produces heterogeneous compute pools backed by different versions of NPU chips.
Unfortunately, due to the lack of system and architecture support for managing NPU heterogeneity in the cloud, it is unclear how to best utilize heterogeneous NPUs to maximize the energy and cost efficiency for LLM services.

% In this paper, we first investigate the benefits of utilizing heterogeneous NPU chips by conducting a characterization study of various generations of real NPU chips, and demonstrate the benefits on energy/cost efficiency and performance.
%To best utilize the inevitably heterogeneous NPU cluster, 
In this paper, we first conduct a characterization study of various generations of NPU chips, using Google TPU chips as an example, to demonstrate the potential benefits in energy and cost efficiency and performance by utilizing heterogeneous NPU chips. 
To realize these benefits, we present \pname{}, an auto-scaling framework to automatically exploit heterogeneous NPUs for cloud platforms.
\pname{} manages heterogeneous NPU resources with a new \vpod{} abstraction, which abstracts the core hardware parameters of different NPU versions and provides compatibility with existing ML frameworks.
It makes the best-fit \vpod{} allocations for different LLM inference requests using an intuitive and lightweight roofline-based analysis.
It supports fine-grained dynamic NPU resource provisioning by adjusting both the \vpod{} configuration (i.e., scaling up/down) and the number of \vpod{}s (e.g., scaling in/out).
To validate the benefits of \pname{} at scale, we implement it with a production-level NPU simulator, as well as a real-system prototype based on different generations of Google TPUs.
Our evaluation with popular LLMs shows that \pname{} can significantly improve energy/cost efficiency and service-level objective (SLO) satisfaction rate of LLM services by best utilizing heterogeneous NPU resources.

\end{abstract}


\section{Introduction}
\label{sec:intro}

\begin{figure}[t]
    \centering
    % \vspace{1ex}
    \includegraphics[width=\linewidth]{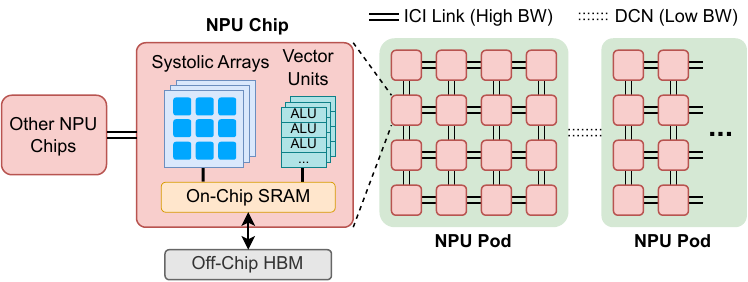}
    % \vspace{-3ex}
    \caption{The NPU pod architecture. We draw 2D pods as examples. A pod can be 2D or 3D. ICI: Inter-Chip Interconnect. DCN: Data Center Network.}
    \label{fig:npuarch}
    % \vspace{-4ex}
\end{figure}

Cloud platforms today have employed hardware accelerators like neural processing units (NPUs) to 
support large language model (LLM) services~\cite{cloudtpu:google,aws_trainium,tenstorrent,v10:isca23}. 
%support machine learning (ML) workloads~\cite{cloudtpu:google,aws_inferentia,tenstorrent,v10:isca23,topology-aware-npu:isca2025}.
These NPUs have evolved at a fast pace to meet the computing demands.  
%of the rapidly evolving ML models, especially with the surge of recent large language models (LLMs).
For instance, Google has produced seven generations of TPUs (a typical example of NPUs) in the past six years~\cite{cloudtpu:google}.
%with each generation often featuring multiple lineups (e.g., efficiency chips~\cite{tpuv5e} vs. performance chips~\cite{tpuv5p}).
Although different NPU versions share similar architectures (Figure~\ref{fig:npuarch}), they have different compute capabilities and performance behaviors (Table~\ref{tab:npu_specs}).
Their rapid development inevitably creates heterogeneous compute pools in data centers.

%Given the rapid development of new generations of NPU chips, enabling the reuse of old generations of NPUs will improve both cloud resource utilization and sustainability. \hl{the motivation of reusing NPUs needs to be elaborated.} 

\noindent{\underline{\textbf{Opportunities with heterogeneous NPUs.}}}
%As NPU chips are continuously evolving, 
% As we continue to develop new NPU chips, 
{As the NPU cluster has inevitably become heterogeneous,}
we need to ensure that different versions of NPUs can be managed and utilized efficiently. 
Our characterization study of different versions of NPU chips (see \S\ref{sec:bkg:study}) shows that no single NPU version is optimal for all workloads. The ``best-fit'' NPU version depends on whether the LLM workload is compute- or memory-bound, as well as the cost metric we optimize for (e.g., energy or monetary cost).
By mapping LLM services to their best-fit NPU resources, we can significantly improve the energy/cost efficiency.
Furthermore, as new NPU generations are in high demand, it is often feasible to use older NPUs to meet the performance goals.
Reusing old NPUs can not only address the accelerator resource shortage, but also amortize their embodied carbon and benefit datacenter sustainability.

% Even though new NPU generations provide more powerful compute capabilities and are usually in high demand, they are not always the most cost-efficient choices for an ML workload. Therefore, reusing old generations of NPU chips can not only address the shortage of accelerator resources but also best utilize existing NPU chips for improving the global resource efficiency. 

%%%% TODO: maybe put this later in intro (with evaluation results)
% Instead of retiring old NPUs with the pace of deploying new NPU generations, reusing old NPUs prolongs their device lifetime, which further helps amortize their embodied carbon footprint. As hardware accelerators like NPU chips have been dominating the computing infrastructure in modern cloud platforms, their reduced carbon footprint will benefit datacenter sustainability globally. 

\newcommand{\cmark}{\ensuremath{\checkmark}}
\newcommand{\xmark}{\ensuremath{\times}}
\newcommand{\pmark}{\ensuremath{\triangle}} % partial support

\begin{table*}[t]
    \centering
    \caption{Comparison of {\pname{}} with prior works on heterogeneous cluster scheduling and auto-scaling.
    \textit{Hardware Abstraction}: the structure of allocated hardware resources, which serves as the unit for scheduling. \textit{LLM Sharding-Aware}: the allocation mechanism considers LLM parallelism configurations (e.g., tensor/pipeline/expert parallelisms). \textit{Seq. Len.-Driven}: the scheduling considers LLM request sequence lengths. \textit{Instance Coalescing}: allocated instances will be dynamically merged (see \mbox{\S\ref{sec:design:scheduling}}). \textit{Fail-Over}: when the requested hardware type cannot be allocated, another type (e.g., a different GPU/NPU version) can be allocated. \cmark: supported; \pmark: partially supported; \xmark: not supported.}
    % \vspace{-3ex}
    \label{tab:related_works}
    \scriptsize
    \setlength{\aboverulesep}{0pt}
    \setlength{\belowrulesep}{1pt}
    \setlength{\tabcolsep}{1pt}
    \renewcommand{\arraystretch}{1}
    \resizebox{\textwidth}{!}{
    \begin{tabular}{l p{6em} c p{5em} c c c c c c c}
        \toprule
        &
        \multirow{3}{5.2em}{\centering\textbf{Target Hardware}} &
        \multirow{3}{5.8em}{\centering\textbf{Target Workload}} &
        \multicolumn{3}{c}{\textbf{Allocation Mechanism}} &
        \multicolumn{5}{c}{\textbf{Scheduling Mechanism}} \\
        \cmidrule(lr){4-6} \cmidrule(lr){7-11}
        &
        &
        &
        \begin{tabular}{@{}c@{}}\textbf{Hardware}\\\textbf{Abstraction}\end{tabular} &
        \begin{tabular}{@{}c@{}}\textbf{LLM Sharding-}\\\textbf{Aware}\end{tabular} &
        \textbf{Algorithm} &
        \begin{tabular}{@{}c@{}}\textbf{Seq. Len.-}\\\textbf{Driven}\end{tabular} &
        \begin{tabular}{@{}c@{}}\textbf{Instance}\\\textbf{Coalescing}\end{tabular} &
        \begin{tabular}{@{}c@{}}\textbf{Fail-}\\\textbf{Over}\end{tabular} &
        \begin{tabular}{@{}c@{}}\textbf{Scale-}\\\textbf{Out/In}\end{tabular} &
        \begin{tabular}{@{}c@{}}\textbf{Scale-}\\\textbf{Up/Down}\end{tabular} \\
        \midrule

        Fernandez et al.~\cite{autoscale_webapp} &
        CPU &
        Generic &
        CPU VM &
        \xmark &
        Profiling-based &
        \xmark & \xmark & \xmark & \cmark & \xmark \\

        iBalloon~\cite{iBalloon} &
        CPU &
        Generic &
        CPU VM &
        \xmark &
        Profiling-based &
        \xmark & \xmark & \xmark & \xmark & \cmark \\

        Autopilot~\cite{autopilot:eurosys2020} &
        CPU &
        Generic &
        CPU VM &
        \xmark &
        Heuristic-based &
        \xmark & \xmark & \xmark & \cmark & \cmark \\

        FIRM~\cite{firm:osdi20}, AWARE~\cite{aware:atc2023} &
        CPU &
        Generic &
        CPU VM &
        \xmark &
        Profiling-based &
        \xmark & \xmark & \xmark & \cmark & \cmark \\

        Kubernetes~\cite{kubernetes_hpa} &
        CPU/GPU &
        Generic &
        \begin{tabular}{@{}l@{}}CPU/GPU VM\end{tabular} &
        \xmark &
        Heuristic-based &
        \xmark & \xmark & \xmark & \cmark & CPU-only \\

        GKE~\cite{gke_gpu_autoscaling,googlecloud:tpu_autoscaling} &
        \begin{tabular}{@{}l@{}}CPU/GPU/\\TPU\end{tabular} &
        Generic &
        \begin{tabular}{@{}l@{}}VM,\;container,\\TPU instance\end{tabular} &
        \xmark &
        Heuristic-based &
        \xmark & \xmark & \xmark & \cmark & CPU-only \\

        \midrule

        Optimus~\cite{peng2018optimus} &
        CPU/GPU &
        ML Training &
        N/A &
        \xmark &
        Profiling-based &
        \xmark & \xmark & \xmark & \cmark & \xmark \\

        Gavel~\cite{gavel:osdi20} &
        GPU &
        ML Training &
        N/A &
        \xmark &
        Profiling-based &
        \xmark & \xmark & \xmark & \xmark & \xmark \\

        Heet~\cite{mo2024heet} &
        GPU &
        ML Training &
        N/A &
        \xmark &
        Profiling-based &
        \xmark & \xmark & \xmark & \cmark & \xmark \\

        JABAS~\cite{jabas:eurosys25} &
        GPU &
        ML Training &
        N/A &
        \xmark &
        Profiling-based &
        \xmark & \xmark & \xmark & \cmark & \xmark \\

        Sailor~\cite{sailor:sosp25} &
        GPU &
        ML Training &
        GPU VM &
        \cmark &
        Profiling-based &
        \xmark & \xmark & \cmark & \cmark & \cmark \\

        \midrule

        HAS-GPU~\cite{has-gpu} &
        GPU &
        ML Serving &
        N/A &
        \xmark &
        Profiling-based &
        \xmark & \xmark & \xmark & \cmark & Intra-GPU \\

        Splitwise~\cite{splitwise:isca24} &
        GPU &
        LLM Serving &
        N/A &
        \xmark &
        Profiling-based &
        \xmark & \xmark & \xmark & \pmark & \xmark \\

        ServerlessLLM~\cite{fu2024serverlessllm} &
        GPU &
        LLM Serving &
        N/A &
        \xmark &
        Heuristic-based &
        \xmark & \xmark & \xmark & \cmark & \xmark \\

        SpotServe~\cite{SpotServe} &
        GPU &
        LLM Serving &
        GPU VM &
        \pmark &
        Profiling-based &
        \xmark & \xmark & \pmark & \cmark & \cmark \\

        BlitzScale~\cite{blitzscale:osdi25} &
        GPU &
        LLM Serving &
        GPU VM &
        \xmark &
        Heuristic-based &
        \xmark & \xmark & \xmark & \cmark & \xmark \\

        M\'elange~\cite{griggs2024melange} &
        GPU &
        LLM Serving &
        N/A &
        \xmark &
        Profiling-based &
        \cmark & \xmark & \xmark & \xmark & \xmark \\

        DynamoLLM~\cite{dynamollm:hpca25} &
        GPU &
        LLM Serving &
        GPU VM &
        \cmark &
        Profiling-based &
        \pmark & \cmark & \xmark & \cmark & \cmark \\

        \midrule

        \textbf{\pname{}} &
        \textbf{NPU} &
        \textbf{LLM Serving} &
        \textbf{\vpod{}} &
        \textbf{\cmark} &
        \textbf{Roofline-based} &
        \textbf{\cmark} &
        \textbf{\cmark} &
        \textbf{\cmark} &
        \textbf{\cmark} &
        \textbf{\cmark} \\

        \bottomrule
    \end{tabular}
    }

    % \vspace{1ex}

    % {\raggedright \footnotesize * Due to the fundamental architectural differences between NPU and CPU/GPU, the hardware abstraction, performance/energy characteristics, and resource allocation and scheduling mechanisms are all different for NPUs. Hence, managing heterogeneous NPU chips requires new systematic support for resource allocation and runtime scheduling.}
    % \begin{minipage}{\linewidth}
    %     \raggedright \footnotesize 
    %     * Due to the fundamental architectural differences between NPU and CPU/GPU, the hardware abstraction, performance/energy characteristics, and resource allocation and scheduling mechanisms are all different for NPUs. Hence, managing heterogeneous NPU chips requires new systematic support for resource allocation and runtime scheduling.
    % \end{minipage}
    % \vspace{-4.5ex}
\end{table*}

\noindent\textbf{Challenges with NPU heterogeneity.} 
Modern cloud platforms lack systematic support for efficiently utilizing heterogeneous NPU chips. This is for three major reasons.

First, cloud platforms lack system software support for heterogeneous NPUs.
Although modern cloud vendors such as Google TPU Cloud~\cite{cloudtpu:google} support NPUs in their compute instances or virtual machines (VMs), they require end users to explicitly specify the NPU version.
Given that different NPU versions have diverse computing capabilities (Table~\ref{tab:npu_specs}) and cost efficiency (\S\ref{sec:bkg:study}), it is difficult for end users to decide which NPU version fits their workloads the best, and how much NPU resource they should allocate.
%Instead of relying on users to choose a specific NPU version, we need a new approach for heterogeneous NPU resource allocation and scheduling. 
% It is highly desirable to automate the NPU resource allocation and scheduling process to minimize the effort required for both end users and cloud platform operators. 

Second, given the diverse performance and cost efficiency characteristics of heterogeneous NPU chips, it is inherently challenging to determine the ``best-fit'' NPU allocation for an LLM workload that satisfies the performance requirement (i.e., service-level objectives, or SLOs) while maximizing energy/cost efficiency.
Each LLM model and inference request may prefer a different NPU allocation (e.g., number of chips, NPU pod topology, model sharding, and batch size), and searching for the best allocation involves complex trade-offs between performance and energy/cost.
%Prior works leveraged %profiling~\cite{dynamollm:hpca25,griggs2024melange} or %analytical models~\cite{DistServe:osdi24,LLMCompass},  
%they mainly focused on GPU clusters. 
%with up to eight all-to-all interconnected GPUs per node.
%However, an NPU cluster has a fundamentally different architecture -- each NPU pod can scale to thousands of chips arranged in a 2D/3D torus. As a result, prior approaches either incur prohibitive profiling overhead or cannot capture the impact of software optimizations (e.g., ML compilers).

%\hlcommon{We need a new approach to efficiently determine the best-fit NPU allocation for a heterogeneous NPU cluster.}

Third, cloud platforms lack runtime scheduling support for heterogeneous NPU chips.
For LLM services, the workload demand often varies over time (see \Cref{fig:traces_stats}).
While cloud platforms have developed auto-scaling techniques~\cite{autopilot:eurosys2020,aware:atc2023,googlecloud:tpu_autoscaling} to enable automated resource configuration, they were designed with the assumption that managed computing resources are homogeneous. 
None of them can directly work for heterogeneous NPUs (see $\S$\ref{sec:bkg:challenges}). A new auto-scaling mechanism is highly desirable for heterogeneous NPUs. 
%We need a new auto-scaling mechanism to allocate and schedule heterogeneous NPUs based on workload demands.
% Therefore, simply relying on these auto-scaling frameworks to manage heterogeneous NPUs will inevitably cause suboptimal performance and unbalanced resource allocation (e.g., the newer NPU versions are over-demanded while the older NPU versions are underutilized). 

Most recently, researchers have been working on the management of heterogeneous GPU clusters~\cite{yi2020optimizing,jia2022whale,um2024metis,helix:asplos25,hetis:sc25}. They utilized system profiling techniques to build performance models and learn the optimal GPU configurations (e.g., GPU version and clock frequency) for a specific LLM service.
Due to the fundamental architectural differences between NPU and CPU/GPU, the hardware abstraction, performance/energy characteristics, and resource allocation/scheduling mechanisms are all different for NPUs. Managing heterogeneous NPU chips requires new systematic support for resource allocation and runtime scheduling (see $\S$\ref{sec:background} and \Cref{tab:related_works}).

\noindent\underline{\textbf{Our solution.}}
In this paper, we present \pname{}, an auto-scaling framework for managing heterogeneous NPU generations for LLM inference services.
% \pname{} identifies the best-fit NPU allocations for different LLM inference requests to optimize their energy/cost efficiency while satisfying SLO constraints, and dynamically adjusts the NPU allocation based on the workload changes.
% \pname{} is designed to automate the deployment of ML workloads on heterogeneous NPUs in a transparent manner, with the goal of optimizing the overall energy/cost efficiency based on ML workload characteristics.

To ease the management of heterogeneous NPUs, we first develop an abstraction named \vpod{} (``virtualized'' NPU pod). It represents a group of NPU chips interconnected with high-speed links.
% which allows cloud platforms to decouple the ML workload deployment from the underlying physical NPU chips.
As different NPU versions usually share the same architecture, \vpod{} abstracts core hardware parameters (see Figure~\ref{fig:vpod_abstraction}) and provides a unified interface for different compute capabilities.
% Therefore, it enables flexible and dynamic resource allocation of different generations of NPU chips.
% \pname{} will map a \vpod{} to one or multiple physical NPU chips, according to the compute demands of ML workloads.
To simplify \vpod{} management and provide compatibility with existing machine learning (ML) frameworks, each \vpod{} is mapped to the same NPU version.
% Different \vpod{}s can be mapped to different NPU versions to exploit heterogeneity.
\pname{} exploits NPU heterogeneity by creating multiple \vpod{}s.
% (different \vpod{}s can be mapped to different NPU versions).

With \vpod{} abstraction, we develop an NPU allocation mechanism that automatically learns the most energy/cost-efficient \vpod{} allocations, while satisfying the SLO requirement of LLM services.
\pname{} utilizes the ML compiler to calculate the arithmetic intensity (FLOPs per byte accessed from HBM) of LLM requests. 
Based on this, we develop lightweight roofline models to analyze the performance and energy/cost efficiency of each possible \vpod{} allocation.
\pname{} identifies Pareto-optimal \vpod{} allocations that represent the best trade-offs between request latency and energy/cost efficiency. 
%\hlcommon{The best-fit allocation results are generated offline, and the results are shared across all services that use the same LLM model.}\revnote{\#Common-4\\\#A1\\\#D1}
We validate that \pname{} can accurately identify best-fit \vpod{} allocations with reasonable analysis overhead ($\S$\ref{sec:design:allocation}).
% \pname{} selects the best-fit NPU version for an ML workload by identifying its bottlenecking resource demand based on our extended roofline-based analytical model. Unlike the classic roofline model that focuses on capturing the arithmetic intensity of an ML workload, which is independent of the hardware it runs on, our extended models will examine the theoretical peak performance, energy efficiency, and monetary cost efficiency of an ML model on different generations of NPU chips, by having the peak computational performance and memory capability of an NPU version as constraints in the roofline analysis. Our study with various ML models shows that an optimal NPU choice depends on the specific cost metric (e.g., energy vs. monetary cost), and using older or less-powerful NPU versions may not compromise the service quality, since different ML models have different resource demands (e.g., compute vs. memory-bound). We validate that our roofline analytical models can help \pname{} accurately decide optimized \vpod{} configurations with acceptable overheads (see the detailed evaluation in $\S$\ref{sec:design:allocation}). 

\pname{} employs a closed-loop controller to {actuate the best-fit allocations. It auto-scales the {\vpod{}} allocations to precisely match workload demands at runtime.}
LLM requests have diverse sequence lengths (see \Cref{fig:traces_stats}), each of which can have a different best-fit \vpod{} allocation.
\pname{} creates heterogeneous \vpod{}s to accommodate diverse sequence lengths.
\vpod{}s of the same configuration are organized into groups. 
%\vpod{}s of the same configuration are organized into \vpod{} groups, such that each group is the best-fit for specific sequence lengths.
As the request rate varies over time, \pname{} dynamically resizes each \vpod{} group and coalesces underloaded groups. % (i.e., serving very few requests).
When a best-fit \vpod{} cannot be allocated (e.g., the requested NPU version is unavailable), \pname{} will fail over to another NPU version with minimal SLO degradation.
% to the next Pareto-optimal \vpod{} allocation that has the minimal efficiency degradation.
% The fail-over \vpod{} can be replaced later when the best-fit NPU version is available.

% In addition to optimized \vpod{} configuration, \pname{} maximizes the scheduling flexibility by enabling reconfiguration of \vpod{s} at runtime (i.e., in-place auto-scaling of \vpod{s}). For instance, when more NPU resources become available, \pname{} allows transparent reconfiguration of \vpod{s} with a different version of NPU chips. To achieve real-time reconfiguration, \pname{} supports preemption and restoration of the execution of an ML model across different NPU versions. It tracks the execution flow of an ML model via automated compiler instrumentation. Therefore, its execution can be preempted at any point, and restored with another NPU binary at the corresponding equivalence point. At the same time, the tensor sharding and memory layouts will be adjusted based on the new \vpod{} configuration. 

We implement the auto-scaling framework of \pname{} with an NPU cluster simulator to evaluate \pname{} at scale, as we cannot access real NPU chips of different generations at a large scale.
The simulator models \vpod{} allocation, \vpod{} creation overhead, request scheduling, and common LLM inference engine optimizations~\cite{vllm}.
It invokes a production-level NPU simulator as the backend to simulate the execution of each batch of requests on an NPU pod.
We validate our simulator with a real-system NPU cluster prototype developed on different generations of Google TPUs.
To the best of our knowledge, this is the first NPU cluster simulator that supports different generations of NPU chips. We have open-sourced our NPU simulator (see Appendix~\ref{app:ae} for more details). 

% Our evaluation shows that
Our evaluation using popular LLMs and service workload traces shows that \pname{} improves energy/cost efficiency by 1.13$\times$/1.43$\times$ and SLO satisfaction rate by 1.36$\times$ over the state-of-the-art (SOTA) NPU auto-scaling method.
As \pname{} offers an efficient way to utilize heterogeneous NPUs, it facilitates NPU reuse and improves datacenter sustainability.
Our contributions are as follows:

\begin{itemize}[leftmargin=*,topsep=1pt,itemsep=2pt]
    \item We conduct a thorough study to quantify the cost efficiency and performance benefits of exploiting NPU heterogeneity.
    % \vspace{2pt}

    \item We present \pname{}, an auto-scaling framework to exploit heterogeneous NPUs for energy/cost-efficient LLM serving.
    % \vspace{2pt}

    \item We propose a new \vpod{} abstraction for managing the allocation and scheduling of heterogeneous NPU chips.
    % \vspace{2pt}

    \item We develop a new NPU allocation scheme that employs a lightweight roofline-based analysis to identify the best-fit NPU allocation based on LLM serving workload demands.
    % \vspace{2pt}

    \item We develop a runtime NPU auto-scaler to dynamically adjust the heterogeneous NPU resource allocation to precisely match changing workload demands.
    % \vspace{2pt}

    \item We build an NPU cluster simulator that supports different generations of NPU chips.
    % \footnote{The simulator codebase is public on \url{https://github.com/platformxlab/NeuSim}. See Appendix~\ref{app:ae} for more details.} 
    % \vspace{3pt}

    \item We evaluate the performance and efficiency of \pname{} with diverse LLM models and production-level traces.
    % \vspace{2pt}
    
    % \item We build a roofline-based analytic model with different cost metrics to learn the best-fit NPU version for resource allocations based on the characteristic of ML models.

    % \item We propose a transparent dynamic NPU instance reconfiguration mechanism, which facilitates the in-place auto-scaling on heterogeneous NPUs with low overhead.

    % \item We implement the auto-scaling framework \pname{} based on Kubernetes and validate its benefits with diverse LLM services running on heterogeneous NPU compute pools. 

    % \item \hl{We can claim the NPU simulator for large-scale heterogeneous NPU study is also a contribution.}

\end{itemize}

{
\begin{table}[t]
    \centering
    \caption{Specifications of different NPU chip versions. The data for NPU-A/B/C/D are derived from TPUv4/5e/5p/6e. ``*'' means the parameter is not officially disclosed, and it is inferred from public data and our experiments.}
    % \vspace{-3ex}
    \footnotesize
    % \small
    \begin{tabular}{|c|c|c|c|c|}
    \hline
        & \textbf{NPU-A} & \textbf{NPU-B} & \textbf{NPU-C} & \textbf{NPU-D} \\\hline
        Release Year & 2021 & 2023 & 2023 & 2024 \\\hline
        Process Node & 7nm & 7nm* & 7nm* & 4nm* \\\hline
        TFLOPS (bf16) & 275 & 197 & 459 & 918 \\\hline
        Max. TFLOP/Joule (bf16) & 1.07* & 0.72* & 1.16* & 3.50* \\\hline
        Systolic Array Width & 128 & 128 & 128 & 256 \\\hline
        SRAM Size (MB) & 128 & 64* & 128* & 128* \\\hline
        HBM Bandwidth (GB/s) & 1200 & 819 & 2765 & 1640 \\\hline
        HBM Size (GB) & 32 & 16 & 95 & 32 \\\hline
        HBM Type & HBM2 & HBM2E & HBM2E & HBM2E* \\\hline
        Max. HBM GB/Joule & 11.06* & 6.94* & 18.51* & 13.52* \\\hline
        ICI BW/link (GB/s) & 50 & 50 & 100 & 112 \\\hline
        \# of ICI links/chip & 6 & 4 & 6 & 4 \\\hline
        ICI Torus Topology & 3D & 2D & 3D & 2D \\\hline
        {DCN BW/chip (Gbps)} & {5} & {50} & {50} & {100} \\\hline
        Max. \# of chips/pod & 4096 & 256 & 6144 & 256 \\\hline
    \end{tabular}
    \label{tab:npu_specs}
    % \vspace{-5ex}
\end{table}
}

\section{Background and Motivation}
\label{sec:background}
% In this section, we first present the technical background for LLM serving and NPUs.
% Then, we investigate the potential benefits of utilizing heterogeneous NPUs in cloud platforms. 

\subsection{Large Language Model Services in the Cloud}
\label{sec:bkg:llm}

Large language model (LLM) services are pervasive in the cloud~\cite{openai_chatgpt,github_copilot,anthropic_claude,google_gemini,multimodal-agent-survey}.
% , powering applications ranging from code assistants to enterprise chatbots and agentic applications. 
Serving an LLM request involves two stages~\cite{DistServe:osdi24,vaswani2017attention}: the \textit{prefill} stage processes input tokens in parallel to generate the first output token; the \textit{decode} stage outputs subsequent tokens iteratively. The service-level objectives (SLOs) for LLM serving are defined by two key metrics: the prefill latency (time-to-first-token or TTFT) and the decode time-per-output-token (TPOT).

Due to their substantial computational and memory requirements, LLMs are typically deployed across multiple NPU chips using different parallelism strategies to partition model parameters and computation~\cite{megatronlm,sequence_parallelism,gshard}.
The parallelism configuration is typically selected based on the request pattern (e.g., input/output sequence lengths) and the hardware capabilities. The chosen parallelism configuration affects both the serving performance and cost efficiency. 

\subsection{System Architecture of NPUs}
\label{sec:bkg:npu_arch}

\begin{figure*}[t]
    \centering
    \includegraphics[width=\linewidth]{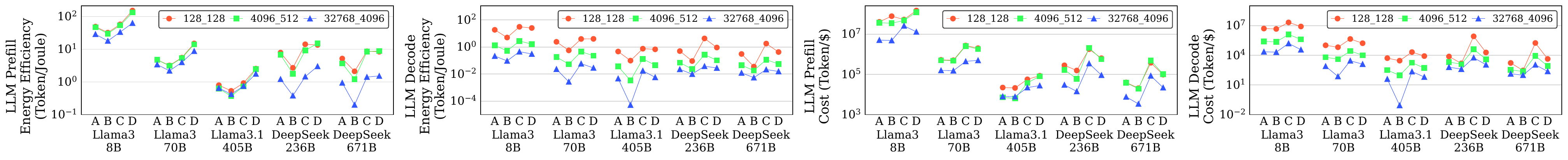}
    % \vspace{-6ex}
    \caption{Energy and economic cost efficiency of different NPU versions. The legend labels are \texttt{input\_output} sequence lengths. The cost is calculated based on \$/chip-hour prices of Google TPUs~\cite{tpu_pricing}.}
    \label{fig:motiv_energy_cost}
    % \vspace{-3.9ex}
\end{figure*}

\noindent
\textbf{NPU hardware architecture.}
Neural processing units (NPUs) are specialized accelerators for machine learning (ML).
A typical example is Google TPU~\cite{tpuv4:isca23}, the SOTA NPU architecture deployed in production.
As shown in \Cref{fig:npuarch}, an NPU chip consists of specialized compute units for ML computations, including systolic arrays for matrix multiplications and SIMD vector units for generic vector operations.
Each NPU chip has an off-chip high-bandwidth memory (HBM) to store the ML model weights and input/output data, and an on-chip SRAM to exploit data locality and hide HBM access latency.

NPU chips can be interconnected via high-speed inter-chip interconnect (ICI) links to form an \textit{NPU pod} (\Cref{fig:npuarch}). A pod is typically arranged as a 2D/3D torus~\cite{tpuv4:nsdi24}.
% , which is optimized for all-reduce bandwidth.
Each chip can perform remote direct memory access (RDMA) to another chip's memory.
% To further scale out, 
Multiple pods communicate via the traditional data center network (DCN), such as InfiniBand and Ethernet, at a much slower speed than ICI.

\noindent
\textbf{NPU software stack.}
To run an ML workload on an NPU pod, the user defines the ML model using ML frameworks like JAX~\cite{jax} or PyTorch~\cite{pytorch2}. The model is represented as a computation graph that specifies tensor operators and the dependencies between operators.
The ML compiler~\cite{xla} determines the parallelism strategies and generates the per-chip graph. Then, it performs optimizations such as operator fusion and tiling, and generates the NPU binary.

% In practice, an ML compiler typically assumes static input tensor shapes for enabling aggressive optimizations~\cite{tensorrt-llm,xla,pytorch2}. For LLM serving, variable sequence lengths are bucketized, and a computation graph is compiled offline for each bucket.

\subsection{NPU Heterogeneity in the Cloud}
\label{sec:bkg:hetero_NPU}
In modern cloud data centers, the heterogeneity of NPU chips is becoming common. This is for two major reasons.
%Heterogeneity of NPU chips is common in modern cloud data centers for two major reasons.

First, NPU chips have evolved rapidly to accommodate the ever-increasing demand of ML workloads.
Taking Google TPUs as an example (\Cref{tab:npu_specs}), the computation power, measured in tera-floating-point operations per second (TFLOPS), increased by 3.34$\times$ over 3 years; the HBM bandwidth increased by 2.3$\times$; the HBM capacity increased by 3$\times$; and the ICI bandwidth increased by 2$\times$.
As the old NPU chips will not immediately retire upon the arrival of new chips, cloud platforms inevitably accumulate multiple NPU generations.

Second, within a single generation, multiple NPU chip variants are designed to optimize for different demands.
For example, NPU-B and NPU-D in Table~\ref{tab:npu_specs} are designed to be ``efficiency'' chips. They feature a smaller HBM capacity and bandwidth with a high compute-to-memory ratio.
In general, they are a good fit for compute-intensive workloads.
NPU-A and NPU-C are ``performance'' chips that have a larger HBM capacity and bandwidth as well as massive FLOPS.
In general, they are more optimized for memory-intensive workloads.
% For example, NPU-A and NPU-C are more optimized for memory-intensive workloads, featuring larger HBM capacity and lower compute-to-memory (i.e., TFLOPS-to-HBM GBps) ratio, while NPU-B and NPU-D feature a higher compute-to-memory ratio.

% Cloud platforms today offer NPUs to users as either IaaS (bare-metal NPU VMs~\cite{tpuvm}) or SaaS (e.g., Google Colab~\cite{colab}, Vertex AI~\cite{vertex_ai}), and the allocated NPUs can be on-demand or preemptible~\cite{preemptible_vm}.

% \vspace{ex}
\subsection{Opportunities of NPU Heterogeneity}
\label{sec:bkg:study}

% The inevitable heterogeneity of NPU chips raises an immediate question: can older or less powerful NPU chips be utilized efficiently to reduce the environmental and economic cost of AI workloads?

% As discussed, different NPU versions have diverse performance characteristics.
% To understand how to best utilize heterogeneous NPUs, we want to answer two key questions:
% \begin{itemize}
%     \item Which NPU version is the best-fit for an ML workload, and how much benefit can we exploit?
%     \item When the latest NPU chips are not available, can we reuse older NPU chips to achieve the same performance goals?
% \end{itemize}
% In this section, we quantify the potential benefits of exploiting NPU heterogeneity by profiling the cost efficiency and performance of different NPU chips running LLM serving workloads, including both dense models (e.g., Llama3~\cite{llama3}) and sparse mixture-of-experts (MoE) models (e.g., DeepSeek~\cite{deepseek}).

We quantify the potential benefits of exploiting NPU heterogeneity for LLM serving.
We investigate both dense models (e.g., Llama3~\cite{llama3}) and mixture-of-experts (MoE) models (e.g., DeepSeek~\cite{deepseek}).
We employ the SOTA prefill/decode disaggregation~\cite{DistServe:osdi24,splitwise:isca24}, as the two phases have distinct demands.
We set the latency SLO to be 5$\times$ the single-request latency on the minimum number of NPU-C or NPU-D chips, whichever is faster~\cite{dynamollm:hpca25} (denoted as ``5$\times$ SLO'').
We study both energy and monetary cost.
% \revnote[1em]{\#D3}
{We sweep all SLO-compliant configurations (number of chips, batch size, and parallelism configuration) with a production-level NPU simulator that has been validated against real TPUs (\S\ref{sec:methodology}), and report the most efficient one.} If an NPU version cannot meet the target SLO, we use the configuration with the best attainable latency. 
% We analyze different SLO targets in \Cref{fig:motiv_perf_cost_tradeoff}.

% \begin{figure}
%     \centering
%     \begin{subfigure}[t]{\linewidth}
%         \centering
%         \includegraphics[width=0.65\linewidth]{figures/motiv_energy_vary_seqlen_prefill.pdf}
%         % \caption{LLM prefill.}
%     \end{subfigure}
%     \vfill
%     \begin{subfigure}[t]{\linewidth}
%         \centering
%         \includegraphics[width=0.65\linewidth]{figures/motiv_energy_vary_seqlen_decode.pdf}
%         % \caption{LLM decode.}
%     \end{subfigure}
%     \vspace{-6ex}
%     \caption{Energy efficiency of different NPU versions. The legend labels are \texttt{input\_output} sequence lengths.}
%     \label{fig:motiv_energy}
%     \vspace{-3ex}
% \end{figure}

\noindent
\textbf{Energy efficiency benefits.}
Energy efficiency is important for cloud providers, as electricity bills and power provisioning strongly correlate with the total cost of ownership (TCO) and operational carbon emission~\cite{tpuv4i:isca21,carbon_life_cycle}.
\Cref{fig:motiv_energy_cost} shows that the best-fit NPU version achieves 3.5--7.9$\times$ higher energy efficiency than the worst-fit for prefill and 3.8--339.7$\times$ for decode.
This is determined by both the workload's resource demands (e.g., FLOPs/sec, HBM bandwidth, and ICI bandwidth) and the NPU chip's per-component energy efficiency (affected by varying factors like process node, microarchitecture, and circuit-level optimizations).
% We analyze the resource demands of workloads in \Cref{fig:motiv_roofline_flops_per_joule} by extending the roofline model~\cite{roofline} to incorporate the energy efficiency instead of computational performance (see \S\ref{sec:design:roofline} for details).
% For example, 
LLM prefill is compute-bound, so the energy efficiency is determined by the computational energy efficiency (TFLOP/Joule). Hence, NPU-D is the most energy-efficient (see \Cref{tab:npu_specs}).
% In contrast, NPU-B has the lowest TFLOP/joule, making it the least energy-efficient.
% NPU-A and NPU-C have a similar TFLOP/joule, so they can achieve a similar token/joule for most prefill workloads.
LLM decode is memory bandwidth-bound, so NPU-C becomes the most energy-efficient due to its high HBM GB/Joule.

% \begin{figure}[t]
%     \centering
%     \begin{subfigure}[t]{\linewidth}
%         \centering
%         \includegraphics[width=0.65\linewidth]{figures/motiv_cost_vary_seqlen_prefill.pdf}
%         % \caption{LLM prefill.}
%     \end{subfigure}
%     \vfill
%     \begin{subfigure}[t]{\linewidth}
%         \centering
%         \includegraphics[width=0.65\linewidth]{figures/motiv_cost_vary_seqlen_decode.pdf}
%         % \caption{LLM decode.}
%     \end{subfigure}
%     \vspace{-6ex}
%     \caption{Economic cost efficiency of different NPUs. The legend labels are \texttt{input\_output} sequence lengths. The cost is calculated with \$/chip-hour prices of Google TPUs~\cite{tpu_pricing}.}
%     \label{fig:motiv_cost}
%     \vspace{-4ex}
% \end{figure}

\noindent
\textbf{Monetary cost savings.}
% We also quantify the monetary cost (token/dollar) in \Cref{fig:motiv_energy_cost} based on~\cite{tpu_pricing}. This is a common optimization goal for third-party users who rent NPU instances and deploy their own LLM inference services in the cloud.
In \Cref{fig:motiv_energy_cost}, we also quantify monetary cost~\cite{tpu_pricing} (token/dollar), a common optimization goal for third-party users who rent NPU instances and deploy their own LLM inference services in the cloud.
% , and the cloud providers directly offer the dollar per chip-hour price of NPU instances.
% The monetary cost of the best-fit and the worst-fit NPUs can differ by 3.3--34.2$\times$ for prefill and 5.5--2,602$\times$ for decode.
Similar to the energy efficiency differences across NPU versions, the monetary cost of the best- vs. worst-fit NPUs can differ by 3.3--34.2$\times$ for prefill and 5.5--2,602$\times$ for decode.

\observation{1}{
The best-fit NPU version depends on an ML workload's bottlenecking resource (e.g., compute, memory) and the specific cost metric (e.g., energy vs. monetary cost). Exploiting heterogeneity helps satisfy diverse optimization goals for various ML workloads.
% By selecting the best-fit NPU version based on the bottlenecking resource of the ML workload, we can achieve orders-of-magnitude cost savings. The optimal NPU choice depends on the specific cost metric (e.g., energy vs. monetary cost). Exploiting NPU heterogeneity enables us to satisfy diverse optimization goals.
}

% \begin{figure}[t]
%     % \vspace{0.5ex}
%     \centering
%     \begin{subfigure}{\linewidth}
%         \centering
%         \includegraphics[width=\linewidth]{figures/prefill_seqlen_vs_energy_by_config_llama3_1-405b.pdf}
%         % \vspace{-0.7ex}
%         % \caption{Prefill.}
%     \end{subfigure}
%     % \vspace{0.2em}
%     \begin{subfigure}{\linewidth}
%         \centering
%         \includegraphics[width=\linewidth]{figures/decode_seqlen_vs_energy_by_config_llama3_1-405b.pdf}
%         % \vspace{-1ex}
%         % \caption{Decode.}
%     \end{subfigure}
%     % \vspace{-5.9ex}
%     \caption{The optimal energy efficiency of LLM requests with various sequence lengths. The legends ``\{NPU type\}\_\{tensor parallelism degree\}\_\{pipeline parallelism degree\}'' indicate the optimal configuration for the specific sequence length.
%     % Data parallelism degree is always 1 for LLM serving.
%     Due to space limitation, we only show Llama3.1-405B here.}
%     \label{fig:motiv_fine_grained_cost}
%     % \vspace{-3.2ex}
% \end{figure}

\begin{figure}[t]
    \centering

    \includegraphics[width=\linewidth]
    {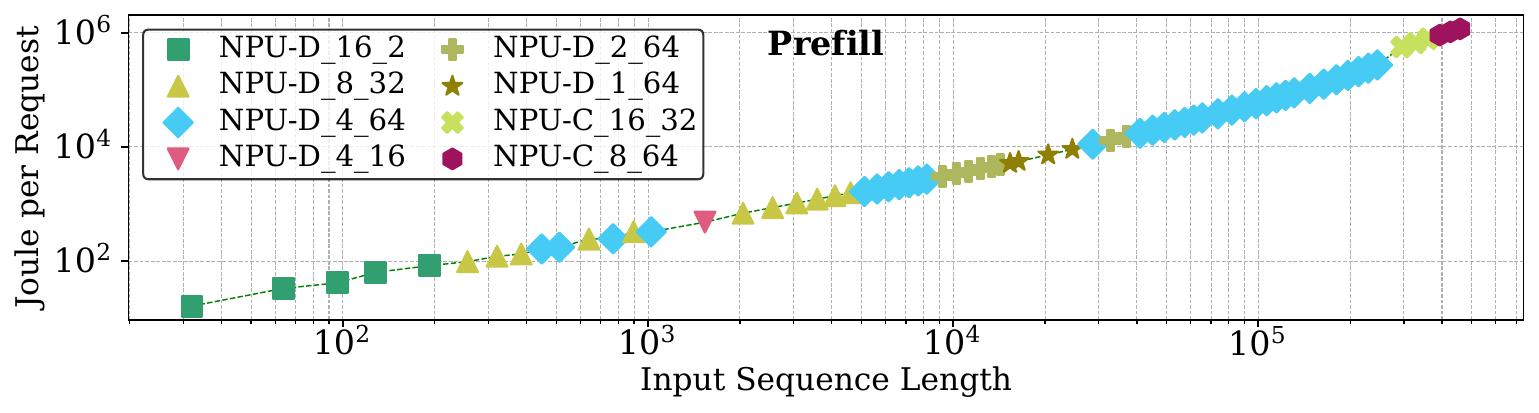}

    \includegraphics[width=\linewidth]
    {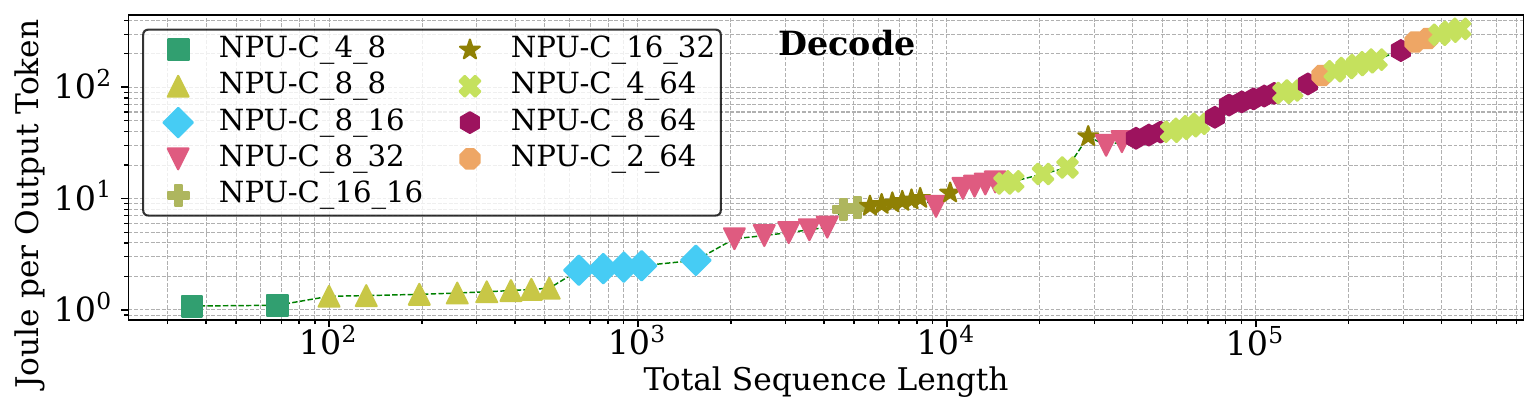}

    \caption{The optimal energy efficiency of LLM requests with various sequence lengths. The legends ``\{NPU type\}\_\{tensor parallelism degree\}\_\{pipeline parallelism degree\}'' indicate the optimal configuration for the specific sequence length. Due to space limitation, we only show Llama3.1-405B here.}
    \label{fig:motiv_fine_grained_cost}
\end{figure}

% \noindent
% \textbf{Diverse resource demands of LLM requests.}
An LLM service needs to serve requests with diverse sequence lengths (see \Cref{fig:traces_stats}).
Each request can have a distinct resource bottleneck. A one-size-fits-all NPU pod configuration is inherently sub-optimal.
In \Cref{fig:motiv_fine_grained_cost}, the most energy-efficient configuration (defined by the NPU version, pod shape, parallelism strategies, and batch size) varies significantly across various sequence lengths.
Exploiting NPU heterogeneity at fine granularity enables us to optimize efficiency for each request with its ``best-fit'' NPU allocation.

\observation{2}{
Different LLM inference requests can favor different NPU versions and NPU pod configurations.
It is desirable to provision a mix of heterogeneous NPU pods at fine granularity to satisfy diverse resource demands. 
}

\begin{figure}[t]
    \centering
    \includegraphics[width=\linewidth]{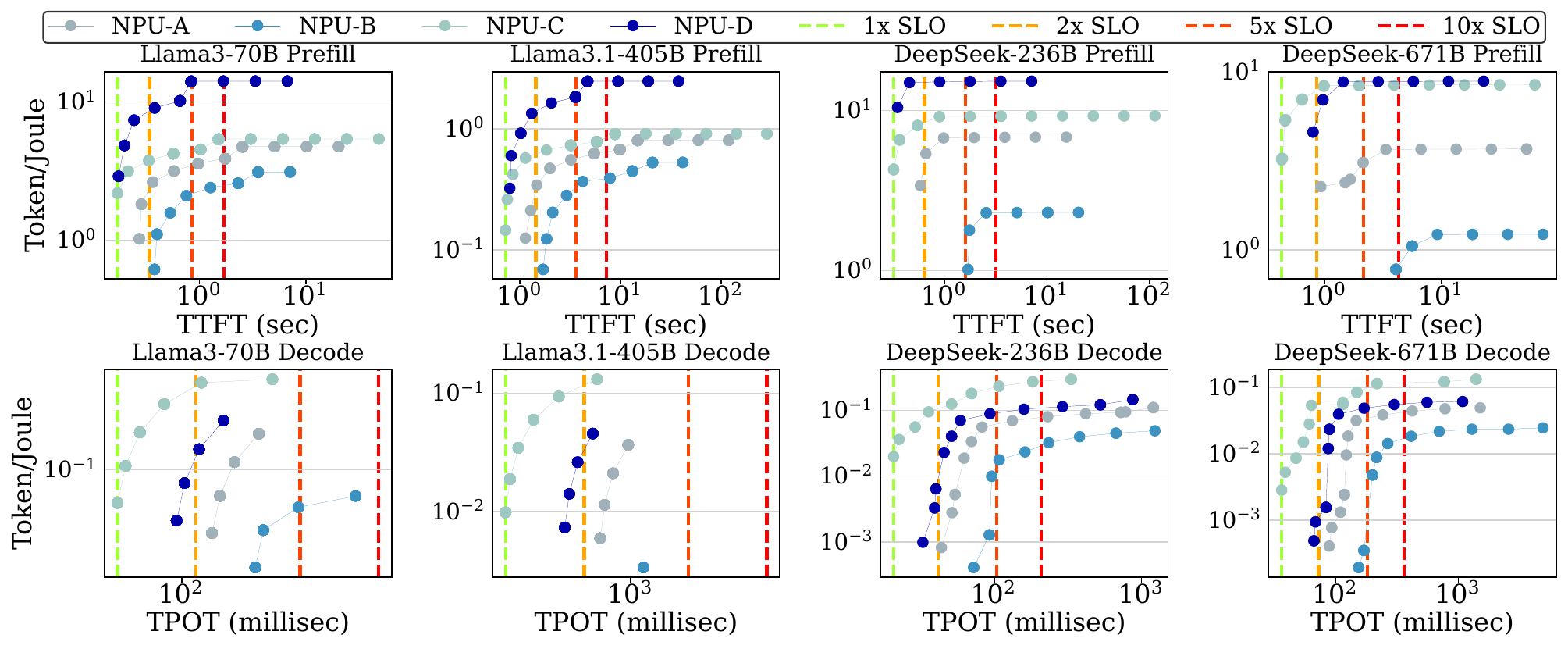}
    % \vspace{-6.6ex}
    \caption{Optimal energy efficiency vs. latency across NPU versions. Monetary costs and Llama3-8B have similar trends (omitted due to space limitations).}
    \label{fig:motiv_perf_cost_tradeoff}
    % \vspace{-3ex}
\end{figure}

\noindent
\textbf{Performance fungibility.}
As new chips are scarce, it is not always possible to allocate the best-fit NPU version.
Therefore, it is often possible for an alternative NPU version to meet the performance goals. Such \textit{performance fungibility} helps alleviate the shortage of the latest NPUs.
% and provides flexibility for NPU allocation based on available resources at runtime.
As shown in \Cref{fig:motiv_perf_cost_tradeoff},
% compares the energy efficiency of different NPU versions at various latency SLOs.
% Except for the strictest 1$\times$ SLO (i.e., requiring that the performance of older chips must be no worse than the latest, most powerful chips), there is always at least one alternative NPU version that can satisfy $\geq 2 \times$ SLO.
% Except for the strictest 1$\times$ SLO, 
there is always at least one NPU version that can satisfy $\geq 2 \times$ SLO.
In practice, a strict 1$\times$ SLO is rare, as it forbids batching even on the most powerful NPUs~\cite{orca:osdi22}.
% This offers the flexibility to trade off between allocation responsiveness (e.g., how fast a new instance is provisioned) and cost efficiency.

\observation{3}{
It is often feasible to use older NPU generations without compromising service quality. Reusing old NPU versions helps mitigate supply shortages of new NPU chips and improve resource availability.
}

\subsection{Challenges with Heterogeneous NPUs}
\label{sec:bkg:challenges}

Despite the performance and efficiency benefits, exploiting NPU heterogeneity is challenging due to three major reasons.

% Exploiting NPU heterogeneity can bring significant performance, energy, and cost efficiency benefits.
% However, this is challenging due to three major reasons.

First, cloud platforms today lack system software support for heterogeneous NPUs.
They manage different NPU versions as separate, homogeneous resource pools, and rely on end users to specify the NPU version~\cite{cloudtpu:google,borg:google}.
Given the diverse computing capabilities and cost efficiency of different NPU versions, it is cumbersome for users to decide which NPU version and how many NPU chips to allocate.
% Even though the user specifies an NPU version, the cloud platform lacks the flexibility to provision different NPU versions based on workload demands \textbf{(T2)}.
% To best exploit the benefits of NPU heterogeneity, it requires a new approach for heterogeneous NPU resource allocation and scheduling.

% Second, we need to determine how to allocate heterogeneous NPU resources for ML workloads to maximize their performance and energy/cost efficiency (\textbf{(T1)} and \textbf{(T2)}).
Second, given the diverse characteristics of heterogeneous NPUs, it is intrinsically challenging to find the ``best-fit'' NPU allocation due to the large search space defined by parameters including NPU version, pod shape, model sharding, and batch size.
% Searching for the best configuration involves trade-offs between conflicting metrics (e.g., performance vs. energy/cost).
%\hlcommon{Recent studies used profiling to find the optimal configuration for GPUs~\mbox{\cite{dynamollm:hpca25,griggs2024melange}} (see \mbox{\Cref{tab:related_works}}).
%However, they rely on intensive system profiling of every instance configuration (e.g., GPU version, clock frequency, and LLM parallelism degrees), sequence length, and batch size to build performance models for learning the optimal allocations, which incurs prohibitive overhead as we scale to hundreds of NPU chips per instance and over 100K sequence length.}
While we can employ analytical models to predict LLM inference latency~\cite{LLMCompass}, they typically model execution at the operator level and cannot capture the impact of different software stack optimizations. % (e.g., ML compilers). 
Moreover, due to differences in hardware architecture and computing stacks, it is hard to apply them directly to heterogeneous NPUs.

Third, cloud platforms lack runtime resource management for heterogeneous NPUs.
For production LLM services, the workload demand varies over time (see \Cref{fig:traces_stats}).
They are commonly deployed with auto-scaling to dynamically adjust the allocated resources~\cite{autopilot:eurosys2020,aware:atc2023,googlecloud:tpu_autoscaling,vertex_ai}.
However, current auto-scaling techniques assume the underlying NPU pool is homogeneous and only adjust the number of allocated NPU pods (i.e., scale-in/out).
A potential solution is to also adjust the configuration of each NPU pod (i.e., scale-up/down).
While there have been scale-up/down techniques for CPU/GPU VMs~\cite{autopilot:eurosys2020,aware:atc2023}, none of them work for heterogeneous NPUs due to fundamental architectural differences.

The unique properties of NPU architecture require careful design of the corresponding system stack to enable the efficient use of heterogeneous NPU chips for LLM serving. As shown in \mbox{\Cref{tab:related_works}}, an efficient auto-scaling system for LLM serving requires systematic support, which includes appropriate hardware abstraction, allocation mechanism, parallelism support of LLMs, scheduling mechanism, exception (fail-over) handling, and others. Unfortunately, we cannot simply borrow or reuse existing auto-scaling design and implementation targeting CPU/GPU architectures, due to the special requirements/consideration for hardware abstraction, performance/energy trade-offs, resource allocation, and scheduling mechanisms in heterogeneous NPU management. We will discuss how we overcome these challenges throughout \mbox{\S\ref{sec:design}}.

\section{Design and Implementation}
\label{sec:design}

%%%%%%%%%%%% draft outline %%%%%%%%%%%%%%%%
% design goals:
% (1) provide users with the flexibility to reuse heterogeneous NPUs in different scenarios
% (1.1) Batch workloads (training or inference): SLO is job completion time, do not care about NPU generations as long as job can finish on time, may allow delay to match renewable
% (1.2) Interactive workloads (e.g., jupyter notebook session): 
% (1.3) Online inference services: 
% (2) maximize utilization by harvesting heterogeneous NPUs
% (3) minimize performance impacts
% (4) user transparency

% 1. New resource abstraction and pricing models needed to incentivize users to reuse NPU chips.
% Use the virtual NPU pod (vPod) abstraction (consisting of many virtual NPU slices (vSlices), each vSlice is a set of ICI-connected NPU chips, and vSlices are interconnected via DCN) to facilitate resource management, including allocation and harvesting.
% Pricing model: ``reusable NPU'' (enable reuse of old chips), ``harvest NPU'' (enable harvesting), ``carbon-aware NPU'' (enable workload shift to utilize renewable energy).

% 2. vPod Allocation

% 3. vPod Harvesting

% 4. Dynamic vPod Reconfiguration (fungibility)

% 5. Integration (how to change ML compiler, runtime (Pathways), cloud OS, etc.)
%%%%%%%%%%%%%%%%%%%%%%%%%%%%%%%%%%%%%%%%%

% \begin{figure}[t]
%     \centering
%     \includegraphics[width=0.8\linewidth]{figures/system_overview.pdf}
%     \caption{System overview.}
%     \label{fig:system_overview}
% \end{figure}

%In this section, we introduce our NPU reuse framework.
We design \pname{}, a fine-grained auto-scaling framework for heterogeneous NPUs to achieve the following goals:
\begin{enumerate}[leftmargin=*,label=(\arabic*)]
    \item \textbf{Efficiency:} \pname{} should map LLM requests to their best-fit NPU pods to maximize energy/cost efficiency;
    \item \textbf{Performance:} \pname{} should meet the SLOs of LLM requests while improving efficiency;
    \item \textbf{Flexibility:} When the best-fit NPUs are not available, \pname{} should automatically fail over to another NPU version with minimal SLO and efficiency degradation;
    \item \textbf{Compatibility:} \pname{} should be compatible with existing ML software stack and cloud infrastructure.
\end{enumerate}

% (1) \textbf{Efficiency:} \pname{} should map LLM requests to their best-fit NPU pods to maximize energy/cost efficiency; (2) \textbf{Performance:} \pname{} aims to meet SLOs of LLM requests; (3) \textbf{Flexibility:} When the best-fit NPUs are not available, \pname{} should automatically fail over to another NPU version with minimal SLO and efficiency degradation; (4) \textbf{Compatibility:} \pname{} should be compatible with existing ML software stack and cloud infrastructure.

%\begin{itemize}[leftmargin=*]
%    \item \textbf{Efficiency:} \pname{} should map LLM requests to their beenergyst-fit NPU pods to maximize energy/cost efficiency.
%    \item \textbf{Performance:} \pname{} aims to meet SLOs of LLM requests. % while optimizing for energy/cost efficiency.
%    \item \textbf{Flexibility:} When the best-fit NPUs are not available, \pname{} should automatically fail over to another NPU version with minimal SLO and efficiency degradation.
%    % \item \textbf{Minimal overhead:} \pname{} should introduce minimal performance overhead to the user workload.
%    \item \textbf{Compatibility:} \pname{} should be compatible with existing ML software stack and cloud infrastructure.
%    % existing ML software stack to simplify LLM service deployment and minimize changes to the existing cloud infrastructure.
    % \item \textbf{Usability:} Deploying ML workloads on heterogeneous NPUs should be as easy as on homogeneous NPUs.
    % \item \textbf{Minimal overhead:} \pname{} should introduce minimal performance overhead to user workloads.
%\end{itemize}

% \vspace{-1ex}
\subsection{\pname{} Overview}
\label{sec:design:overview}

%\hl{This paper is talking about how to build an auto-scaling framework for NPUs. It has a few major components: roofline model for facilitating our analysis for autoscaling, ML compiler support for learning workload characteristics, vPoD abstraction for managing heterogeneous NPUs, dynamic allocation and scheduling for autoscaling implementation, exception handling....}

\begin{figure}
    \centering
    \includegraphics[width=0.73\linewidth]{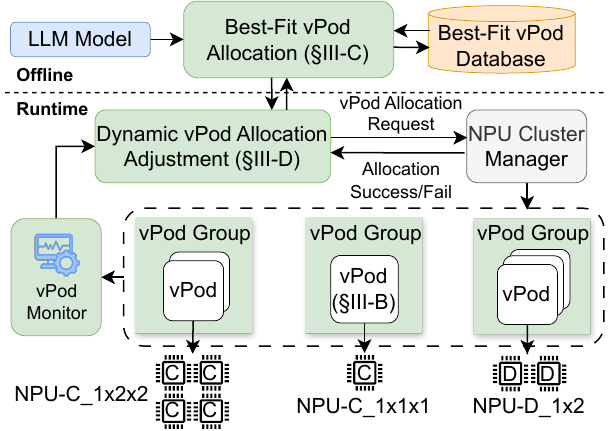}
    % \vspace{-3.2ex}
    \caption{System architecture of \pname{}. \texttt{NPU-C\_1x2x2} is an \texttt{1x2x2} NPU-C pod. \texttt{NPU-C\_1x1x1} is a single NPU-C chip. \texttt{NPU-D\_1x2} represents two interconnected NPU-D chips.}
    \label{fig:system_overview}
    % \vspace{-3.9ex}
\end{figure}

% \begin{comment}
% % \pname{} provides an end-to-end framework to manage LLM service deployment on heterogeneous NPUs.
% \Cref{fig:system_overview} shows the system architecture of \pname{}.
% To simplify the management of heterogeneous NPUs, we introduce a new \vpod{} abstraction to represent the key properties of an NPU pod (\S\ref{sec:design:abstraction}).
% % To allocate NPU resources for an LLM service, 
% \pname{} learns the workload characteristics with the ML compiler to derive the ``best-fit'' \vpod{} allocation for each representative input/output sequence length (\S\ref{sec:design:allocation}). % sampled from historical traces or datasets (see \Cref{tab:eval_traces}). %  using a roofline cost-efficiency model
% % The sequence lengths can be sampled from historical traces or datasets (see \Cref{tab:eval_traces}).
% % \pname{} supports co-optimizing for diverse cost metrics, such as maximizing the energy efficiency while achieving a specific SLO latency. 
% The best-fit allocations can be generated offline, and the results are shared across all workloads using the same backbone LLM model.
% At runtime, \pname{} auto-scales the \vpod{} allocations to match the workload demand (\S\ref{sec:design:scheduling}).
% \pname{} organizes \vpod{}s of the same configuration into \vpod{} groups.
% It creates, evicts, and selectively coalesces \vpod{} groups and adjusts the number of \vpod{}s in each group.
% \end{comment}

\Cref{fig:system_overview} shows the system architecture of \pname{}.
To simplify the management of heterogeneous NPUs, we introduce a new \vpod{} abstraction in \S\ref{sec:design:abstraction}.
% To allocate NPU resources for an LLM service, 
We describe how \pname{} learns workload characteristics with ML compilers to derive the ``best-fit'' \vpod{} allocation that can be shared across all workloads using the same backbone LLM model in \S\ref{sec:design:allocation}. % sampled from historical traces or datasets (see \Cref{tab:eval_traces}). %  using a roofline cost-efficiency model
% The sequence lengths can be sampled from historical traces or datasets (see \Cref{tab:eval_traces}).
% \pname{} supports co-optimizing for diverse cost metrics, such as maximizing the energy efficiency while achieving a specific SLO latency. 
% The best-fit allocations can be generated offline, and the results are shared across all workloads using the same backbone LLM model.
Then we show how \pname{} auto-scales the \vpod{} allocations to match the workload demand at runtime in \S\ref{sec:design:scheduling}.

\subsection{\vpod{} Abstraction for Heterogeneous NPUs}
\label{sec:design:abstraction}

\begin{figure}[t]
    \centering
    \includegraphics[width=0.9\linewidth]{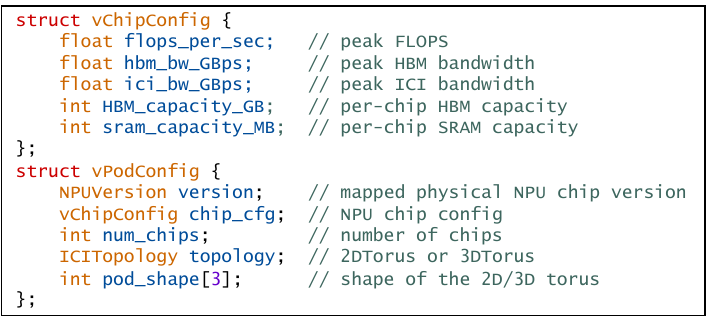}
    % \vspace{-3ex}
    \caption{\vpod{} configuration parameters.}
    \label{fig:vpod_abstraction}
    % \vspace{-4ex}
\end{figure}

The \vpod{} abstraction serves three purposes:
(1) It provides a unified representation for different NPU pod configurations.
(2) It abstracts out the parameters needed for \pname{} to allocate and schedule NPU resources.
(3) It provides compatibility with the ML workloads and existing cloud infrastructure.
% (3) It provides a generic interface for integrating future NPU generations.
% \hl{elaborate each point}

\noindent
\textbf{\vpod{} abstraction.}
A \vpod{} represents a group of NPU chips interconnected via high-speed ICI links.
\Cref{fig:vpod_abstraction} shows its configurable parameters.
% including both pod-level and chip-level parameters.
At the pod level (\texttt{vPodConfig}), a \vpod{} specifies the chip count and ICI topology.
% It can represent any physical NPU pod that is organized in a 2D/3D torus (a popular topology for NPU clusters~\cite{tpucloud,aws_trainium}).
% In practice, each dimension of the torus is typically a power of 2 to minimize resource fragmentation.
At the chip level (\texttt{vChipConfig}), it contains the on-chip and off-chip memory capacities and key performance parameters, including peak FLOPS, HBM bandwidth, and ICI bandwidth.
% The \texttt{version} field tracks which physical NPU chip version this \vpod{} is mapped to.
% After the \vpod{} is instantiated, this field can be queried by the ML framework and compiler to perform device-specific optimizations and generate the binary program.
The \texttt{version} field tracks which physical NPU chip version this \vpod{} is mapped to,
allowing ML frameworks to perform device-specific optimizations.
% which allows the ML framework and compiler to perform device-specific optimizations.
% It can be queried by the ML framework and compiler to perform device-specific optimizations.

All NPU chips in a single \vpod{} are homogeneous. {This aligns with the organization of physical NPU pods, as different NPU versions have incompatible ICI links and will not be mixed in the same physical pod.}
% and \pname{} manages heterogeneous NPUs by creating \vpod{} instances with different configurations.
This also offers the best compatibility with existing ML frameworks and compilers, which simplifies model deployment.
% For users who wish to manually split their workload on heterogeneous NPUs, they can create multiple \vpod{}s.
% For example, LLM prefill and decode have different resource demands~\cite{DistServe:osdi24}, so they can be deployed on \vpod{}s of different configurations.

\begin{figure}[t]
    \centering
    % \includegraphics[width=\linewidth]{figures/npureuse-mapping.drawio.pdf}
    % \vspace{-6.9ex}
    \includegraphics[width=\linewidth]{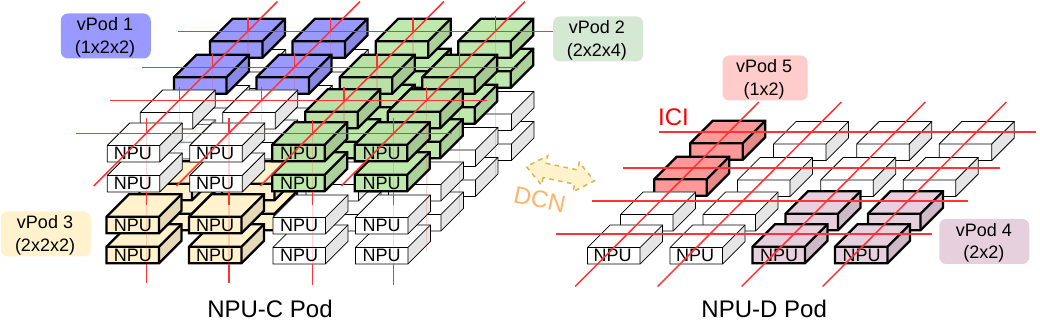}
    % \vspace{-3.9ex}
    \caption{Mapping of \vpod{}s to different slices of physical NPU pods. NPU-C supports 3D torus, and we draw a \texttt{4x4x4} slice as an example. An NPU-C pod can scale to \texttt{16x16x24}. NPU-D supports 2D torus, and the largest pod is \texttt{16x16}. Each NPU chip is connected to its neighbors via ICI links. {Different NPU versions have incompatible ICI links, so they are in different NPU pods and must communicate over DCN.} 
    %For clarity, we do not draw all ICI links.
    }
    \label{fig:vpod_mapping}
    % \vspace{-4ex}
\end{figure}

\noindent
\textbf{\vpod{} deployment.}
To create a \vpod{}, \pname{} submits \vpod{} allocation requests to the NPU cluster manager (e.g., Borg Scheduler~\cite{borg:google} or Kubernetes kube-scheduler~\cite{kubernetes_scheduler}), which finds the available NPU resources to map this \vpod{}.
{To minimize the chance of fragmentation, {\pname{}} restricts the {\vpod{}} topology to powers of two in each dimension, which aligns with the underlying physical NPU pods~\cite{tpuv4:nsdi24,tpuv4:isca23}},
% \Cref{fig:vpod_mapping} shows an example of how \vpod{}s can be mapped to the physical NPUs.
as shown in \Cref{fig:vpod_mapping}.
{Each {\vpod{}} is mapped to a dedicated mesh of NPU chips, which provides physical isolation for compute, memory, and ICI.
% If the \vpod{} creation fails, \pname{} can automatically fail over to another \vpod{} configuration (\S\ref{sec:design:scheduling}). 
% At runtime, {\pname{}} monitors the health of all NPU chips. Upon a failure, it evicts the corresponding {\vpod{}} and submits a new allocation request to recreate it.
}
After \vpod{} creation, \pname{} initializes LLM inference engine (e.g., vLLM~\cite{vllm}, JetStream~\cite{jetstream}). % to load the model weights onto the NPU chips.

\subsection{Best-Fit \vpod{} Allocation}
\label{sec:design:allocation}

% As discussed in \S\ref{sec:bkg:study}, different ML workloads and inference requests have different ``best-fit'' \vpod{} configurations.
% As discussed in \S\ref{sec:bkg:study}, to identify the most energy/cost efficient \vpod{} configuration for a workload, we need to identify its bottlenecking resource \textbf{(T1)}.
% The bottlenecking resource is determined by multiple factors: (1) the \vpod{} configuration, (2) the model architecture, (3) the input/output sequence length, (4) the model partitioning across the NPU chips in the \vpod{} (e.g., data, tensor, pipeline, and expert parallelisms~\cite{megatronlm,gshard}), and (5) the batch size supported by this \vpod{}.
% These factors are interdependent: a workload that is compute-bound on a \vpod{} (e.g., with an older NPU version) may become memory-bound on another \vpod{} (e.g., with a newer NPU version that features higher FLOPS).

To allocate a \vpod{} for an LLM model, we need to determine (1) the \vpod{} configuration, (2) the LLM parallelism configuration, and (3) the maximum batch size.
These parameters are interdependent: the \vpod{} configuration will limit how the model can be partitioned; parallelism configuration and batch size determine the computation task on each chip, which significantly affects performance and energy/cost efficiency. % (computation, memory, and network) 
We must co-optimize all factors.
% to achieve the best energy/cost efficiency while satisfying the SLO constraint.

% To address this challenge, a na\"ive way is to profile all possible allocations~\cite{griggs2024melange}, which is prohibitively expensive given the large number of configurable parameters. An alternative way is to build dedicated analytical models~\cite{DistServe:osdi24,LLMCompass} to estimate the end-to-end execution time, energy, and monetary cost of each allocation.
% These tools require significant engineering efforts to develop and are often specialized for dedicated ML models and accelerator architectures; they are typically not generalizable to the evolving workloads and NPU hardware.

Our key insight is that the energy/cost efficiency is determined by a workload's bottlenecking resource \textbf{(T1)}, and we can identify it by learning the workload's arithmetic intensity (i.e., FLOPs per byte accessed from HBM). %, which characterizes the compute-to-memory resource demand ratio.
Based on this, we can use a lightweight roofline model~\cite{roofline} to predict the energy/cost efficiency and the execution time of an allocation.
With these predictions, we can find all Pareto-optimal allocations that represent the optimal trade-offs between execution time and energy/cost efficiency, and pick the most efficient allocation that satisfies the SLO.
% \hlB{While the roofline model may be inaccurate for exact numerical estimations (e.g., due to complex memory access patterns), it can reliably capture the relative performance and efficiency trends across millions of allocations.}
% Compared to existing profiling-based methods~\mbox{\cite{dynamollm:hpca25,griggs2024melange}}, which incur high profiling overhead, our roofline model can capture the performance trend given hardware specifications and identify best-fit ones with low profiling cost.

% \hlcommon{Recent studies used profiling to find the optimal configuration for GPUs~\mbox{\cite{dynamollm:hpca25,griggs2024melange}} (see \mbox{\Cref{tab:related_works}}).
%However, they rely on intensive system profiling of every instance configuration (e.g., GPU version, clock frequency, and LLM parallelism degrees), sequence length, and batch size to build performance models for learning the optimal allocations, which incurs prohibitive overhead as we scale to hundreds of NPU chips per instance and over 100K sequence length.}

% This approach avoids expensive profiling while achieving high accuracy compared to the ideal allocation.
% This approach avoids expensive profiling or building complex, dedicated analytical models, while achieving high accuracy compared to the ideal allocation.
% The arithmetic intensity is sufficient for us to predict the ``best-fit'' allocation.

\noindent
\textbf{Learning the arithmetic intensity with ML compiler.}
The arithmetic intensity depends on the model architecture, parallelism degrees, sequence length, batch size, and compiler optimizations.
To systematically cover these factors, we extend the XLA compiler~\cite{xla,hlo-opt} with three analysis passes to perform a fast design space exploration (DSE).
% over all feasible \vpod{} allocations.
% We extend the XLA compiler~\cite{xla,hlo-opt} by adding three analysis passes.
%Given a model graph and a sample request, \hlC{the compiler outputs a mapping from candidate {\vpod{}} allocation parameters (i.e., NPU version, pod shape, LLM parallelism degrees, and batch size) to their arithmetic intensities.}\revnoteReverse[-1em]{\#C1}

First, the \textit{\vpod{} configuration generation pass} enumerates all candidate \vpod{} configurations {(i.e., the NPU chip version and the pod shape)}. 
To reduce the search space size, we limit the number of chips per pod to 1024 (which is typically sufficient for serving a single LLM model replica). % which yields 10 \vpod{} configurations per NPU version.

Second, for each \vpod{} configuration, the \textit{parallelism configuration pass} determines all valid combinations of parallelism degrees (e.g., tensor/pipeline/expert parallelisms). % {of the LLM on that \vpod{}, based on the pod shape and the per-chip HBM capacity}.
% For each combination, this pass also generates per-chip computation graphs for all batch sizes that are
For each combination, it also generates per-chip computation graphs for batch sizes (bs) that are
power-of-2 (1$\leq$bs$<$16) or multiple-of-16 (16$\leq$bs$\leq$1024)~\cite{vllm} by default.
% power-of-2 (from 1 to 16) or multiple-of-16 (larger than 16), up to 1024 by default~\cite{vllm}.
These batch size buckets are sufficient to cover most request patterns and are user-configurable.
% and we also allow users to configure them based on their workload requirements.
% Graphs that exceed the HBM capacity are filtered out.
For each graph, we apply standard graph- and operator-level optimizations (e.g., operator fusion, tiling) to ensure an accurate estimate of the arithmetic intensity.

Finally, the \textit{arithmetic intensity analysis pass} invokes high-level operation (HLO) cost analysis~\cite{hlo-cost-analysis} on each operator to extract its FLOP count and HBM traffic.
% It outputs the arithmetic intensity and total FLOP count for each graph, which
They are fed into the roofline model for performance, energy, and cost predictions.

\noindent
\textbf{Predicting performance and efficiency with roofline model.}
% characterize the NPU chip's hardware capability and
% We extend the roofline model~\cite{roofline} to predict the performance and efficiency of an allocation based on its arithmetic intensity.
% We fundamentally rethink the roofline model as a tool for quantifying the theoretical work-done-per-cost, where ``work done'' is the total FLOPs and ``cost'' is an arbitrary metric, such as time (seconds), energy (joules), or monetary price (dollars).
We rethink roofline models as tools for quantifying work-done-per-cost, where ``work done'' is total FLOPs and ``cost'' is an arbitrary metric, such as time, energy, or monetary price.
We use a roofline model because it is lightweight and can accurately capture the performance trend based on the hardware specifications and model architecture with low overhead. 
%  (see Figure~\mbox{\ref{fig:roofline_model_accuracy}} and ``Allocation time overhead'' at the end of \mbox{\S\ref{sec:design:allocation}})
We do not use profiling-based approaches~\mbox{\cite{dynamollm:hpca25,griggs2024melange}}, as they incur high performance overhead due to the need to profile a large number of \mbox{\vpod{}} configurations (every combination of batch size, sequence length, LLM parallelism configuration, and NPU pod shape needs to be profiled).
% Compared to existing profiling-based methods~\mbox{\cite{dynamollm:hpca25,griggs2024melange}}, which incur high profiling overhead, our roofline model can capture the performance trend given hardware specifications and identify best-fit ones with low profiling cost.

% \hlcommon{Recent studies used profiling to find the optimal configuration for GPUs~\mbox{\cite{dynamollm:hpca25,griggs2024melange}} (see \mbox{\Cref{tab:related_works}}).
%However, they rely on intensive system profiling of every instance configuration (e.g., GPU version, clock frequency, and LLM parallelism degrees), sequence length, and batch size to build performance models for learning the optimal allocations, which incurs prohibitive overhead as we scale to hundreds of NPU chips per instance and over 100K sequence length.}

\begin{figure}[t]
    \centering
    \includegraphics[width=\linewidth]{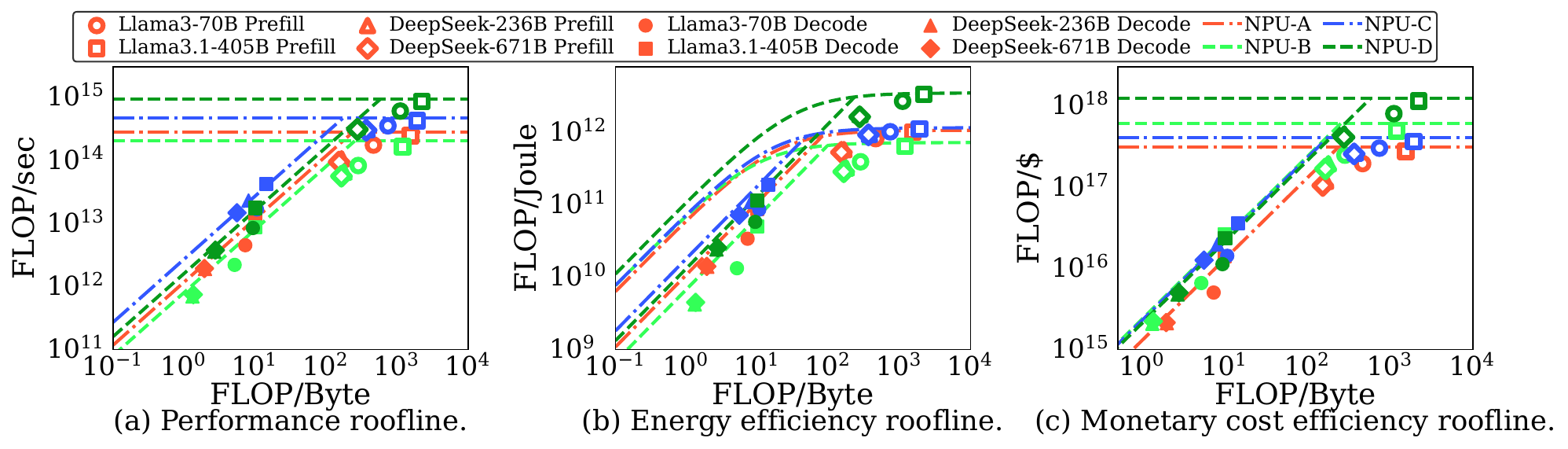}
    % \vspace{-6ex}
    \caption{Cost-efficiency roofline models with different cost metrics. Marker shapes represent LLM models. Colors represent NPU versions. We only plot input/output sequence length 4K/512 due to space limitations.}
    \label{fig:roofline_models}
    % \vspace{-4.5ex}
\end{figure}

\begin{figure}[t]
    \centering
    \includegraphics[width=\linewidth]{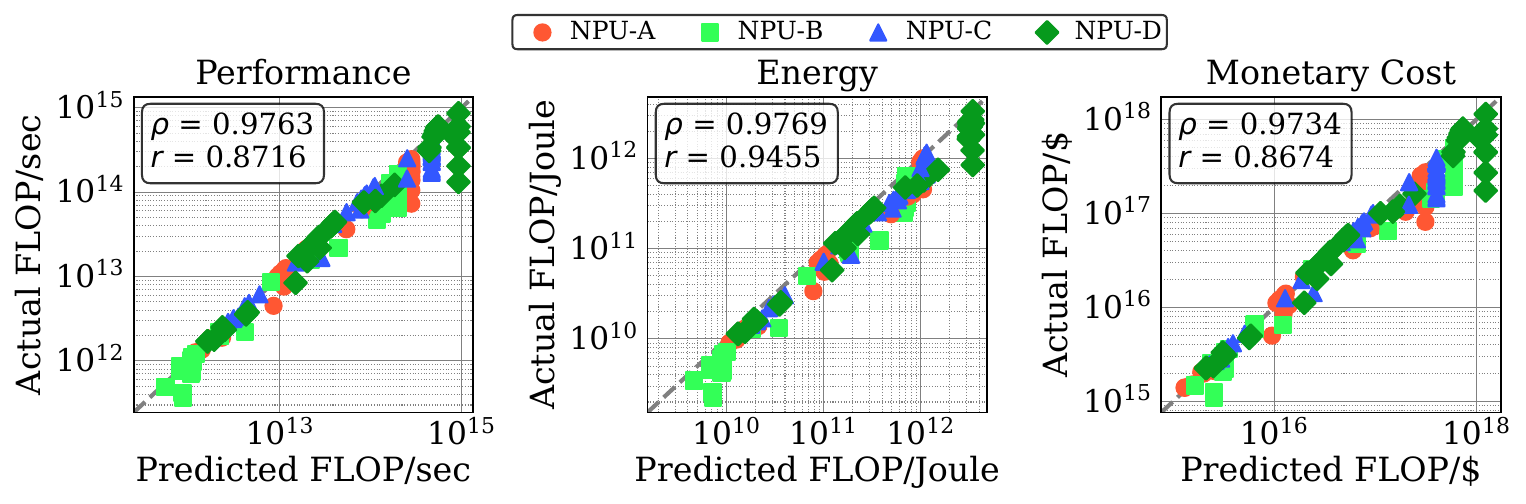}
    % \vspace{-3.6ex}
    \caption{Roofline prediction accuracy for Llama3.1-405B and input/output sequence lengths 4K/512. $\rho$ is Spearman's rank correlation. $r$ is Pearson correlation. Each point represents the predicted/actual value of an allocation.}
    \label{fig:roofline_model_accuracy}
    % \vspace{-4ex}
\end{figure}

\begin{figure}[t]
    \centering
    \includegraphics[width=\linewidth]{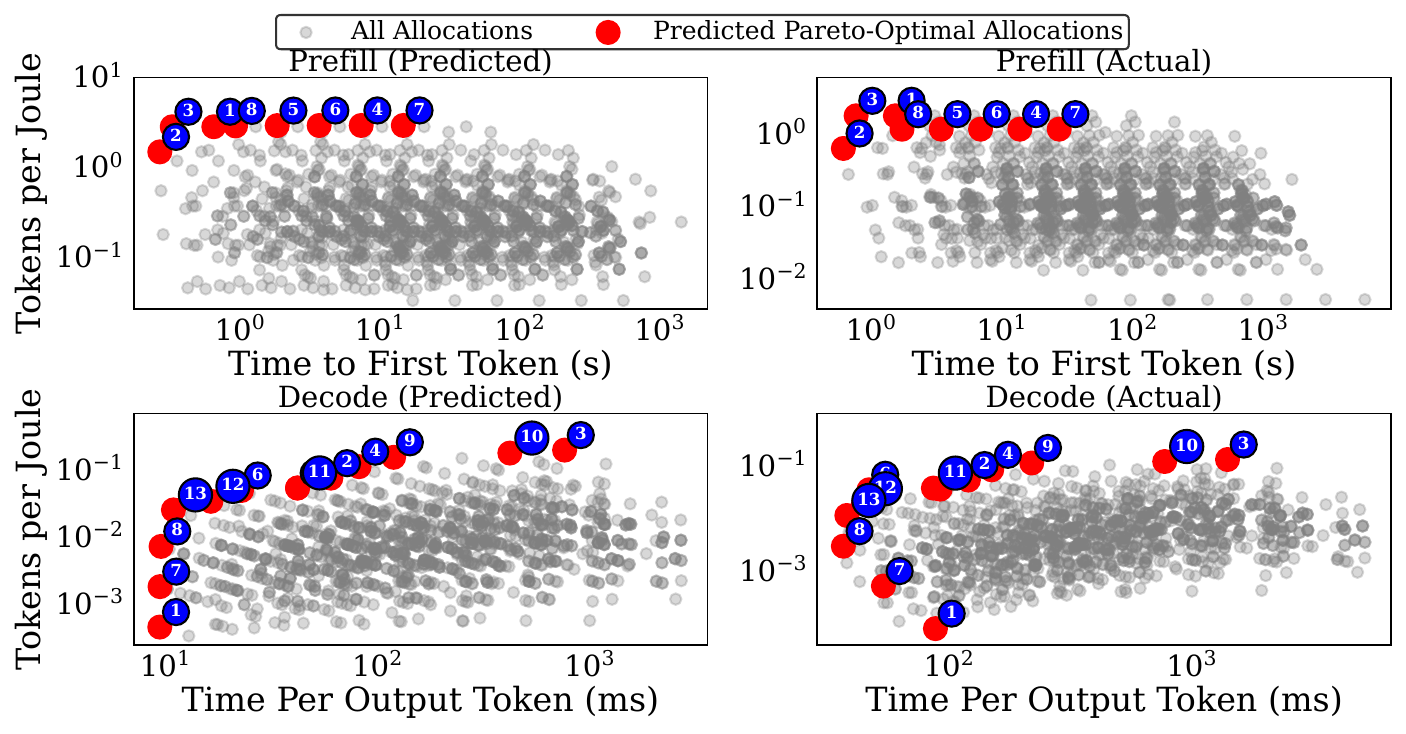}
    % \vspace{-3.6ex}
    \caption{Predicted vs. actual latency and energy efficiency of all allocations. The points with the same label in the predicted and actual graphs correspond to the same allocation. We only plot DeepSeekV3-671B with input/output sequence length 4K/512 due to space limitations.}
    \label{fig:roofline_pareto}
    % \vspace{-3.5ex}
\end{figure}

\begin{figure}[t]
    \centering
    \includegraphics[width=\linewidth]{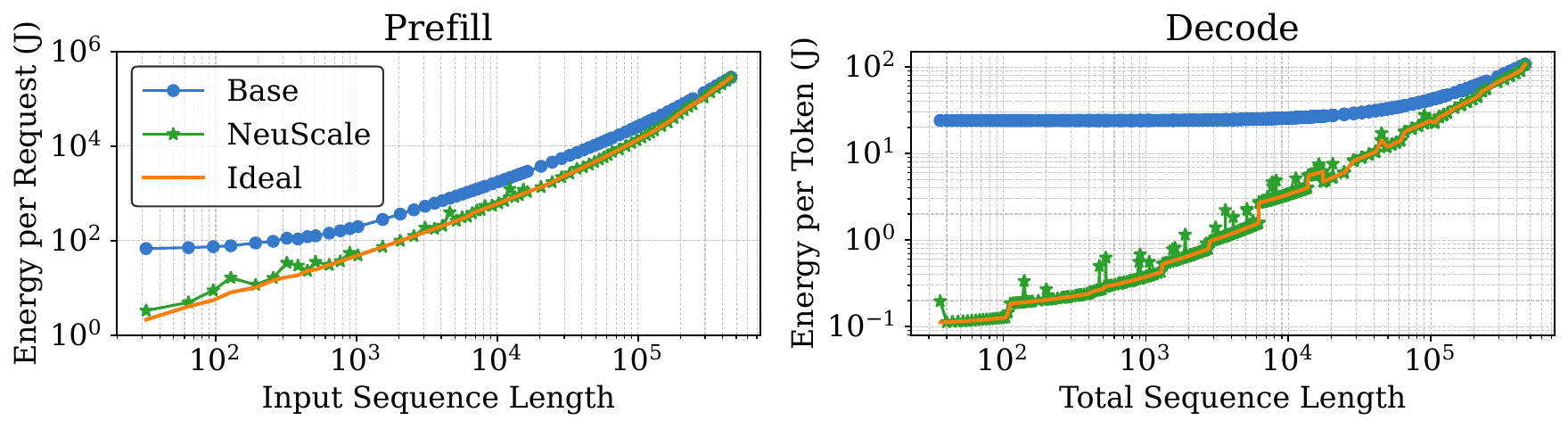}
    % \vspace{-3.6ex}
    \caption{Allocation accuracy (Llama3-70B). \texttt{Base} uses the best-fit allocation for the largest sequence length in order to guarantee SLO (see \S\ref{sec:eval:setup}).}
    \label{fig:eval_alloc_accuracy}
    % \vspace{-2ex}
\end{figure}

\Cref{fig:roofline_models} shows the roofline models for different cost metrics.
% Each roofline model characterizes an NPU version using a 2D graph. The X-axis is the arithmetic intensity in FLOP/byte.
% % This captures an intrinsic property of a workload that is, in general, independent of the hardware it runs on.
% The Y-axis is the efficiency in FLOP/cost.
An NPU chip is represented by two rooflines: the horizontal line is the peak computational efficiency (FLOP/cost); the sloped/curved line is the peak HBM bandwidth efficiency (byte/cost).
{These rooflines are derived from the hardware specifications provided by the vendor (e.g., peak FLOPs, memory bandwidth, and energy efficiency) and the instance pricing from the cloud platform.}
% \Cref{fig:roofline_models}a is the conventional performance roofline model using time (seconds) as the cost metric.
The energy efficiency roofline (\Cref{fig:roofline_models}b) is a curve instead of a straight line.
This is because for each FLOP, we need to fetch data from HBM, so the peak FLOP/Joule also depends on HBM accesses. 
The monetary cost roofline (\Cref{fig:roofline_models}c) is based on dollars per chip-hour, following the pay-as-you-go model~\cite{pay-as-you-go} in the cloud, and hence proportional to the performance roofline (\Cref{fig:roofline_models}a).

% As a result, the monetary cost efficiency roofline is a scaled version of the performance roofline based on the \$/chip-hour rate.

{For each candidate {\vpod{}} allocation generated by the ML compiler, we can plot it as a vertical line at its arithmetic intensity.}
The intersection of this vertical line with the roofline predicts the performance/efficiency.
% for that allocation on the specific NPU version.
\pname{} uses the performance roofline to predict the request latency, and the energy/monetary cost roofline to predict the cost.
% \hlF{As ICI communications only occur in a few operators (e.g., collective operators like reduce-scatter and all-gather), we model them separately.
% %\footnote{\hlF{For custom, human-written kernels that may overlap computation with collective communications, the standard practice in modern ML compiler workflows is to integrate custom cost models\mbox{\cite{tvm,xla}}.}}. 
% We calculate the ICI latency and cost of these operators by dividing the ICI traffic by the ICI link bandwidth or efficiency. The ICI energy is directly added to the roofline energy prediction of the request. The ICI latency and monetary cost of an operator are added only if ICI latency is higher than the roofline compute and memory latency of this operator.}
For the cost metric, first-party cloud platforms may prioritize energy efficiency, as electricity bills are highly correlated to TCO~\cite{carbon_life_cycle}.
Third-party users who rent \vpod{}s may optimize monetary costs, as instances are billed per chip-hour.

\noindent
\textbf{Generating the best-fit \vpod{} allocations.}
{\pname{}} takes the performance and cost of all candidate {\vpod{}} allocations and identifies the Pareto-optimal ones.
{\pname{}} picks the most energy/cost-efficient allocation that satisfies the SLO constraint of the sample input request.

To generate the best-fit {\vpod{}} allocations for all representative sequence lengths, {\pname{}} performs offline DSE with sample requests grouped into sequence length buckets.
This only needs to be performed once, and the results are shared across all workloads that use the same LLM model.
The generated best-fit database size is proportional to the number of sequence length buckets and the number of NPU versions.
By default, {\pname{}} creates sequence length buckets with 128-token increments up to 1024 tokens. Beyond that, we double the increment at each power-of-2 boundary, e.g., 256-token increments up to 2048 and 512-token increments up to 4096~\cite{vllm}. With a maximum of 1M sequence length, this yields $48\times 2 =96$ (total for prefill and decode) buckets. For each bucket, the database stores a best-fit \vpod{} configuration (i.e., all fields in \mbox{\Cref{fig:vpod_abstraction}}, plus metadata such as the maximum batch size, LLM parallelism degrees, and the estimated throughput/latency, taking up to 128 bytes) for each NPU version. With 4 NPU versions, the database is only 48 KB per model, and the query latency is negligible ($\leq$20$\mu$s).

\noindent
\textbf{Roofline prediction accuracy.}
\Cref{fig:roofline_model_accuracy} shows the accuracy of our roofline models.
% The Spearman correlation $\rho$ is higher than 0.97 ($\rho=1$ indicates the predictions preserve all relative ranking)
The Spearman correlation is $\rho>$0.97 ($\rho=1$ means predictions preserve all relative ranking),
% despite the slightly lower Pearson correlation $r$ ($r=1$ implies a perfect linear relationship).
despite a slightly lower Pearson correlation $r$ ($r=1$ implies a perfect linear relationship).
% s, implying that the predicted allocation is always the best-fit
% This implies that if an allocation is predicted to be better than another allocation, it is highly likely to be actually better.
This implies that if an allocation is predicted to be better than another, it is highly likely to be actually better, which is sufficient for DSE.
% that the actual performance/efficiency is better.
% This is sufficient for DSE, even if the predicted absolute values are not perfectly accurate.
% This is confirmed in \Cref{fig:roofline_model_accuracy}. which shows the prediction accuracy for the cost efficiency of ML models running different input request sizes on different NPU versions.

\noindent
\textbf{\pname{} allocation accuracy.}
% We first examine whether \pname{} can accurately capture the Pareto-optimal allocations.
% shows the comparison between the predicted and actual Pareto curve.
% Because our roofline model reliably preserves the relative ranking of different allocations, 
\Cref{fig:roofline_pareto} shows the predicted Pareto-optimal allocations (left) and their actual latencies and energy efficiencies (right).
The predicted Pareto curve covers all actual Pareto-optimal allocations despite some false positives. 
%which can be filtered out by a few profiling runs.
% we have manually inspected the allocations on the predicted and actual Pareto curves to confirm that they are the same despite the inaccuracies of the predicted absolute values.
In practice, testing an allocation on real hardware is necessary to guarantee SLO and is a common safeguard before deployment~\cite{google:mlops,todd:sre_for_ml}.
% deploying production-level ML services
% % Hence, our goal is to minimize the profiling cost by testing as few candidate allocations as possible.
As we only need to test a few Pareto-optimal allocations, we can minimize this testing cost.
% In most cases, the allocations on the predicted and actual Pareto curves are the same despite the inaccuracies of the predicted absolute values.
% \pname{} only needs to profile up to 11 allocations on the Pareto curve.
% Furthermore, as \pname{} correctly ranks all NPU versions by their cost efficiencies, it facilitates fail-over allocation \textbf{(T3)}: when the most efficient NPU version is unavailable, we can choose the next NPU version with minimal cost efficiency degradation.
\Cref{fig:eval_alloc_accuracy} shows the end-to-end energy efficiency for different allocation policies (see the detailed setup in \S\ref{sec:eval:setup}). \pname{} successfully finds the best-fit allocation for most sequence lengths and achieves near-ideal energy efficiency.

% \noindent
% \textbf{\pname{} allocation accuracy.}
% We analyze representative ML workloads with our roofline model in \Cref{fig:roofline_models}.
% For most workloads, we correctly predict the best-fit NPU version, even though the predicted absolute cost efficiency value is often higher than the actual value.
% This is because the roofline prediction largely preserves the relative cost-efficiency rankings between different NPU versions.
% This is confirmed in \Cref{fig:roofline_model_accuracy}. which shows the prediction accuracy for the cost efficiency of ML models running different input request sizes on different NPU versions.
% The Spearman correlation $\rho$ is consistently higher than 0.97 ($\rho=1$ indicates the predictions perfectly preserve ordering, implying that the predicted best-fit NPU version is always correct), despite the slightly lower Pearson correlation $r$ ($r=1$ indicates a perfect linear relationship).
% As our method correctly ranks all NPU versions by their cost efficiencies, it facilitates fail-over allocation \textbf{(T3)}: when the most efficient NPU version is unavailable, we can choose the next NPU version with minimal cost efficiency degradation.

\noindent
\textbf{Allocation time overhead.}
The analysis passes take up to 40 seconds on a Cloud TPU VM (120 AMD EPYC 7B12 CPU cores) for each sharded graph for the largest DeepSeekV3-671B model.
% This is expected as we skipped the auto-tuning passes and the backend LLVM optimizations and code generation, which are the major time-consuming components in an ML compiler~\cite{t10:sosp24,roller:osdi2022,ansor}.
In practice, the entire DSE can finish in 10 minutes for each sequence length.
%The overhead scales linearly with the number of NPU versions. 
This is acceptable, as the DSE can be parallelized across multiple servers, and performed offline. 
The results are saved in a database and shared by all services that use the same LLM.

% the \vpod{} topology and each NPU chip's memory capacity will limit how the model can be partitioned, as well as the maximum batch size; the model partitioning strategy and the batch size determine the resource demand (computation, memory, and network) on each chip.
% We must co-optimize all factors to achieve the best energy/cost efficiency while satisfying the SLO latency constraint.

% First, a workload's bottleneck can be different across different NPU versions. A workload that is compute-bound on an older NPU version may become memory-bound on a newer NPU version that features higher FLOPS.

% Second, the bottleneck resource is affected by many factors, including both NPU hardware configurations (NPU chip type and pod topology) and ML inference engine configurations~\cite{vllm,tensorrt-llm} (model sharding strategies, parallelism degrees, maximum batch sizes).
% The resulting combinatorial search space can consist of tens of millions of design choices for each inference request size for each ML model.

% Third, the best-fit NPU version depends on the optimization goal. For instance, performance (latency) is dictated only by the bottlenecking resource, while total energy must account for both bottlenecking and non-bottlenecking resources.

% why choose arithmetic intensity as the workload characteristic?

% End-to-end workflow

% 1. how to extract workload characteristic (arithmetic intensity)

% 2. roofline model for cost efficiency

% 3. save for future query

% 4. accuracy evaluation

\subsection{Dynamic \vpod{} Allocation Adjustment}
\label{sec:design:scheduling}

% \begin{figure}[t]
%     \centering
%     \includegraphics[width=0.8\linewidth]{figures/npureuse-autoscaling_controller.drawio.pdf}
%     \caption{\pname{} auto-scaling control feedback loop. The left side lists the runtime statistics used for auto-scaling.}
%     \label{fig:autoscale_controller}
% \end{figure}

\begin{figure}[t]
    \centering
    % \includegraphics[width=0.9\linewidth]{figures/npureuse-vpod_autoscale_process.drawio.pdf}
    % \vspace{-3ex}
    \includegraphics[width=\linewidth]{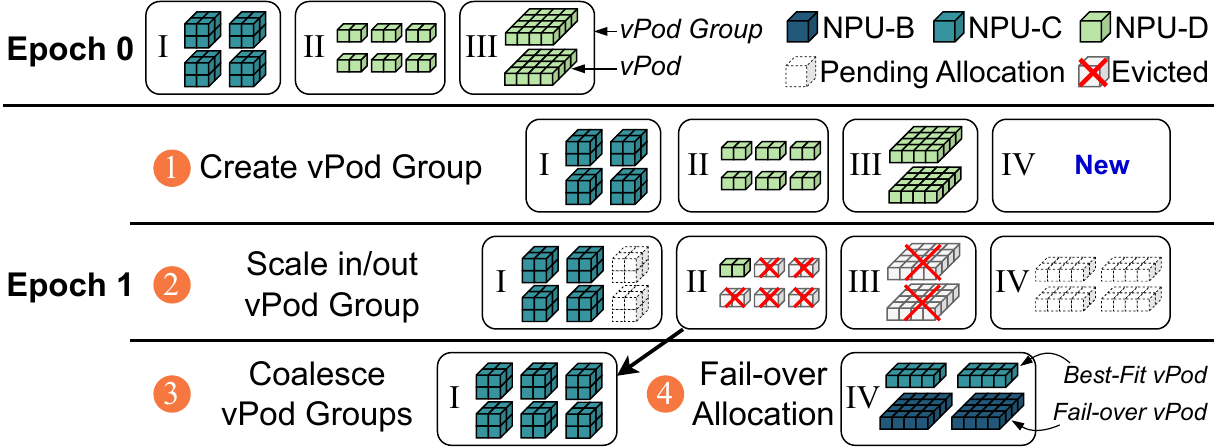}
    % \vspace{-3.2ex}
    \caption{\vpod{} auto-scaling process in \pname{}.}
    \label{fig:autoscale_process}
    % \vspace{-4.5ex}
\end{figure}

At runtime, \pname{} organizes \vpod{}s into \textit{\vpod{} groups}. All \vpod{}s in the same group have the same configuration, and each group is the best fit for specific sequence lengths.
% \Cref{fig:autoscale_controller} shows the control loop of \pname{}.
The \textit{\vpod{} monitor} tracks the runtime statistics of all \vpod{}s in the last time window (30 minutes by default). At every epoch (5 minutes by default), it reports the statistics to the \textit{\vpod{} allocation recommender}, which adjusts the number of \vpod{} groups and the number of \vpod{}s in each group.
Globally, \pname{} tracks the sequence lengths of all requests in the last time window.
For each \vpod{}, \pname{} collects its incoming request rate as a time series of measurements sampled every 10 seconds.
% The per-\vpod{} request rate is aggregated into the time-series request rate for the entire \vpod{} group.
\Cref{fig:autoscale_process} shows the end-to-end auto-scaling process.

%\hlC{In each epoch, {\pname{}} first determines the target {\vpod{}} allocations, including the {\vpod{}} groups and the {\vpod{}} count of each group (steps \mbox{\circleo{1}} to \mbox{\circleo{3}}). Then, it submits {\vpod{}} creation and eviction requests to the NPU cluster manager to adjust the current allocation. If an allocation fails, it allocates fail-over {\vpod{}}s or triggers additional evictions (step \mbox{\circleo{4}}).}\revnote[-4em]{\#C2\\\#C4}
% \hl{Yikang: add something to give readers the sense that steps 1-4 are conducted sequentially.}

\noindent
\circleo{1} \textbf{Creating a \vpod{} group.}
The recommender queries the DSE module (\S\ref{sec:design:allocation}) to determine the best-fit \vpod{} allocation for each unique sequence length. It has negligible overhead if the allocations are already generated offline.
Otherwise, the recommender picks the best-fit allocation for the closest sequence length in the database and triggers the background DSE for the new sequence length.
\pname{} compares the new set of \vpod{} allocations with existing \vpod{} groups to determine if a new group needs to be created.

\noindent
\circleo{2} \textbf{Scaling in/out a \vpod{} group.}
\pname{} determines the \vpod{} count in a group ($N_{pod}$) based on the peak request rate ($R_{peak}$) in the last time window that would be served by this group and the maximum throughput of each \vpod{} ($T_{pod}$): $N_{pod} = \left\lceil{R_{peak}}/{T_{pod}}\right\rceil$.
% where $t$ is a user-configurable utilization target (0.6 by default)~\cite{kubernetes_hpa}.
\pname{} evicts a \vpod{} group if $N_{pod}$ is scaled down to 0.
% To avoid thrashing, the number of \vpod{}s can be at most doubled in every epoch and halved every 30 minutes (user-configurable).
To avoid thrashing, the \vpod{} count can at most double in every epoch and halve every 30 minutes (user-configurable).
{To handle bursty workloads, {\pname{}} monitors the peak aggregated queue size of each {\vpod{}} group. If this size exceeds the group's aggregated maximum batch size,
% which indicates severe queue backlogging,
{\pname{}} triggers an emergency {\vpod{}} allocation adjustment. This reactive scale-out bypasses the doubling constraint to quickly absorb workload spikes. To minimize underutilization once a burst diminishes, we apply an exponentially weighted moving average to the $R_{peak}$ signal.
% , allowing the system to scale in rapidly. % based on more recent request rates
}
% \hlcommon{how to handle burst: peak aggregated queue size all \vpod{}s in a group, if greater than aggregated max batch size of all \vpod{}s, this indicates queue backlogging. immediately triggers the \vpod{} allocation adjustment process. this emergency scale-out is not limited by the double count constraint. We apply ewma to the peak request rate signal in the last time window, so that as the burst ends and gets absorbed/processed, we can quickly scale in based on the most recent request rate.}
% \pname{} scales the number of \vpod{}s in each group to avoid overloading or underutilization.
% It checks all collected statistics and makes the most conservative scaling decision to guarantee SLO.
% First, at each epoch, \pname{} scales the \vpod{} group size to $N_{rr}$ (\Cref{eqn:hpa_req_rate}) to ensure the group can accommodate the peak load.

\noindent
\circleo{3} \textbf{Coalescing underutilized \vpod{} groups.}
When the sequence lengths are sparse, each request may have a different best-fit \vpod{} allocation, leading to underutilization if a separate \vpod{} group is created for each request. To mitigate this, \pname{} merges a \vpod{} group into the group serving the next larger sequence length if the \vpod{} group's normalized peak throughput $N_L = R_{peak}/T_{pod}$ falls below a configurable threshold (0.5 by default), and the target \vpod{} group has sufficient spare capacity to absorb all its requests without allocating a new \vpod{}. \pname{} repeatedly scans \vpod{} groups in ascending order of $N_L$ and performs merges until no further \vpod{} groups can be merged.

\iffalse
When the sequence lengths are sparse, each request may have a different best-fit \vpod{} allocation, leading to severe underutilization if we na\"ively create a \vpod{} group for each request.
% This will lead to severe resource underutilization and energy/cost inefficiency if we na\"ively create a separate \vpod{} group for each request. 
% \pname{} selectively coalesces underutilized \vpod{} groups.
To mitigate this issue, \pname{} merges a \vpod{} group into another group that serves the closest sequence lengths if the peak request of this group is much lower than the maximum throughput of a single \vpod{}.
For each \vpod{} group, \pname{} computes its normalized peak load $N_L=R_{peak}/T_{pod}$.
$N_L<t$, where $t$ is a configurable coalescing factor (0.5 by default), indicates this \vpod{} group only has a single \vpod{}, and it is underutilized.
% and may be coalesced with another group.
For each \vpod{} group $G_s$ with $N_L<t$, \pname{} finds the \vpod{} group $G_t$ that serves the next larger sequence length than $G_s$. If $G_t$ has spare capacity to accommodate all requests from $G_s$ without provisioning a new \vpod{}, \pname{} will evict $G_s$ and route all its future requests to $G_t$.
{{\pname{}} iteratively examines underutilized groups in ascending order of $N_L$. If a group is coalesced, \pname{} updates the loads and restarts the search from the new lowest $N_L$; otherwise, it evaluates the group of the next lowest $N_L$. The process terminates when no remaining groups can be coalesced.}
\fi

\noindent
\circleo{4} \textbf{Fail-over \vpod{} allocation.}
When a best-fit \vpod{} cannot be allocated {(e.g., due to resource scarcity, fragmentation, or exceeded quota)}, \pname{} automatically fails over to the best-fit \vpod{} on another NPU version. %that has the minimal performance and efficiency degradation. %, until a \vpod{} can be successfully provisioned.
The fail-over \vpod{} belongs to the same group as the best-fit ones to simplify request routing and \vpod{} group scaling.
In each epoch, \pname{} will replace as many fail-over \vpod{}s as possible.
{If the fail-over {\vpod{}} allocation also fails (indicating resource scarcity or a quota limit), {\pname{}} will evict \vpod{}s from the group with the lowest $R_{peak}$ to provision the new \vpod{}.}% provided that more requests will be served by the new {\vpod{}} (i.e., the $R_{peak}$ of the new {\vpod{}} is higher than the evicted ones). This ensures that \pname{} can serve as many requests as possible with limited NPU resources.}

\noindent
\textbf{\pname{} request routing.}
\pname{} employs a BERT-based model to predict the output sequence length based on the input prompt~\cite{qiu2024seqlenprediction,tao2025seqlenprediction,piotrowski2025seqlenprediction}.
It routes requests to their best-fit \vpod{} groups with a sequence length-to-\vpod{} group mapping, which is updated when \vpod{} groups are created or evicted.
In each group, \pname{} employs optimizations such as load balancing and continuous batching~\cite{vllm}.
To handle output length mispredictions, \mbox{\pname{}} re-estimates the remaining output length of a request whenever its sequence length reaches the next bucket boundary (see \mbox{\S\ref{sec:design:allocation}} for the bucket definitions).
When either the new estimate or the current sequence length exceeds twice the best-fit sequence length for the current {\vpod{}}, \pname{} triggers a non-blocking request migration~\mbox{\cite{llumnix:osdi24}} to move the request to the \vpod{} group responsible for longer sequences.
The misprediction penalty and request migration overhead are tolerable (see our evaluation in \mbox{\S\ref{sec:sens_output_predict_accuracy}}).

\section{Methodology}
\label{sec:methodology}

\begin{table}[t]
    \centering
    \caption{System-level overhead of managing \vpod{}s, measured on TPU instances on Google Cloud Platform. *DCN BW varies across NPU versions (5--100 Gbps/chip, as listed in \mbox{\Cref{tab:npu_specs}}).}
        % \vspace{-3ex}
    % \footnotesize
    \setlength{\tabcolsep}{2pt}
    \begin{tabular}{lc}
        \toprule
        Source & Overhead \\
        \midrule
        Best-fit \vpod{} lookup & $\leq$20 $\mu$s \\
        Create a \vpod{} instance & 73--108 s \\
        Delete a \vpod{} instance & $\sim$15 s \\
        Download sharded model weights & DCN BW* \\
        NPU runtime \& LLM serving engine startup & $\sim$31 s \\
        % Load Model Weights to TPUs \& Allocate KV Cache & $\sim$15 s \\
        Load model weights to NPU chips & 24 GB/s/chip (PCIe 4.0 x16) \\
        % Initialize KV cache & $\sim$10 s \\
        % Output length prediction & 7.6 ms~\cite{qiu2024seqlenprediction} \\
        Query \vpod{} runtime statistics & 2--5 s \\
        \vpod{} allocation adjustment algorithm & $\sim$100 ms \\
        Request routing latency & 45.9 ms \\
        KV cache transfer between \vpod{}s & see \Cref{fig:vpod_dcn_data_transfer} \\
        % Total & 127--172 s \\
        \bottomrule
    \end{tabular}
       
    \label{tab:vpod_startup_overhead}
    % \vspace{-3ex}
\end{table}

% \begin{table}[t]
%     \centering
%     \setlength{\aboverulesep}{0pt}
%     \setlength{\belowrulesep}{1pt}
%     \caption{\hlA{{\vpod{}} system-level overhead of Llama3-70B on a TPUv4-64 instance, with tensor/pipeline parallelism degrees 8/4.}}
%         \vspace{-3ex}
%     \footnotesize
%     \begin{tabular}{lc}
%         \toprule
%         Overhead Source & Time \\
%         \midrule
%         Create a TPU Instance (8 VMs; 4 TPUv4 chips per VM) & 73--108 s \\
%         Download Sharded Model Weights (18--22 GB per VM) & 18--28 s \\
%         TPU Runtime \& LLM Serving Engine Startup & $\sim$21 s \\
%         Load Model Weights to TPUs \& Allocate KV Cache & $\sim$15 s \\
%         Total & 127--172 s \\
%         \bottomrule
%     \end{tabular}
       
%     \label{tab:vpod_startup_overhead}
%     \vspace{-3ex}
% \end{table}

\begin{figure}[t]
    \centering
    \includegraphics[width=\linewidth]{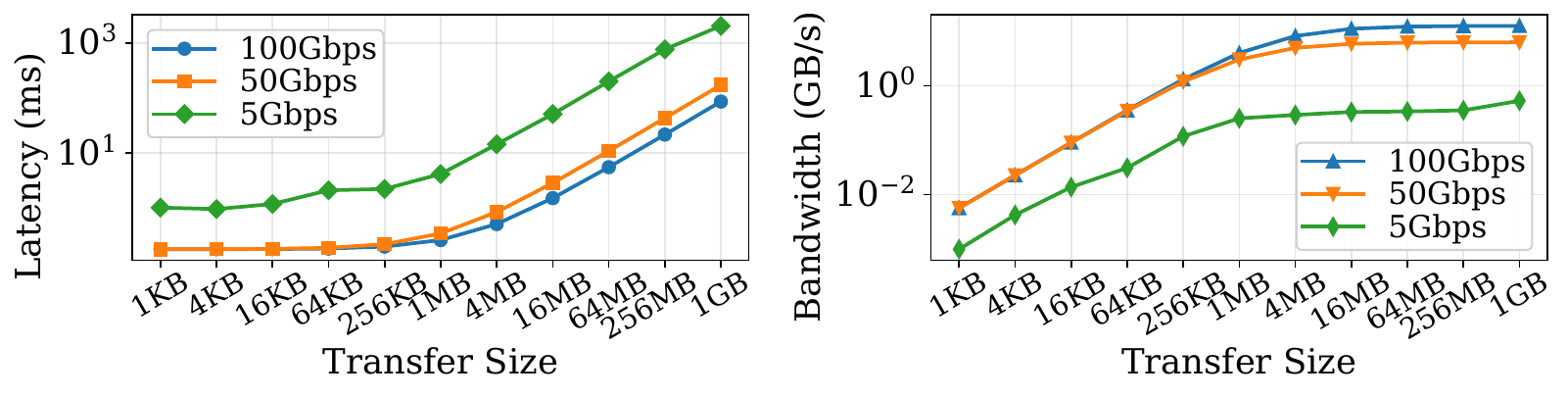}
    % \vspace{-5.8ex}
    \caption{{P2P data transfer latency and bandwidth between two chips over DCN, measured on real TPUs. The two chips are in different \vpod{}s.}}
    \label{fig:vpod_dcn_data_transfer}
    % \vspace{-3.5ex}
\end{figure}

\noindent
\textbf{Real system prototype.}
To validate our design, we developed a real system prototype and deployed it on a real TPU cluster consisting of 32 TPUv4, 64 TPUv5e, and 64 TPUv6e chips (three on-demand TPU instances: \texttt{TPUv4-64}, \texttt{TPUv5e-64}, and \texttt{TPUv6e-64}\footnote{{We did not have access to TPUv5p instances as they are scarce and in high demand. However, TPUv5p shares the same core architecture as TPUv5e.}}).
The offline DSE (\mbox{\S\ref{sec:design:allocation}}) is implemented as a standalone Python module. It first enumerates all \mbox{\vpod{}} configurations, sequence lengths, LLM parallelisms, and batch sizes for a model. Then, it invokes the XLA compiler to generate and compile all execution graphs with extra compiler flags to dump the intermediate HLO cost analysis pass results. Finally, it invokes the roofline-based allocation to generate the best-fit database.
The runtime auto-scaling controller (\mbox{\S\ref{sec:design:scheduling}}) lives in a dedicated CPU VM to manage all TPU VMs (a TPU instance has multiple TPU VMs and 4--8 TPU chips per VM). It bookkeeps the \mbox{\vpod{}}-to-TPU chip mappings and queries the runtime statistics from the vLLM~\mbox{\cite{vllm}} engines running in the TPU VMs. To (de)allocate a \mbox{\vpod{}}, it utilizes \texttt{ssh} to access the corresponding TPU VMs and launch/kill the vLLM~\mbox{\cite{vllm}} engines.
We extended the vLLM router~\mbox{\cite{vllm_router}} to support sequence length-based request routing and Llumnix-style non-blocking request migration~\mbox{\cite{llumnix:osdi24}}.

\noindent
\textbf{Simulator.}
To evaluate \pname{} at scale, we build a simulator to model the auto-scaling framework, including dynamic \vpod{} allocation and inter-\vpod{} request routing.
% \hlcommon{how we model \vpod{} startup time based on VM config and model weight size}
For each \vpod{}, it models common LLM inference optimizations such as continuous batching, KV cache swapping~\cite{vllm}, and request migrations across \vpod{}s in the same group for load balancing~\cite{llumnix:osdi24}. 
We model the major system-level overheads for auto-scaling, as shown in \mbox{\Cref{tab:vpod_startup_overhead}}.
We calibrated the overhead based on measurements on real TPU instances.
We compute model weight downloading and loading time based on the model size, parallelism configuration, and measured network download and PCIe bandwidths of TPU VMs.
% To account for VM provisioning and serving engine startup, we calibrated the simulator based on the measured TPU instance startup time in \mbox{\Cref{tab:vpod_startup_overhead}}. 
% The overhead of downloading model weights and loading them onto NPU chips will increase for larger models, but this is manageable since a larger model will require more chips to serve. 
%as a result, the aggregated DCN/PCIe speed also scales. 
%Other overheads remain similar for different \vpod{} configurations.
To model KV cache transfer overhead over DCN due to request migration and prefill/decode \vpod{} communication,
% due to request migration and prefill/decode \vpod{} communication over DCN,
we quantify the communication overhead in \mbox{\Cref{fig:vpod_dcn_data_transfer}} with different TPU versions (see DCN bandwidth in \mbox{\Cref{tab:npu_specs}}) and fit a link model~\mbox{\cite{ahead:2019,loggp:1995}} based on the measured numbers.

\begin{figure}[t]
    \centering
    \includegraphics[width=\linewidth]{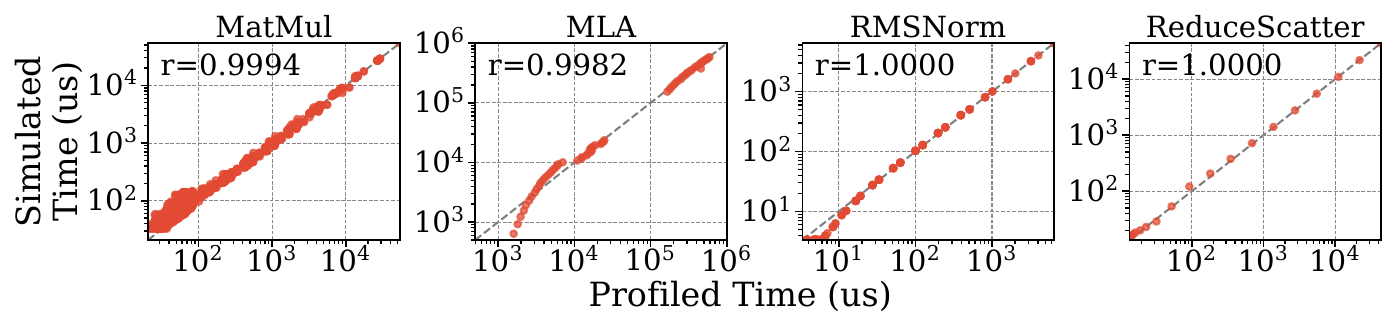}
    % \vspace{-3.7ex}
    \caption{{Simulated vs. profiled execution time of representative operators on a $2\times 4\times 4$ TPUv4 pod. Each data point corresponds to a different input tensor shape and tensor parallelism degree. $r$ is the Pearson correlation.}}
    \label{fig:sim_validate_operator}
    % \vspace{-3.4ex}
\end{figure}

\begin{figure}[t]
    \centering
    \includegraphics[width=\linewidth]{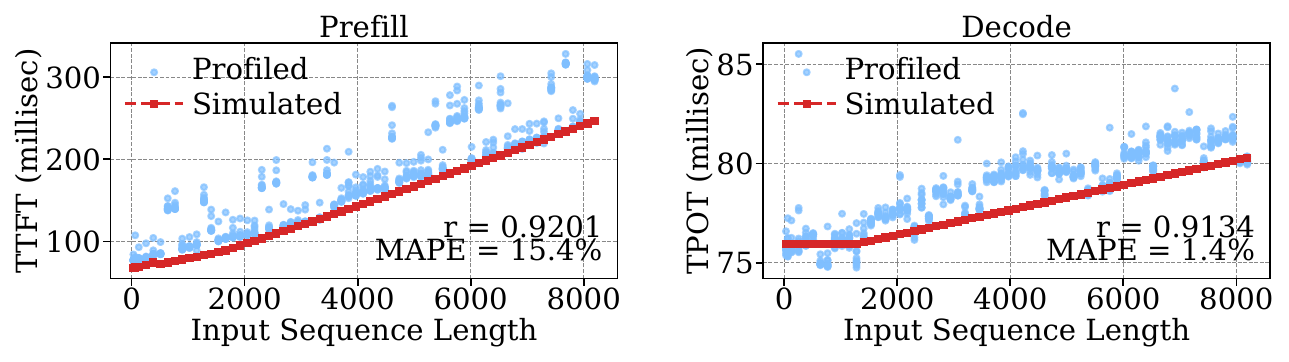}
    % \vspace{-3.5ex}
    \caption{Simulated vs. profiled latency of a single LLM inference request with various sequence lengths.
% Each sequence length is tested 10 times, and all profiled samples are plotted.
$r$ is Pearson correlation. MAPE is the mean absolute percentage error. We show the results of Llama3-70B on a $2\times 4\times 4$ TPUv4 pod.}
    \label{fig:sim_validate_seqlen}
    % \vspace{-3ex}
\end{figure}

% The KV cache transfer between prefill/decode \vpod{}s is overlapped with prefill computation\mbox{\cite{splitwise:isca24}}. The request migration is non-blocking\mbox{\cite{llumnix:osdi24}}. Hence, the KV cache transfer overhead is minimal ($<$5\%) compared to the end-to-end request latency.}

For each request, a backend production-level NPU chip simulator is invoked to get its latency and energy. The backend simulates executing the per-chip computation graph on all major components on an NPU chip (SA, VU, SRAM, HBM, and ICI).
It reports the per-operator statistics for each component, including execution time, FLOPs utilization, and memory/ICI traffic.
For MoE models, based on empirical studies~\cite{moe_imbalance:iclr26}, we adjust the expert load factor such that the most loaded expert receives 7$\times$ as many tokens as an average expert.
We model the power consumption of each component and combine the power model with the performance to quantify energy consumption.
For SA, VU, and SRAM, we implement them in RTL, synthesize and place \& route the components with ASAP7 PDK~\cite{asap7} using Synopsys toolchain to obtain the power estimation.
For HBM/ICI, we derive the power based on public TPU data~\cite{neurometer:hpca21,tpuarch:google:commACM20,tpuv4:isca23,tpuv4:nsdi24}, Google's datacenter optical network~\cite{Jupiter}, and HBM IP datasheets~\cite{hbm_ip:amd,hbm_ip:atria,hbm_ip:rambus}.
Our simulator is publicly available (see Appendix~\ref{app:ae}).

\begin{figure}[t]
    \centering
    \includegraphics[width=\linewidth]{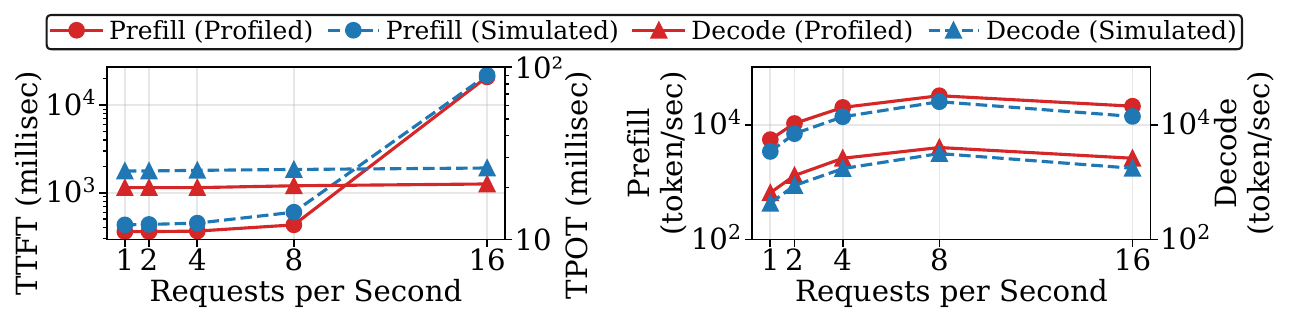}
    % \vspace{-3.5ex}
    \caption{{Simulated vs. profiled TTFT, TPOT, and throughput of {\pname{}} at steady state on a small TPU cluster under different request rates.}}
    \label{fig:e2e_sim_real_validate}
    % \vspace{-3.5ex}
\end{figure}

\noindent
\textbf{Simulator validation.}
We verified our NPU chip simulator with different TPU versions.
% \texttt{TPUv4-64} (\texttt{2x4x4} torus), \texttt{TPUv5e-64} (\texttt{8x8} torus), and \texttt{TPUv6e-64} (\texttt{8x8} torus)
We validated the execution time of all operators in the LLMs studied in the paper.
% including compute-bound operators (e.g., matrix multiplication), memory-bound operators (e.g., RMSNorm), and inter-chip collectives (e.g., reduce-scatter), 
%with different input tensor shapes 
We show representative operators in \Cref{fig:sim_validate_operator}.
% {We report the results of representative operator types in \mbox{\Cref{fig:sim_validate_operator}}}.
We validated the TTFT and TPOT of models with different parallelism degrees, sequence lengths, and batch sizes.
{\mbox{\Cref{fig:sim_validate_seqlen}} shows the validation results for Llama3-70B with tensor parallelism degree 32 and batch size 1 as an example. Our simulator achieves high correlations with profiled results.}
We also validated our power model against the published TPU data~\cite{tpuv4i:isca21,tpuv4:isca23,tpuarch:google:commACM20,neurometer:hpca21}.

{
% We validated our cluster-level simulator against our real-system prototype serving Llama3-70B.
% We generate synthetic workloads with the same sequence length distribution as Azure trace~\cite{azure-llm-trace-2023}. %since the original trace does not contain actual texts.
% We synthesize workload by sampling sequence length from the Azure trace~\cite{azure-llm-trace-2023}. %since the original trace does not contain actual texts.
% We sweep different request rates and report the steady-state performance in \mbox{\Cref{fig:e2e_sim_real_validate}}.
We further validate our cluster-level simulator against our real-system prototype serving Llama3-70B, running both with synthetic workloads sampled from the Azure trace~\cite{dynamollm:hpca25}, sweeping request rates and reporting steady-state performance in \mbox{\Cref{fig:e2e_sim_real_validate}}.
Since our simulator achieves high correlation at operator- and request-level, and it implements the same {\vpod{}} allocation and request routing algorithms as the real-system prototype, it faithfully reproduces the trends observed in the real-system: (1) the prefill throughput saturates at 8 requests/second, and the prefill TTFT increases drastically with a higher request rate; (2) the decode throughput drops at 16 requests/second due to prefill bottlenecks, but the TPOT is not affected since decode instances are not yet saturated.
}

% 32 v4, 64 v5e, 64 v6e chips, run llama3-70b with Azure trace. We constrain the search space of \pname{}'s offline best-fit \vpod{} allocation to our cluster setup. We validated the end-to-end throughput and latency of our simulator against the real system prototype.
% We also profiled the \vpod{} startup overhead (see \mbox{\Cref{tab:vpod_startup_overhead}}) and cross-validated the results with our simulator. \mbox{\Cref{fig:e2e_sim_real_validate}}: at rps=16, prefill becomes bottleneck, so decode throughput also drops without affecting tpot.

\section{Evaluation}
\label{sec:eval}

Our evaluation shows that (1) \pname{} improves energy/cost efficiency by 1.13$\times$/1.43$\times$ and SLO satisfaction rate by 1.36$\times$ over the state-of-the-art auto-scaler baselines (\S\ref{sec:eval:cost_efficiency});
(2) its benefits generalize to diverse LLM inference workloads (\S\ref{sec:sens_workloads}) {and various NPU cluster configurations (\mbox{\S\ref{sec:eval:sens_resource}})}; 
(3) it maintains a higher SLO satisfaction rate than the baseline despite resource scarcity (\S\ref{sec:eval:sens_resource});
(4) {its fine-grained \vpod{} grouping strategy adapts to sequence length variance more effectively than a coarse-grained multi-pool approach (\mbox{\S\ref{sec:eval:ablation}})};
(5) it is robust to output sequence length prediction accuracies (\mbox{\S\ref{sec:sens_output_predict_accuracy}});
(6) On a real heterogeneous TPU cluster, {\pname{}} improves cost efficiency by up to 11.2\% and always maintains high SLO satisfaction rate (\mbox{\S\ref{sec:eval:real_cluster}});
and (7) it facilitates reusing old NPUs and helps improve datacenter sustainability (\S\ref{sec:eval:carbon}).

\subsection{Experimental Setup}
\label{sec:eval:setup}

{
\setlength\tabcolsep{0.15em}
\begin{table}[t]
    \centering
    \caption{LLM inference traces used in our evaluation. Azure~\cite{dynamollm:hpca25} is a production-scale trace for coding services. LVEval~\cite{LVEval} is a long-context Q\&A dataset. The OpenThoughts dataset~\cite{OpenThoughts} contains complex questions in math, coding, and science, and triggers long reasoning outputs.}
    % \vspace{-3ex}
    % \footnotesize
    % \small
    \begin{tabular}{|c|c|c|c|}
    \hline
        & \textbf{Request Rate} & \textbf{Input} & \textbf{Output} \\
        & {(average} & \textbf{Seq. Len.} & \textbf{Seq. Len.} \\
        & req/minute) & (min/avg/max) & (min/avg/max) \\\hline
        \texttt{Azure}~\cite{dynamollm:hpca25} & 1667 & 1/2556/7691 & 1/23/3500 \\\hline
        % \texttt{LVEval-100K}~\cite{LVEval} & 81 & 8.8K/53K/127K & 1/17/278 \\\hline
        \texttt{LVEval}~\cite{LVEval} & 41 & 8.8K/105K/431K & 1/17/278 \\\hline
        \texttt{OpenThoughts}~\cite{OpenThoughts} & 27 & 5/246/12K & 531/11.6K/16.8K \\\hline
    \end{tabular}
    \label{tab:eval_traces}
    % \vspace{-4ex}
\end{table}
}

\begin{figure}[t]
    \centering
    \includegraphics[width=\linewidth]{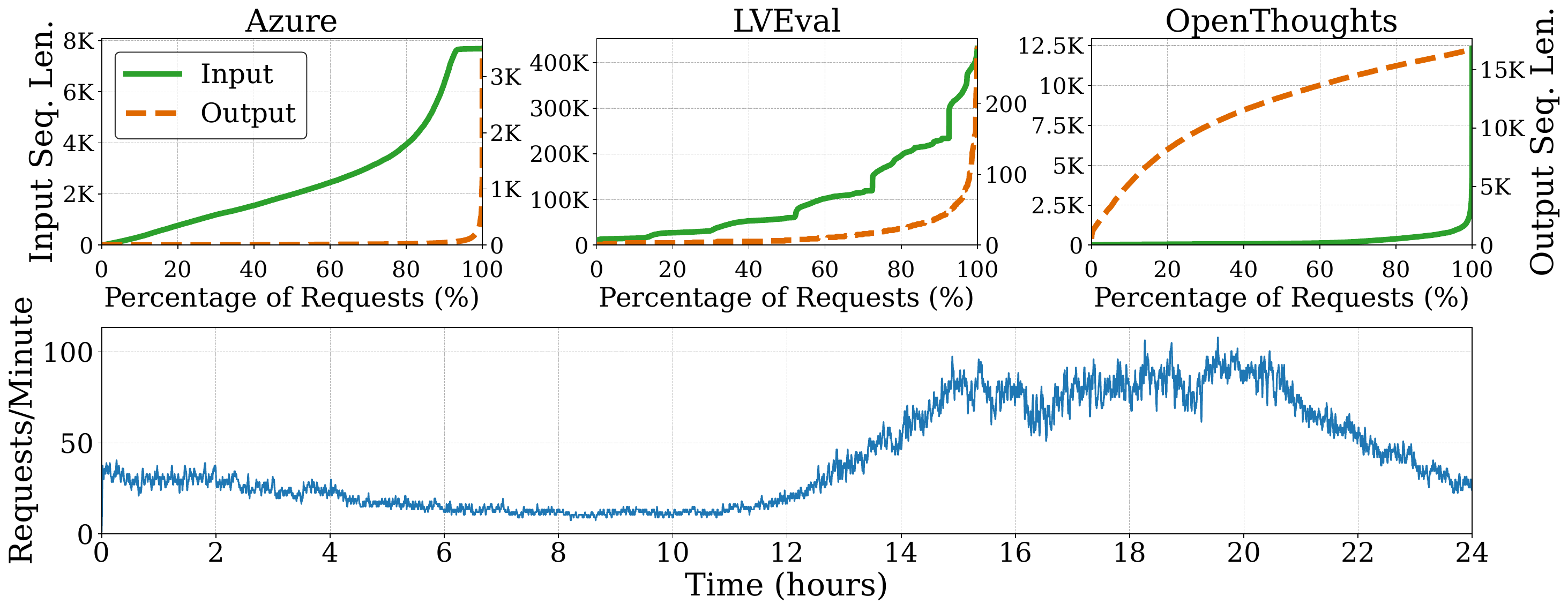}
    % \vspace{-3.5ex}
    \caption{Sequence length and request rate of traces (showing \texttt{LVEval} request rate from \Cref{tab:eval_traces} as an example).}
    \label{fig:traces_stats}
    % \vspace{-3.5ex}
\end{figure}

% \begin{figure}[t]
%     \centering
%     \includegraphics[width=0.7\linewidth]{figures/time_series_req_rate_Azure-LVEval.pdf}
%     \vspace{-1ex}
%     \caption{Request rate of \texttt{LVEval}. Other traces have the same temporal pattern with different request rates (see \Cref{tab:eval_traces}).}
%     \label{fig:traces_req_rate}
% \end{figure}

\noindent
\textbf{Workloads.}
We evaluate both dense models (Llama) and sparse Mixture-of-Experts (MoE) models (DeepSeek).
We replay production-level traces as shown in Table~\ref{tab:eval_traces}. 
%AzurePublicDataset~\cite{dynamollm:hpca25}.
{All models use BF16 weights.}
As the traces exhibit daily periodicity, we replay a 24-hour segment (see \Cref{fig:traces_stats}).
% We replay a representative 24-hour segment from the production-level trace AzurePublicDataset~\cite{dynamollm:hpca25}, as the traces exhibit daily periodicity (see \Cref{fig:traces_stats}).
To better represent workloads with long sequence lengths (e.g., reasoning models and agentic applications), we create synthetic traces by sampling sequence lengths from two recent datasets~\cite{LVEval,OpenThoughts} and issuing requests following the timestamps in the Azure trace.
As the Azure trace only contains the timestamp and sequence length of each request, we simulate a 60\% output sequence length prediction accuracy by default (i.e., when a request needs to be routed, there is a 60\% chance it is routed to the correct \mbox{\vpod{}} group; otherwise, it is routed to a random \vpod{} group whose sequence length is no shorter than the request's input or current sequence length). We study different accuracies in \mbox{\S\ref{sec:sens_output_predict_accuracy}}.
We scale the request rate for each dataset to account for the differences in input/output sequence lengths, while preserving the temporal pattern~\cite{SpotServe,blitzscale:osdi25,trace_upscaler:eurosys24} (see \Cref{tab:eval_traces}).
% We scale the request rate such that the total tokens per minute is similar to that in the original Azure trace, while preserving the temporal pattern~\cite{SpotServe,blitzscale:osdi25,trace_upscaler:eurosys24} (see \Cref{tab:eval_traces}).
We set the SLO following \S\ref{sec:bkg:study}.
% We set the SLO for each request to be 5$\times$ the latency of this single request running on the minimum number of NPU-C or NPU-D chips, whichever is faster~\cite{dynamollm:hpca25}.
We deploy separate \vpod{}s for prefill and decode phases~\cite{DistServe:osdi24}; each phase is managed by an independent auto-scaler.

\noindent
\textbf{Testbed setup.}
We configure the simulator for a cluster containing all four NPU versions from \mbox{\Cref{tab:npu_specs}}. 
%Based on our study in \mbox{\S\ref{sec:bkg:study}}, 
We treat NPU-C/D as the newer, more powerful generations, 
%as they are always preferred by {\pname{}}'s best-fit {\vpod{}} allocation algorithm in \mbox{\S\ref{sec:design:allocation}}. 
NPU-A/B are the older chips. By default, we assume a workload has enough resource quota, 
%(i.e., the maximum number of NPU chips that can be allocated by the auto-scaler), 
and a {\vpod{}} allocation always succeeds. This is typical for cloud services, as service providers
% will estimate the peak load of their workload and
typically choose a conservatively large quota to ensure service quality. We study the impact of insufficient quota in \mbox{\S\ref{sec:eval:sens_resource}}.
We also evaluate our {\pname{}} prototype on a real heterogeneous TPU cluster in \mbox{\S\ref{sec:eval:real_cluster}} (see the cluster setup in \mbox{\S\ref{sec:methodology}}).
% and the impact of allocation failures due to scarcity of newer chips in \mbox{\S\ref{sec:eval:alloc_failover}}.

% the scenario where the requested NPU chips are unavailable by synthetically generating allocation failures.
% By default, we assume the allocation failure rate is 5\%, and we experiment with different failure rates in \S\ref{sec:eval:alloc_failover}.
% For each experimental setup, we run two separate sets of experiments by setting either energy or monetary cost as the optimization target.
% We evaluate both energy and monetary cost as the optimization target, respectively.
% We vary the request rate and total number of NPUs in the cluster in the sensitivity analysis.

\begin{figure}[t]
    \centering
    \includegraphics[width=\linewidth]{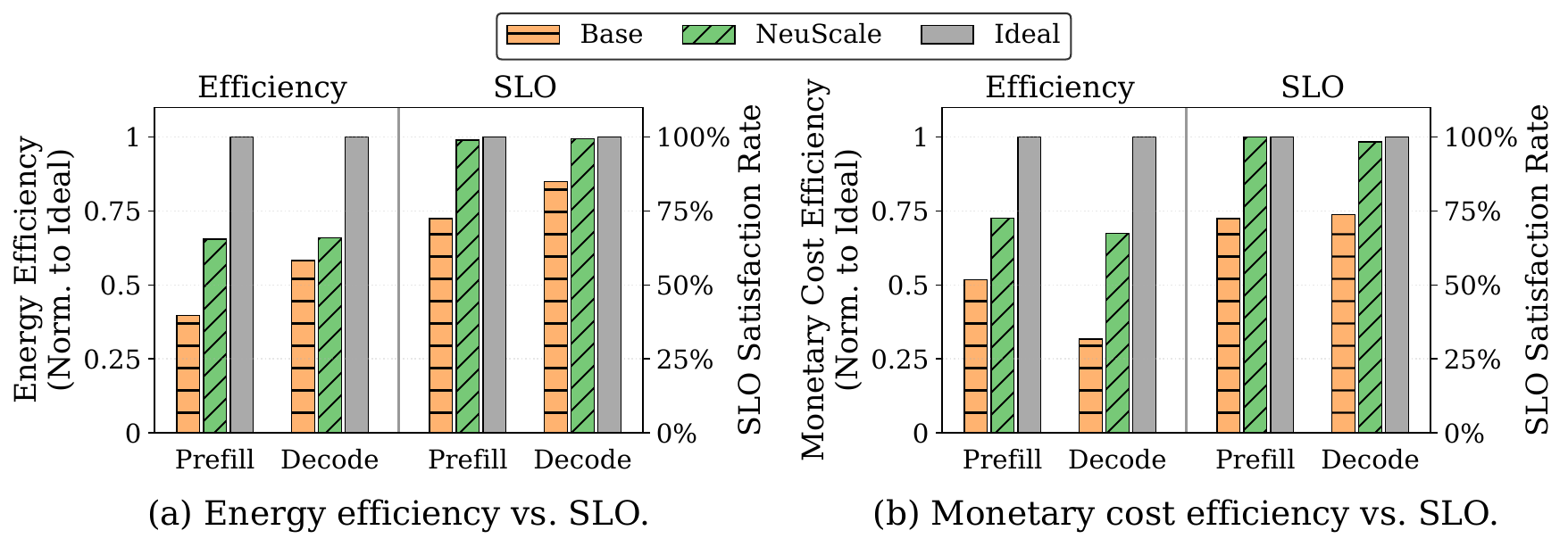}
    % \vspace{-6.3ex}
    % \includegraphics[width=0.85\linewidth]{figures/deepseekv3_671b_azure_dual_axis.pdf}
    % \vspace{-3ex}
    \caption{End-to-end energy/cost efficiency and SLO satisfaction rate on DeepSeekV3-671B (the largest model in our evaluation) with \texttt{Azure}. \S\ref{sec:sens_workloads} covers other workloads.}
    \label{fig:eval_deepseekv3671b_azure}
    % \vspace{-4ex}
\end{figure}

% By default, we assume a workload has enough resource quota (i.e., the maximum number of NPU chips that can be allocated by the autoscaler), and the allocation of new \vpod{}s always succeeds[cite: 820]. This is typical for cloud services, as the service provider will estimate the peak load to configure a conservatively large quota to ensure service quality. Since our study in \S\ref{sec:bkg:study} finds that NPU-C and NPU-D are generally the ``best-fit'' versions for workloads due to their higher performance and efficiency, \pname{} primarily targets these chips. The older NPU-A and NPU-B chips are utilized for failover allocations. We study allocation failovers caused by the resource scarcity of newer NPU versions in \S\ref{sec:eval:alloc_failover}[cite: 820, 1191, 1192]. We also evaluate \pname{} under an insufficient resource quota in \S\ref{sec:eval:sens_resource}. We evaluate both energy and monetary cost as the optimization target, respectively[cite: 821].

\begin{figure*}[t]
    \centering

    \subfloat[Prefill energy consumption.]{
        \includegraphics[width=0.34\linewidth]
        {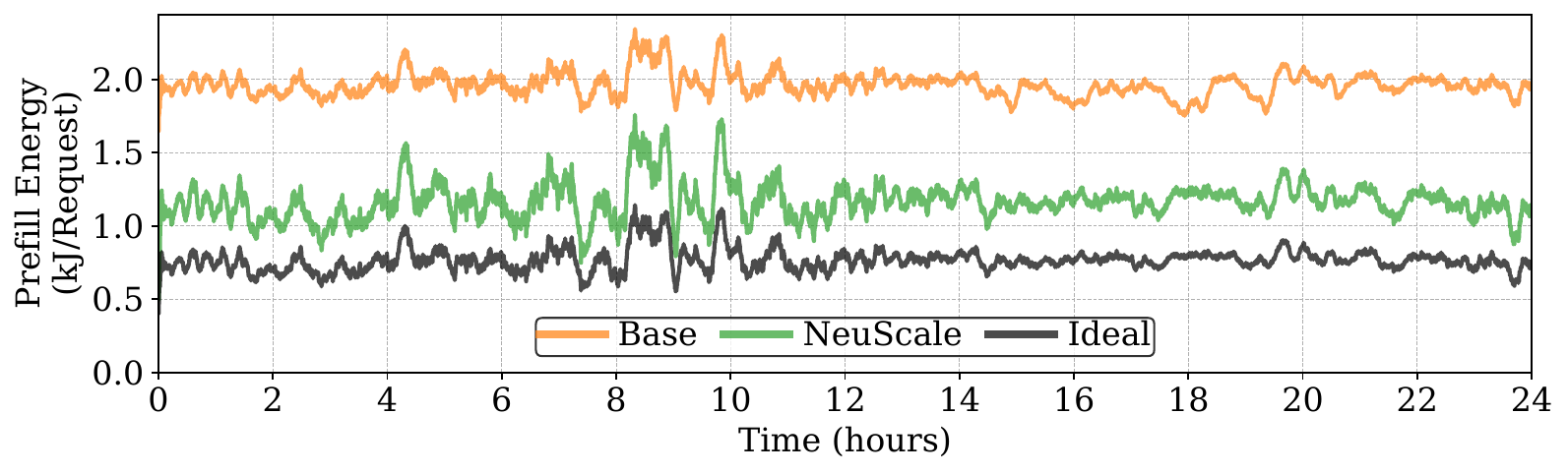}
        \label{fig:eval_time_series_energy_prefill}
    }
    \subfloat[Decode energy consumption.]{
        \includegraphics[width=0.34\linewidth]
        {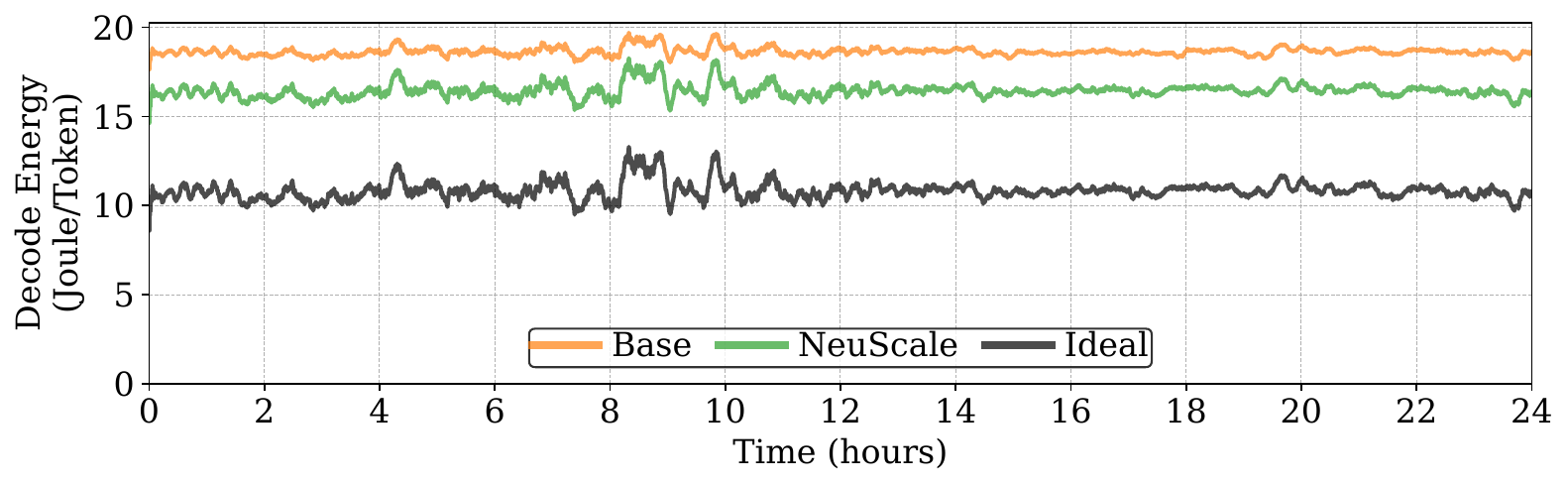}
        \label{fig:eval_time_series_energy_decode}
    }
    \subfloat[Energy vs. sequence length.]{
        \includegraphics[width=0.2394\linewidth]
        {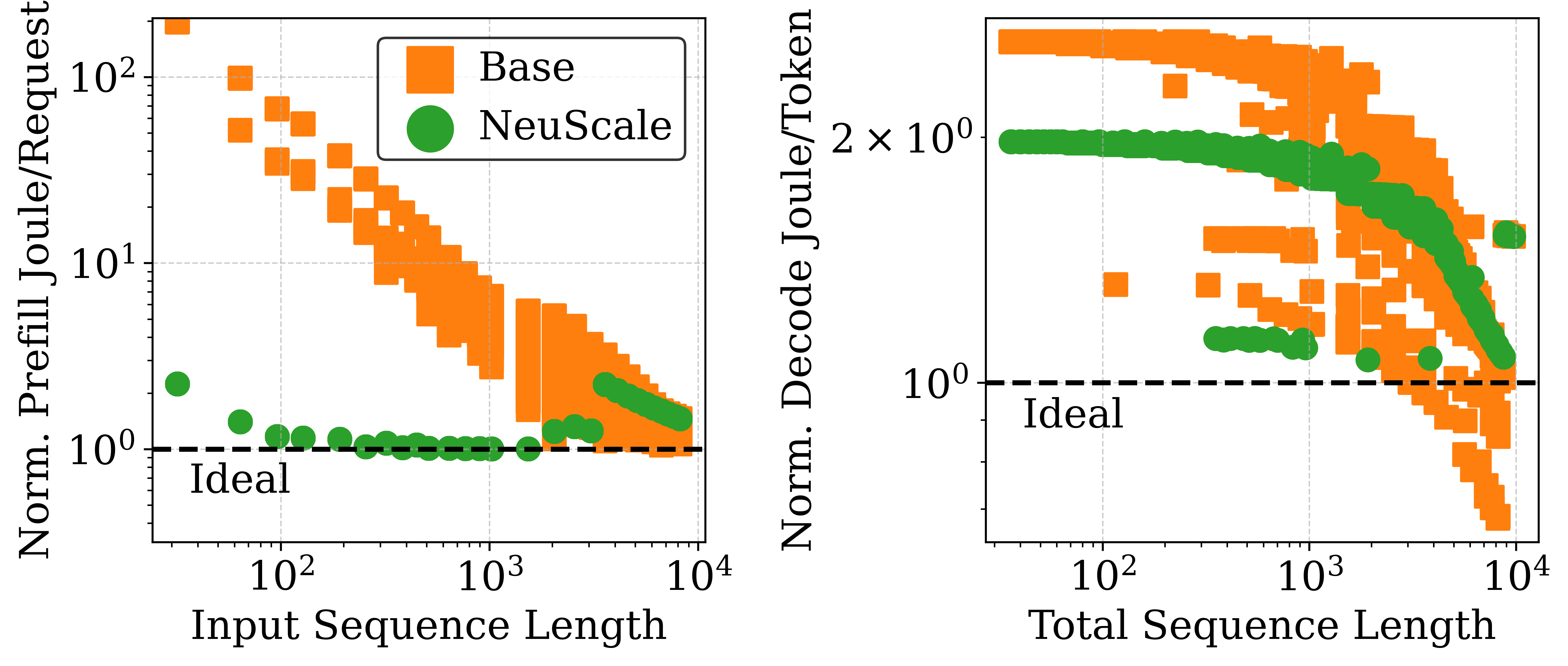}
        \label{fig:eval_energy_vs_seqlen}
    }

    \vspace{0.5em}

    \subfloat[Prefill SLO satisfaction rate.]{
        \includegraphics[width=0.34\linewidth]
        {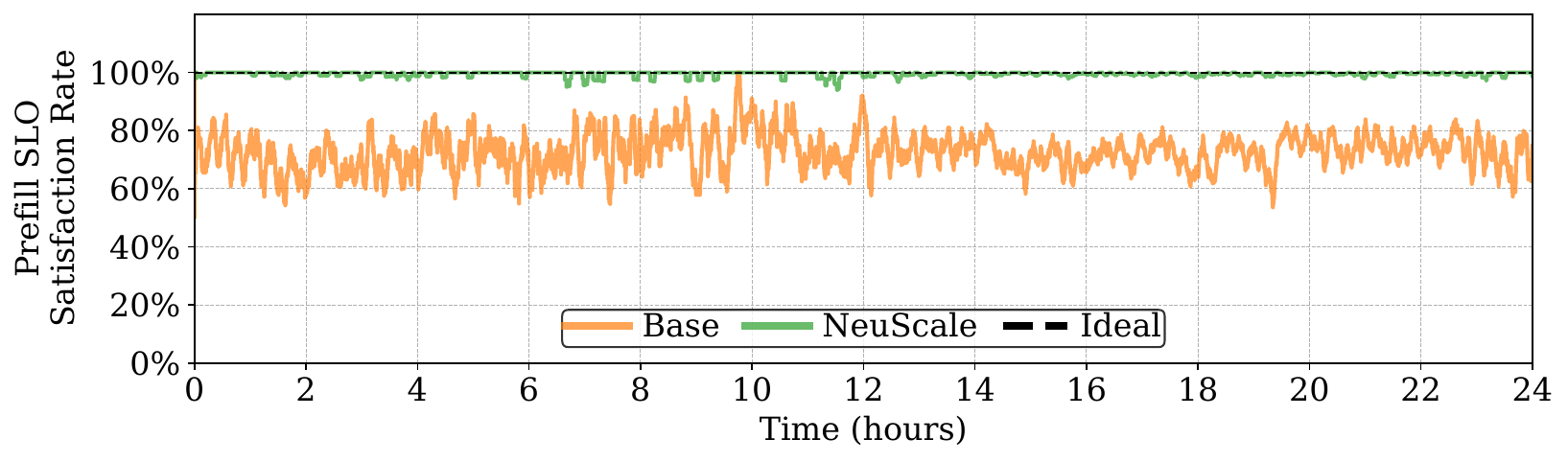}
        \label{fig:eval_time_series_prefill_slo_sat}
    }
    \subfloat[Decode SLO satisfaction rate.]{
        \includegraphics[width=0.34\linewidth]
        {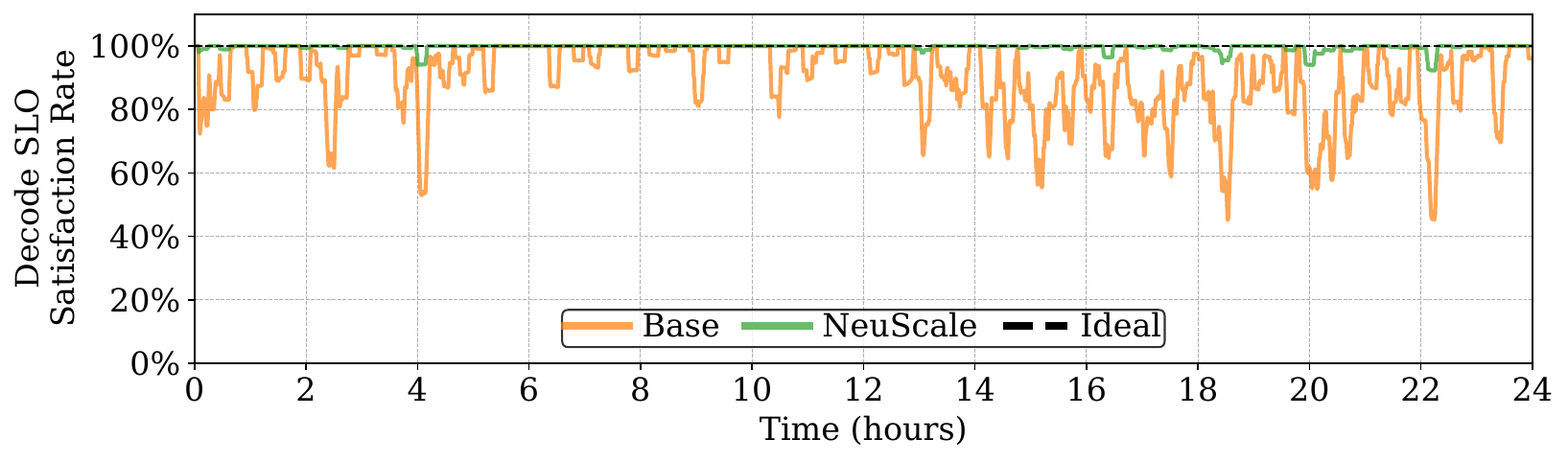}
        \label{fig:eval_time_series_decode_slo_sat}
    }
    \subfloat[Latency vs. sequence length.]{
        \includegraphics[width=0.2394\linewidth]
        {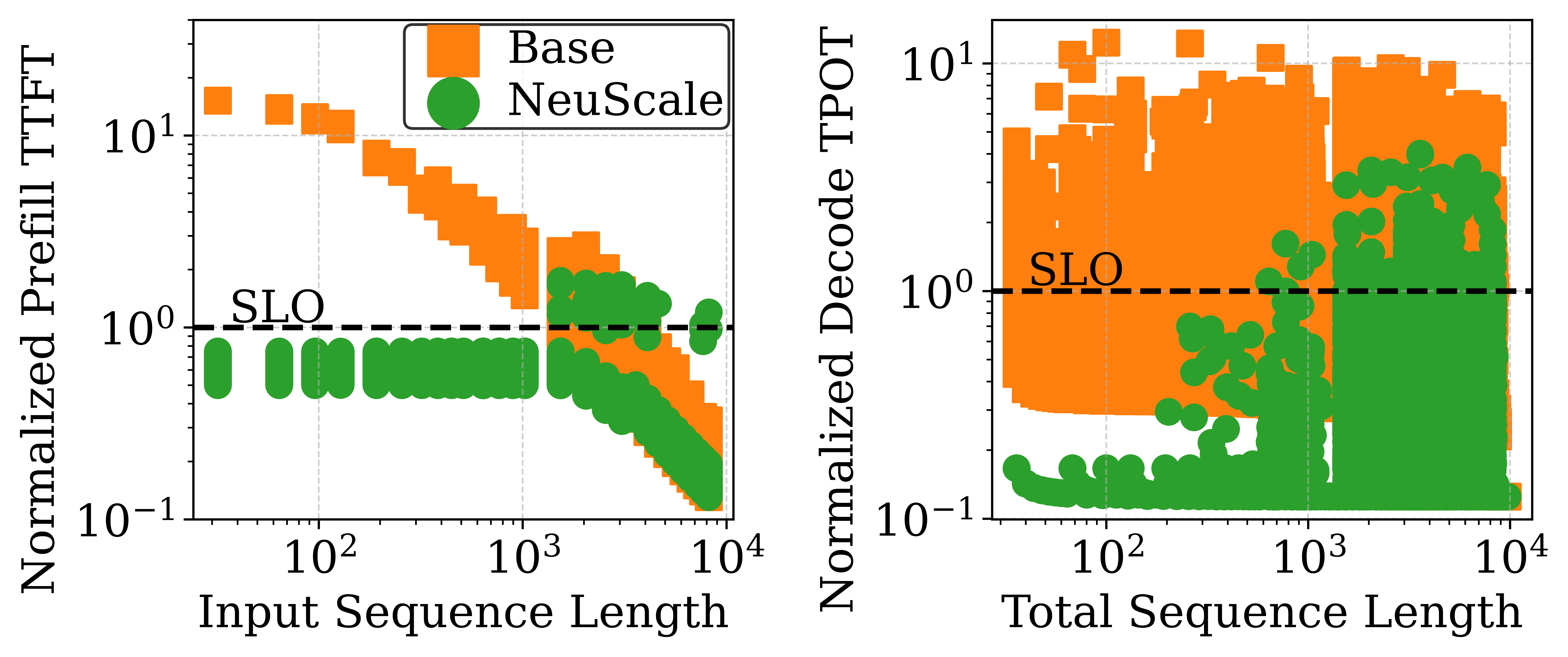}
        \label{fig:eval_latency_vs_seqlen}
    }

    \caption{Left/middle columns: energy consumption and SLO satisfaction rate over time. Right column: energy/latency of each request w.r.t. its sequence length. We use DeepSeekV3-671B with \texttt{Azure} as an example and analyze other workloads in \S\ref{sec:sens_workloads}.}
    \label{fig:eval_time_series_combined}
\end{figure*}

\noindent
\textbf{Baselines.} We evaluate \pname{} with the following designs.
%We compare \pname{} with \base{}, the state-of-the-art NPU auto-scaling approach in today's cloud data centers.
% We empirically determine the auto-scaling time window and epoch to be 30 minutes and 5 minutes.
%We also include an \ideal{} oracle design that can always achieve the optimal efficiency while satisfying SLO for all requests.
%We summarize the designs as follows:
\begin{itemize}[leftmargin=*]
    \item \base{}: the state-of-the-art auto-scaling method for NPUs. It scales only \vpod{} count and uses homogeneous configuration~\cite{autopilot:eurosys2020,kubernetes_hpa}, which is commonly used in modern cloud platforms like Google GKE~\mbox{\cite{googlecloud:tpu_autoscaling}} (see \mbox{\Cref{tab:related_works}}).
    % It uses the best-fit \vpod{} allocation for the largest sequence length of each trace, as provisioning for the peak demand is common in the cloud.
    It uses the best-fit \vpod{} allocation for the largest sequence length of each trace to provision for the peak demand.
    \item \ours{}: our design for auto-scaling heterogeneous \vpod{}s. We generate the best-fit allocations for each unique sequence length for each model before running the experiments.
    \item \ideal{}: an oracle design that always achieves the best efficiency while meeting SLOs. We derive the ideal efficiency for a request by first finding its best-fit \vpod{} allocation. We then run a maximum-sized batch sharing that same sequence length on the best-fit \vpod{} to obtain the ideal energy/cost.
    % all requests are served with optimal cost efficiency and perfect SLO satisfaction. 
    % This resembles instant horizontal and vertical auto-scaling with optimal \vpod{} allocation, perfect batching, no queuing delay, and no \vpod{} startup overhead.
    % and roofline reconfiguration overhead (determined solely by the ICI/DCN inter-chip bandwidth and the minimum data volume that needs to be transferred).
\end{itemize}

\subsection{Efficiency and SLO Satisfaction Rate}
\label{sec:eval:cost_efficiency}

\begin{figure}[t]
    \centering
    \includegraphics[width=\linewidth]{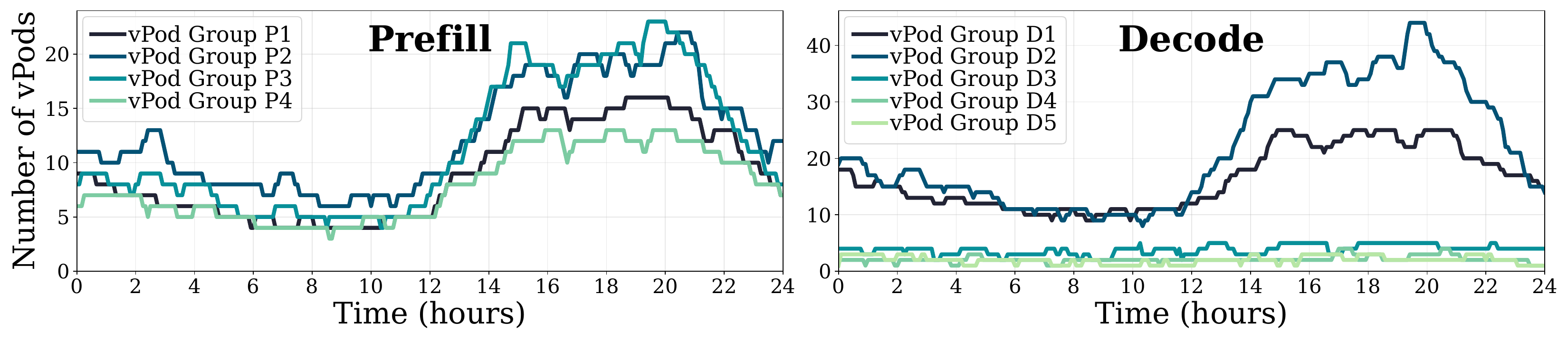}
    % \vspace{-3.9ex}
    \caption{Number of \vpod{}s in each \vpod{} group over time. We plot DeepSeekV3-671B with the \texttt{Azure} trace as an example.}
    \label{fig:eval_scale_curve_comparison}
    % \vspace{-3.5ex}
\end{figure}

We evaluate the end-to-end energy/cost efficiency and SLO satisfaction rate of \pname{}.
We focus on DeepSeekV3-671B, the largest model in our experiments, and the \texttt{Azure} trace (see \S\ref{sec:sens_workloads} for sensitivity analysis with other workloads).
\Cref{fig:eval_deepseekv3671b_azure} summarizes the results.
Overall, \pname{} improves the energy efficiency by up to 1.65$\times$ (1.37$\times$ on average) over \base{}, the monetary cost efficiency by up to 2.13$\times$ (1.73$\times$ on average), and the SLO satisfaction rate by 1.38$\times$ (1.31$\times$ on average).

% Compared to \base{}, \pname{} improves both energy/cost efficiency and SLO satisfaction rate.
% \base{} allocates all \vpod{}s based on the largest sequence length, as provisioning for the ``peak'' resource demand is a common strategy for deploying cloud services.
% This one-size-fits-all strategy suffers from two limitations.
The one-size-fits-all strategy of \base{} suffers from two limitations.
First, it leads to energy/cost inefficiency, as the best-fit \vpod{} for the largest sequence length contains many NPUs. Running small requests on large \vpod{}s underutilizes these NPUs.
Second, large \vpod{}s may employ a high tensor parallelism degree to exploit the aggregated FLOPS and memory bandwidth of all NPUs. Running small requests on them may suffer from ICI bottleneck, making the latency worse than on a smaller \vpod{} and leading to SLO violations.

\pname{} outperforms \base{} by allocating the best-fit \vpod{} for each sequence length.
% Hence, \pname{} achieves significant improvements in both SLO satisfaction rate and energy/cost efficiency.
The gap between \pname{} and \ideal{} is mainly due to reactive auto-scaling at runtime.
Since \pname{} adjusts the \vpod{} allocation based on the peak load and sequence lengths observed in the last 30-minute time window, if the actual current load is lower, some \vpod{}s may not be fully utilized.
If \pname{} cannot find a best-fit \vpod{} instance for a request (e.g., due to unseen sequence lengths or \vpod{} group coalescing), this request may suffer from sub-optimal \vpod{} allocation.
Despite these limitations, \pname{} still significantly improves the energy/cost efficiency over \base{} and can achieve more than 80\% of ideal efficiency for 22 out of 48 workloads in our experiments (see \S\ref{sec:sens_workloads}).

\noindent
\textbf{Benefits breakdown.}
% In \Cref{fig:eval_time_series_combined}, we investigate \pname{}'s energy consumption and SLO satisfaction rate over time, as well as the energy/latency of each request with respect to its sequence length.
As shown in \Cref{fig:eval_time_series_combined},
\pname{} consistently delivers higher energy efficiency and SLO satisfaction rate than \base{} regardless of the changes in request load.
The benefits come from mapping each request to their best-fit \vpod{}s.
As shown in \Cref{fig:eval_latency_vs_seqlen,fig:eval_energy_vs_seqlen}, \pname{} improves the energy efficiency and latency of most requests. In contrast, \base{} is only optimized for the largest requests, and it suffers severe energy and latency overhead for smaller requests.
% \hl{add timeline figure for auto-scaling number of vpods and vpod groups}
We visualize how \pname{} auto-scales \vpod{}s over time in \Cref{fig:eval_scale_curve_comparison}. 
% The number of \vpod{} groups reflects the number of best-fit allocations, which is determined by the request sequence length distribution (see \Cref{fig:traces_stats}).
For prefill, the number of \vpod{}s in each group follows the request rate changes.
For decode, the groups \texttt{D1} and \texttt{D2} are larger than others. They are the best-fit \vpod{}s for requests with total sequence length greater than 2K, which account for more than half of the requests in \texttt{Azure}.

\subsection{Sensitivity Analysis with Various Workloads}
\label{sec:sens_workloads}

\begin{figure}[t]
    \centering
    \includegraphics[width=\linewidth]{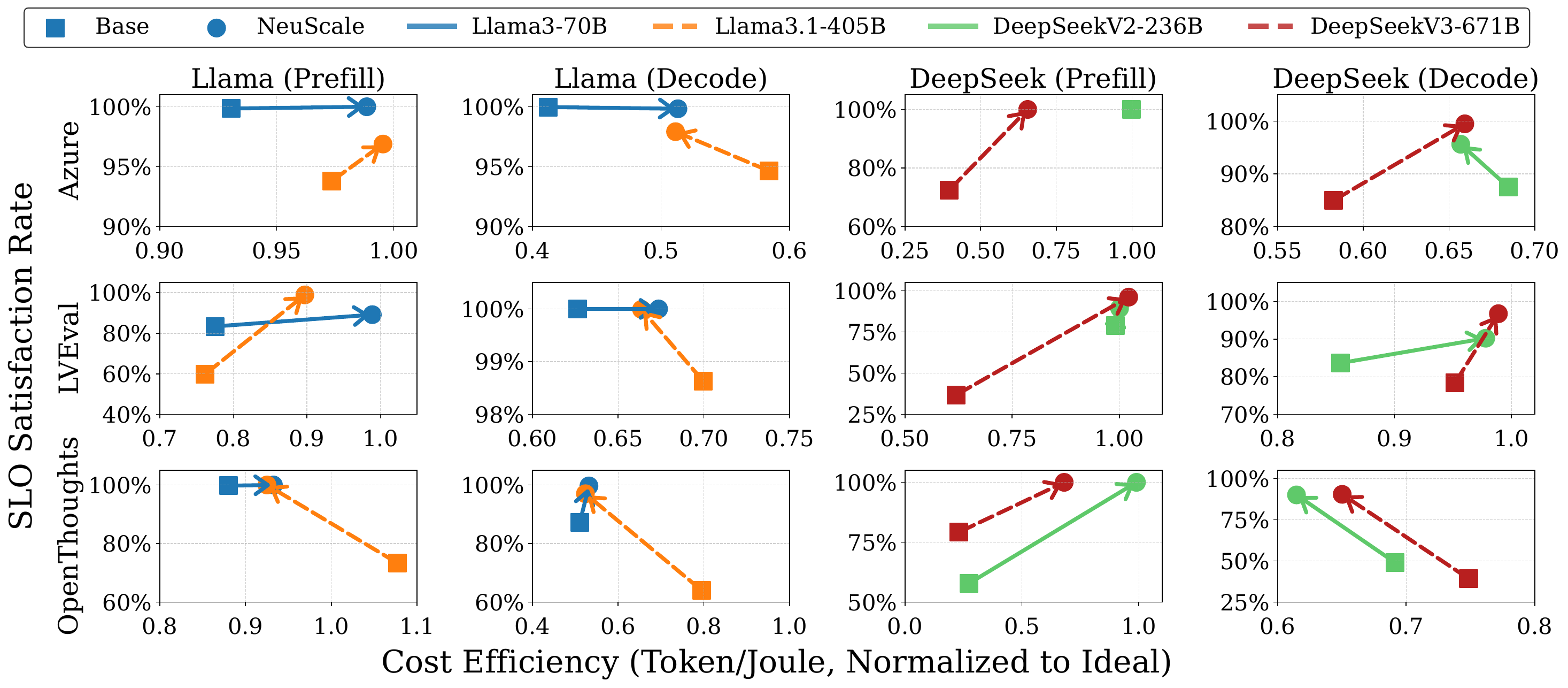}
    % \vspace{-6.5ex}
    \caption{Energy efficiency vs. SLO satisfaction rate. Different colors represent different models. Arrows indicate the improvement from \base{} to \pname{}.}
    \label{fig:eval_energy_vs_slo_sat}
    % \vspace{-3.2ex}
\end{figure}

\begin{figure}[t]
    \centering
    \includegraphics[width=\linewidth]{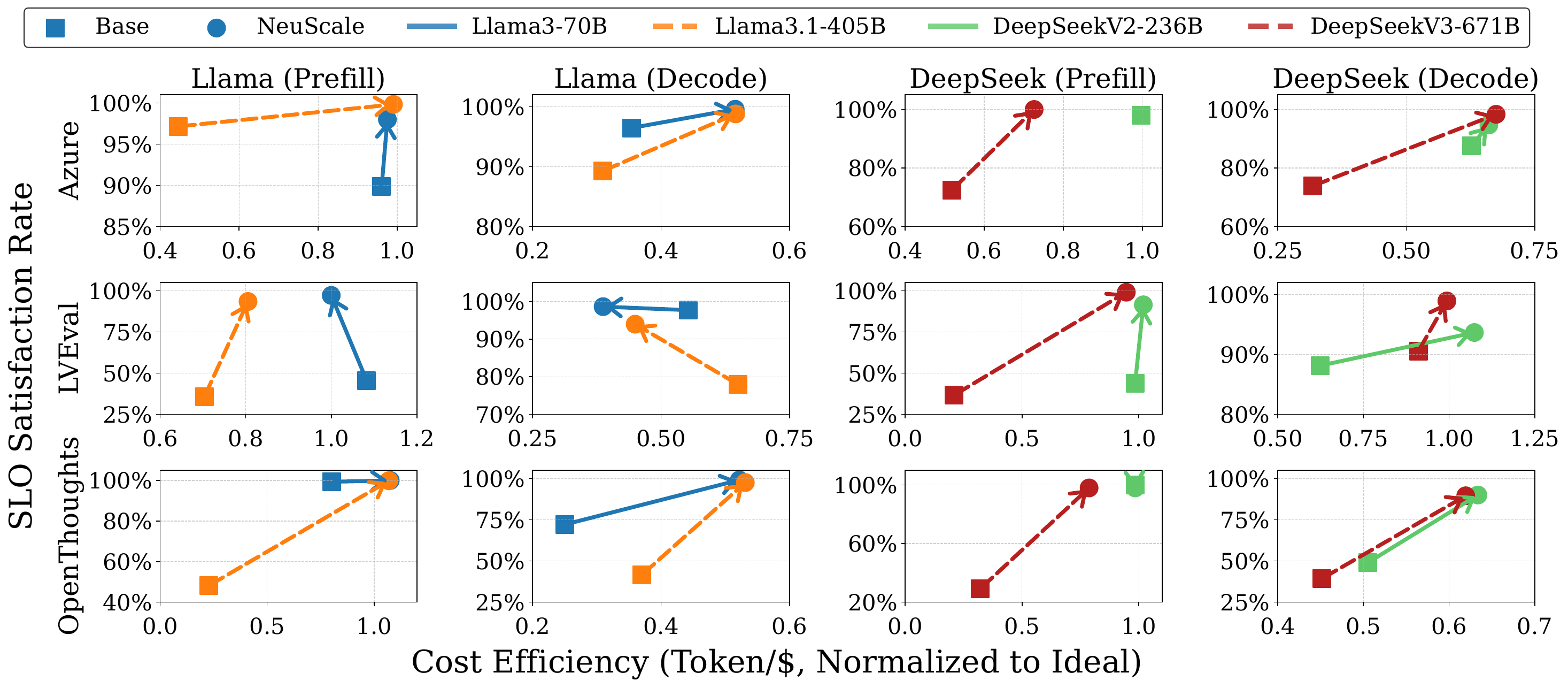}
    % \vspace{-6.3ex}
    \caption{Monetary cost vs. SLO satisfaction rate. Different colors represent different models. Arrows indicate the improvement from \base{} to \pname{}.}
    \label{fig:eval_cost_vs_slo_sat}
    % \vspace{-3ex}
\end{figure}

% \hl{revise: highlight we test diverse settings and they all have diverse slo requirement}
We evaluate \pname{} with different models and traces in \Cref{fig:eval_energy_vs_slo_sat} and \Cref{fig:eval_cost_vs_slo_sat}.
On average, \pname{} improves the energy efficiency, monetary cost efficiency, and SLO satisfaction rate by 1.13$\times$ (up to 3.61$\times$), 1.43$\times$ (up to 5.64$\times$), and 1.36$\times$ (up to 3.44$\times$), as compared to \base{}.
The trend across different models and traces is diverse, as they all have different SLO requirements for different requests.
% , leading to diverse resource bottlenecks.
In general, \pname{}'s benefits are more obvious for larger models and more diverse sequence lengths.
This is because larger models require larger \vpod{}s, making the allocation search space larger. With more diverse sequence lengths, the drawback of \base{}'s one-size-fits-all allocation also becomes more obvious.
% For example, for the \texttt{Azure} trace in \Cref{fig:eval_energy_vs_slo_sat}, the largest model DeepSeekV3-671B gains higher improvements than others.
% \texttt{LVEval} and \texttt{OpenThoughts} have longer sequence lengths than \texttt{Azure}, so \pname{}'s improvements over \base{} are also more obvious.

For most workloads, \pname{} improves SLO satisfaction rate over \base{}.
Note that {\pname{}} prioritizes SLO guarantees; it sometimes leads to lower energy/cost efficiency, as an SLO-compliant {\vpod{}} configuration may not be the most efficient one (e.g., Llama3.1-405B in \texttt{OpenThoughts} in \Cref{fig:eval_energy_vs_slo_sat}).
% This is because the best-fit \vpod{} of the largest sequence length used by \base{} employs a high pipeline parallelism
% This is because \base{} employs a high pipeline parallelism degree for the largest sequence length
% to minimize memory pressure (pipeline parallelism incurs no model weight or activation duplication compared to tensor/expert parallelisms) and improve throughput.
This is because \base{} uses a high pipeline parallelism degree for the largest sequence length, minimizing memory pressure (pipeline parallelism duplicates no model weights or activations) and improving throughput.
While deep pipelining improves efficiency, it adds significant latency for small requests (longer requests are less affected by deep pipelining, since their SLO latency target is higher).
% In contrast, \pname{} creates a specialized \vpod{} per request to meet its SLO while improving efficiency on a best-effort basis.
% While deep pipelining is efficient for both long and short requests, it incurs significant latency overhead for small requests (longer requests are less affected since their SLO latency target is higher).
% While deep pipelining improves efficiency, it incurs significant latency overhead for small requests (longer requests are less affected since their SLO latency target is higher).
% In contrast, \pname{} creates specialized \vpod{}s for each request to meet its SLO while improving efficiency with the best effort.
In contrast, \pname{} finds the best-fit \vpod{} for each request to meet its SLO while improving efficiency with best effort.
% based on its SLO target and resource demands.
% It makes the correct trade-off between performance and efficiency for each request with the best effort.

In a few cases (e.g., DeepSeekV3-671B Prefill with \texttt{LVEval} in \Cref{fig:eval_energy_vs_slo_sat}), \pname{}'s efficiency is slightly higher than \ideal{}, while the SLO satisfaction rate is lower than 100\%.
This implies some requests are executed on more efficient \vpod{}s than their best-fit \vpod{}s (i.e., when the best-fit \vpod{} is coalesced or not allocated in the current epoch), while they suffer from SLO violations.
% For these cases, 
The user can disable \vpod{} group coalescing or adjust the auto-scaling epoch to trade off between SLO satisfaction and efficiency.
% For these cases, the user can disable \vpod{} group coalescing or adjust the auto-scaling epoch and time window to allocate new \vpod{}s more aggressively for achieving the desired trade-off between SLO satisfaction and efficiency.

% \begin{figure}[t]
%     \centering
%     \includegraphics[width=\linewidth]{figures/seqlen_timeseries_synthetic_BurstGPT_trace.pdf}
%     \vspace{-4ex}
%     \caption{\hlcommon{Synthetic bursty trace for sensitivity study. The sequence length follows a normal distribution. The workload bursts from 30 to 80 requests/minute, with average input/output sequence length increasing from 2K/128 to 100K/5K, and the burst period lasts for 30 minutes.}\revnoteReverse[-1em]{\#Common-3}}
%     \label{fig:burst_trace}
%     \vspace{-3ex}
% \end{figure}

\begin{figure}[t]
    \centering
    \includegraphics[width=\linewidth]{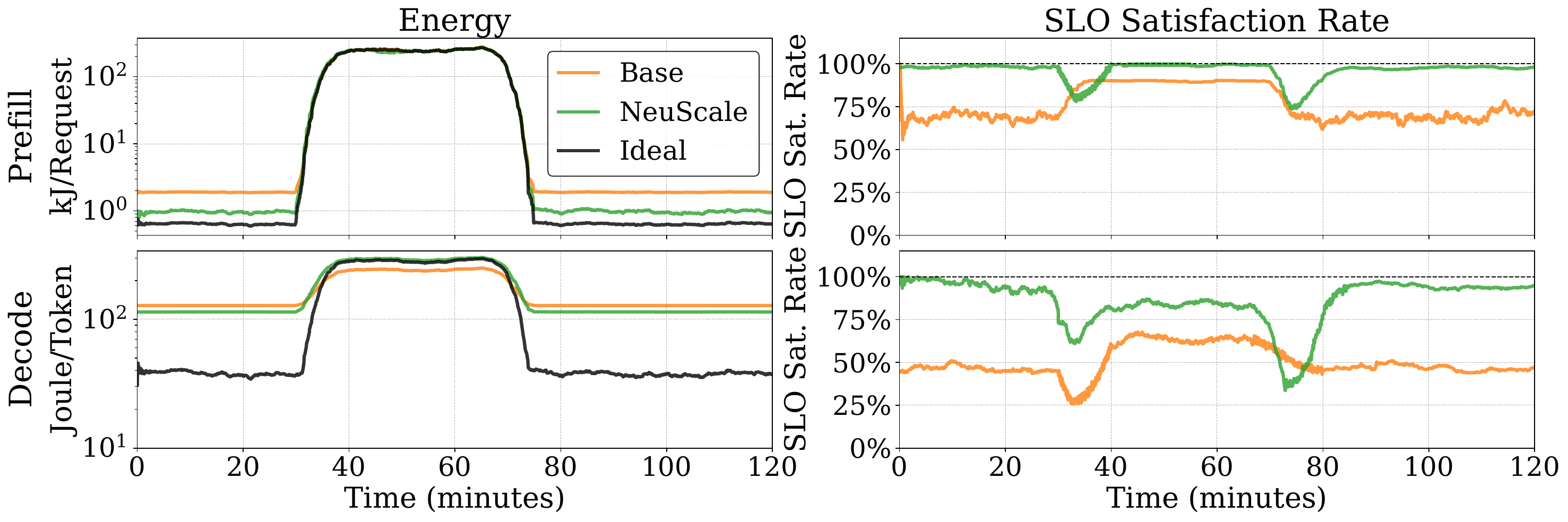}
    % \vspace{-6.4ex}
    \caption{{Energy efficiency and SLO satisfaction rate with a synthetic bursty trace.
% The sequence length follows a normal distribution.
    }}
    \label{fig:eval_burst_trace}
    % \vspace{-3.5ex}
\end{figure}

\noindent
\textbf{Impact of bursty workloads.}
% We create synthetic traces with request spikes in \mbox{\Cref{fig:burst_trace}}.
We create synthetic traces with a 45-minute burst period, during which the workload bursts from 30 to 80 requests/minute, with average input/output sequence length increasing from 2K/128 to 100K/5K.
As shown in \mbox{\Cref{fig:eval_burst_trace}}, when the request pattern shifts abruptly (at $\sim$30 and 70 minutes), \pname{} suffers from $\sim$10 minutes of SLO degradation. However, it quickly recovers from degradation and maintains its advantages over \base{}.
Note that {\base{}}'s SLO satisfaction rate increases during the burst period, since the sequence lengths increase in the burst period, and {\base{}}'s {\vpod{}}s are provisioned for long sequence lengths, which cause SLO violations for some shorter sequences.

% Figure 23: Why can NeuScale recover from low SLO faster? Why does Base’s SLO improve under bursty workloads?

% Q: NeuScale’s auto-scaling responsiveness may be insufficient for bursty workloads. The default 30-minute monitoring window and 5-minute epoch-based scaling may miss rapid demand changes, potentially leading to underutilized resources. The paper acknowledges "if the actual current load is lower, some vPods may not be fully utilized" but doesn't quantify the efficiency loss from this inaccuracy.
% A: 
% - What is the vPod creation latency and how does the auto-scaler perform under bursty workload traces?
% - If vPod creation takes minutes and the epoch is also five minutes then the controller is always behind. The Azure trace used has smooth daily periodicity which is the easiest case for any reactive auto-scaler. A bursty trace with sudden spikes would stress the system and the paper does not include one. 

\subsection{Sensitivity to Resource Availability}
\label{sec:eval:sens_resource}

% \begin{figure}[t]
%     \centering
%     \includegraphics[width=\linewidth]{figures/npu_preference_tokens_combined.pdf}
%     \caption{\hlcommon{Token-level NPU version preference, defined as the percentage of total tokens within each trace whose best-fit allocation maps to NPU-C/D.}}
%     \label{fig:token_npu_version_preference}
% \end{figure}

\begin{figure}[t]
    \centering
    % \includegraphics[width=0.95\linewidth]{figures/npu_version_llama3_1-405b_combined.pdf}
    % \vspace{-3.3ex}
    \includegraphics[width=\linewidth]{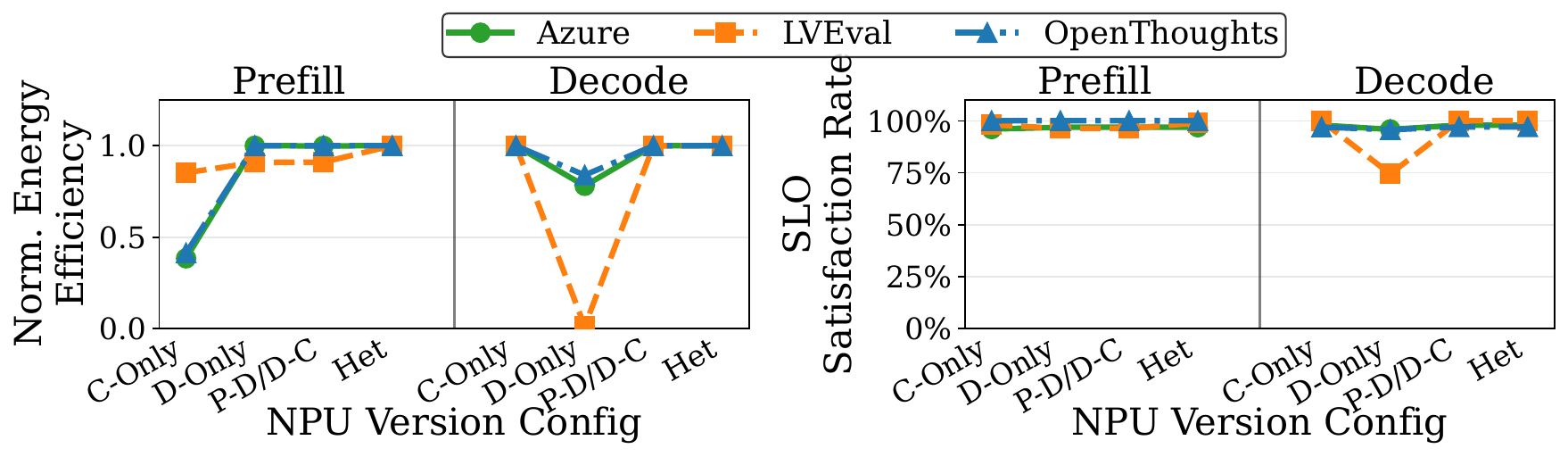}
    % \vspace{-6.5ex}
    \caption{{Energy efficiency (normalized to ``Het'') and SLO satisfaction rate of \pname{} with varying combinations of NPU versions. 
    %``C-Only''/``D-Only'' refer to homogeneous clusters with only NPU-C/NPU-D chips. ``P-D/D-C'' refers to Prefill on NPU-D and Decode on NPU-C. ``Het'' has both NPU-C and NPU-D. 
    We study Llama3.1-405B as an example.}}
    \label{fig:sens_vary_homo_het}
    % \vspace{-4ex}
\end{figure}

This section evaluates \pname{} with various mixes of NPU versions and resource quotas (i.e., limited number of chips).

\noindent
{\textbf{Varying available NPU versions.}
We deploy {\pname{}} across three cluster configurations: (1) homogeneous clusters using exclusively NPU-C or NPU-D (C-Only or D-Only); (2) an ``expert-choice'' setup that maps prefill to NPU-D and decode to NPU-C (P-D/D-C in Figure~\ref{fig:sens_vary_homo_het}), based on the compute vs. memory bandwidth demand of prefill and decode phases; and (3) a heterogeneous cluster with both NPU-C and NPU-D available (``Het'' in Figure~\ref{fig:sens_vary_homo_het}).}
%The trend highly correlates with the best-fit NPU version of each request (see \mbox{\Cref{fig:motiv_fine_grained_cost}}). 
As shown in \mbox{\Cref{fig:sens_vary_homo_het}}, although the compute-optimized NPU-D ``intuitively'' matches the compute-intensive prefill phase, mapping all prefill requests to NPU-D degrades energy efficiency by 11\% for the LVEval trace, as extremely long-context requests prefer NPU-C's larger memory capacity. Mapping decode requests to NPU-C performs well across all traces, aligning with our findings in \mbox{\S\ref{sec:bkg:study}}. While most NPU versions can maintain high SLO satisfaction, D-Only struggles with LVEval because serving long-context requests requires many NPU-D chips, exacerbating ICI overhead. {\pname{}} alleviates the manual effort of NPU selection, automatically mapping requests to their best-fit hardware to improve efficiency and SLO satisfaction.

\begin{figure}[t]
    \centering
    \includegraphics[width=\linewidth]{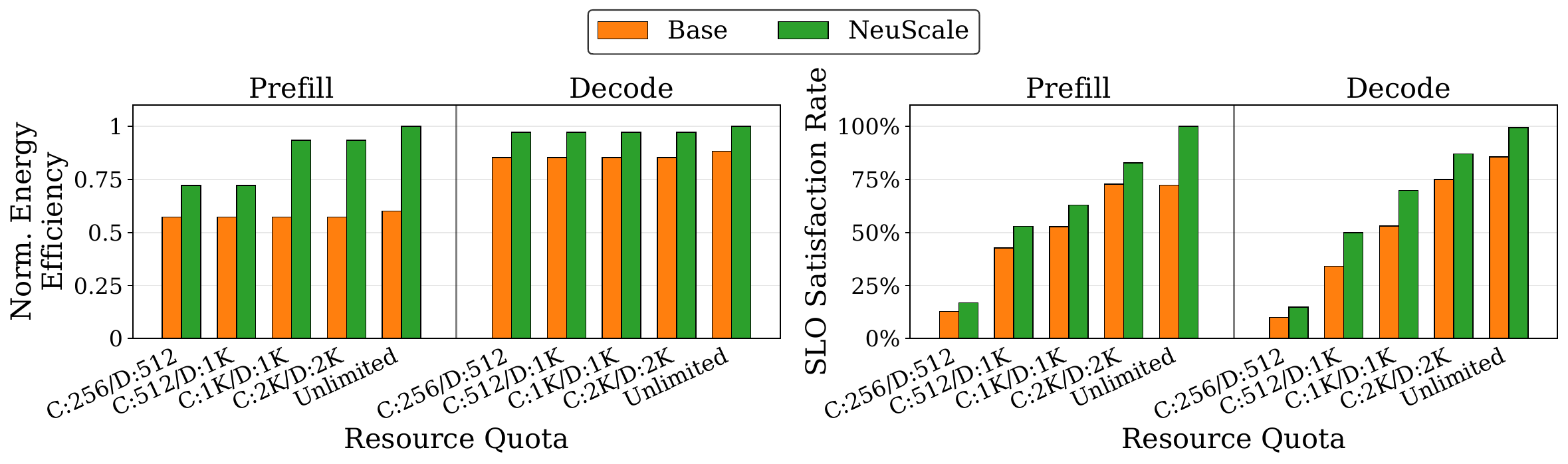}
    % \vspace{-6.3ex}
    \caption{{Energy efficiency (normalized to Unlimited) and SLO satisfaction rate with different resource quotas, on DeepSeekV3-671B with Azure as an example. ``C:256/D:512'' refers to the maximum number of NPU-C/NPU-D.}}
    \label{fig:sens_resource_constrained}
    % \vspace{-3.5ex}
\end{figure}

\noindent
{\textbf{Varying resource quota.}
In \mbox{\Cref{fig:sens_resource_constrained}}, we examine \pname{}} under insufficient resource quota. We restrict the maximum number of NPU-C/D chips. To isolate the impact of fail-over allocation to older chips, we do not allow fail-over to any NPU-A/B. %but fail-over between NPU-C/D is allowed.}
In general, provisioning enough {\vpod{}} groups is essential for efficiency (to ensure each request is mapped to its best-fit \vpod{} group), and having enough \vpod{}s per group is essential for SLO satisfaction (to prevent overloading).
As we lower the resource quota, the prefill phase suffers from up to 1.34$\times$ worse efficiency compared to unlimited resources. Decode is less affected, even though it demands more \vpod{} groups than prefill (see \mbox{\Cref{fig:eval_scale_curve_comparison}}), because the decode energy efficiency of most requests is less sensitive to \vpod{} configurations. Both prefill and decode suffer from low SLO satisfaction rate (as low as 14\%/11\%). While an insufficient resource quota inevitably degrades service quality, {\pname{}} still outperforms {\base{}} in all cases.

% Prefill: have enough vPod groups, but do not have enough vPods per group due to quota limits. So energy efficiency is close to unlimited, but SLO sat rate drops.
% Decode: cannot create enough vPod groups, so both energy efficiency and SLO sat rate drops significantly. 
% Even under resource-constrained cases, NeuScale still outperforms Base.

% By default, \pname{} assumes no resource quota limit, so it only evicts \vpod{} groups when the number of \vpod{}s in that group falls to 0. In resource-constrained scenarios, this may be suboptimal since a large \vpod{} may consume many NPU chips, leaving little room for other \vpod{} groups.
% When a resource quota is given, \pname{} prioritizes allocating \vpod{} groups that have the highest peak request rate in the last time window. If the resource quota is exhausted when creating a new \vpod{}, it will evict \vpod{}s from the \vpod{} group with the lowest historical peak request rate, as long as the new \vpod{} will serve more requests than the evicted \vpod{}s.

% \subsection{Ablation Study}
\subsection{Benefits of Fine-Grained \vpod{} Grouping}
\label{sec:eval:ablation}

\begin{figure}[t]
    \centering
    % \includegraphics[width=0.90\linewidth]{figures/multipool_deepseekv3-671b_combined.pdf}
    % \vspace{-3ex}
    \includegraphics[width=\linewidth]{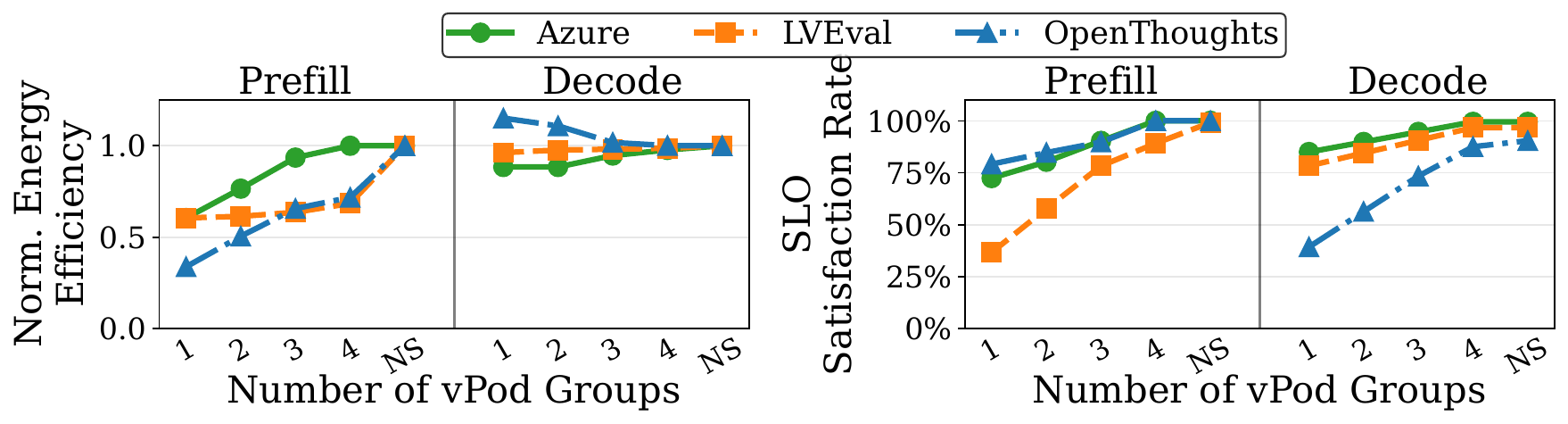}
    % \vspace{-6ex}
    \caption{{Energy efficiency (normalized to NS) and SLO satisfaction rate with different numbers of \vpod{} groups. ``1'' is equivalent to \base{}. ``NS'' is \pname{} (no group number limit). We study DeepSeekV3-671B as an example.}}
    \label{fig:sens_multipool}
    % \vspace{-2.5ex}
\end{figure}

% To understand the benefit of allocating best-fit \vpod{}s for various sequence lengths, we evaluate {\pname{}} against a coarse-grained multi-pool baseline: we partition the sequence lengths evenly into ranges (e.g., the 0-33rd, 34th-66th, and 67th-100th percentiles) and create a \vpod{} group (pool) for each range.
To better understand the benefits of having multiple {\vpod{}} groups, we analyze how many {\vpod{}} groups are required (i.e., the ``granularity'' of {\vpod{}} grouping) to achieve the maximum efficiency with \mbox{\pname{}}.
We create a coarse-grained multi-pool baseline by partitioning the sequence lengths evenly into a fixed number of ranges (e.g., the 0-33rd, 34th-66th, and 67th-100th percentiles) and creating a \vpod{} group (pool) for each range.
This represents an enhanced version of DynamoLLM~\cite{dynamollm:hpca25}. The original DynamoLLM design relies on the user to select the number of pools and the {\vpod{}} configuration for each pool, and it scales the number of \vpod{}s in each pool. To ensure a fair comparison, in our experiments, we employ \mbox{\pname{}}'s allocation algorithm to determine the optimal {\vpod{}} configuration for each pool in DynamoLLM.

As shown in \mbox{\Cref{fig:sens_multipool}}, the benefit of having finer-grained \vpod{} groups depends on the sequence length variance of the workload. Azure has a relatively narrow sequence length distribution, so 4 \vpod{} groups are sufficient. LVEval and OpenThoughts exhibit massive variance in input and output lengths, respectively. Hence, they need more \vpod{} groups for better efficiency and SLO satisfaction rate. 
For OpenThoughts, the energy efficiency is the best with one {\vpod{}} group, because more requests are batched together. However, this also causes significant SLO violations due to interference between requests with diverse sequence lengths. Having more {\vpod{}} groups helps improve performance isolation between these requests, which improves SLO satisfaction significantly.
As different workloads have diverse sequence length patterns, relying on a fixed number of pools is inherently sub-optimal. {\pname{}} overcomes this and can adapt to various workloads.

\subsection{Impact of Output Length Prediction Accuracy}
\label{sec:sens_output_predict_accuracy}

\begin{figure}[t]
    \centering
    \includegraphics[width=\linewidth]{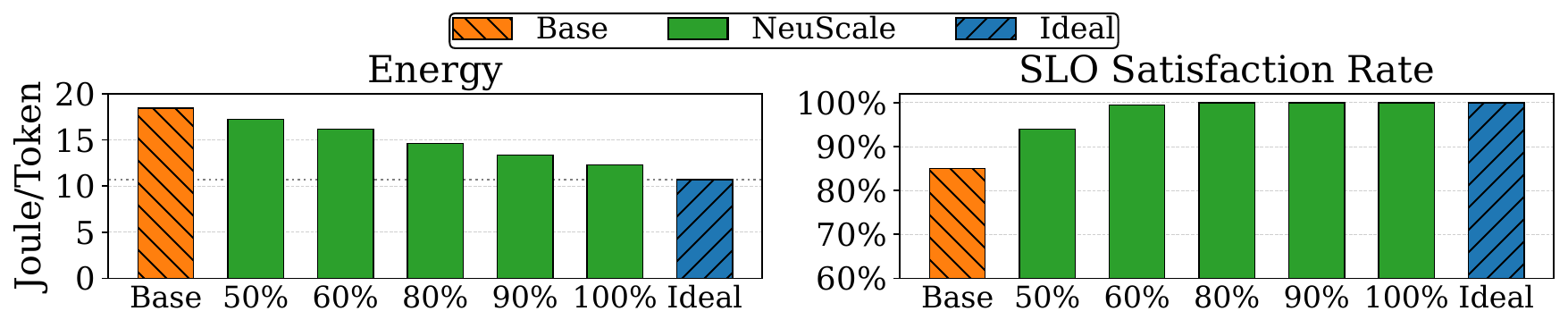}
    % \vspace{-3.5ex}
    \caption{Energy efficiency and SLO satisfaction rate of decode under various output length prediction accuracies. Prefill is not affected. We show DeepSeekV3-671B with Azure trace as an example.
    }
    \label{fig:eval_seqlen_predict}
    % \vspace{-3ex}
\end{figure}

In \mbox{\Cref{fig:eval_seqlen_predict}}, we simulate different output sequence length prediction accuracies and analyze the impact on decode energy efficiency and SLO satisfaction rate.
With 60\% accuracy (the state-of-the-art predictors~\mbox{\cite{qiu2024seqlenprediction,tao2025seqlenprediction,piotrowski2025seqlenprediction}}), \mbox{\pname{}} already achieves significant energy savings over \mbox{\base{}}, and the accuracy is sufficient for near-perfect SLO satisfaction. The misprediction penalty on SLO is largely hidden, as \mbox{\pname{}} proactively re-predicts the output length and migrates requests to their best-fit {\vpod{}} groups accordingly (see \mbox{\S\ref{sec:design:scheduling}}).
% The non-blocking request migration~\mbox{\cite{llumnix:osdi24}} overhead is low (on average $<$5\%) compared to the total decoding time.
Output length prediction is still an active research direction. As the predictor accuracy continues to improve in the future, the energy savings of \mbox{\pname{}} improve from 12\% at 60\% accuracy to up to 33\% at perfect accuracy.

\subsection{Evaluation on Real TPU Cluster}
\label{sec:eval:real_cluster}

\begin{figure}[t]
    \centering
    % \includegraphics[width=0.95\linewidth]{figures/rate_sweep_monetary_Azure4Hr_llama3-8b_bars.pdf}
    % \vspace{-3ex}
    \includegraphics[width=\linewidth]{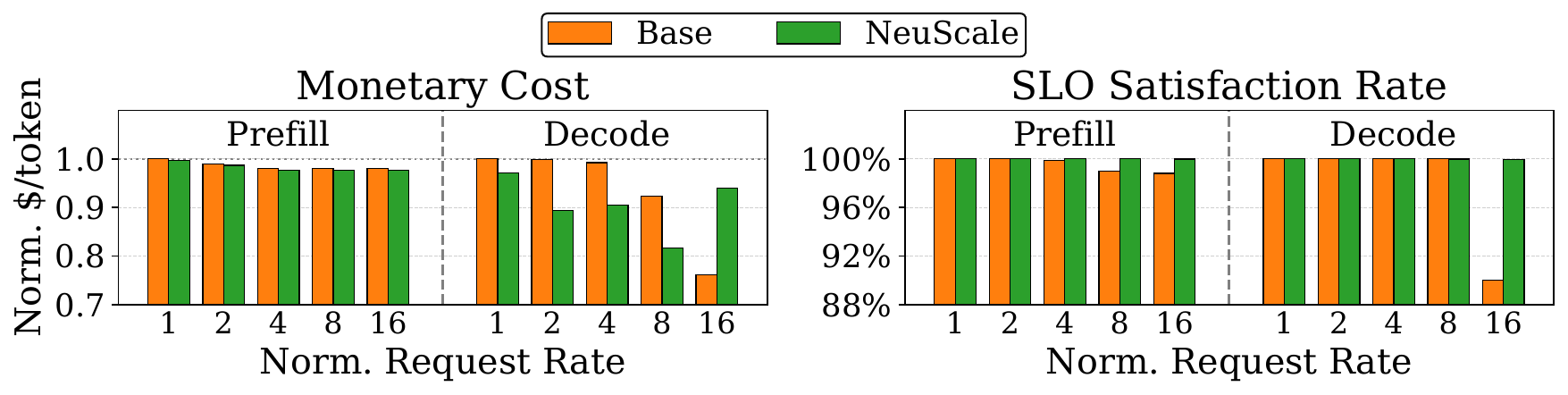}
    % \vspace{-3.4ex}
    \caption{Monetary cost efficiency and SLO satisfaction rate with Llama3-8B on a real TPU cluster.
    }
    \label{fig:real_cluster_eval}
    % \vspace{-3ex}
\end{figure}

In \Cref{fig:real_cluster_eval}, we evaluate \pname{} on a small TPU cluster (see the setup in \S\ref{sec:methodology}) using Llama3-8B and the Azure trace~\cite{dynamollm:hpca25} (a production LLM serving trace as described in \S\ref{sec:eval:setup}).
As the cloud TPU instance does not expose a public API for measuring power, we report monetary cost.
For this workload, TPUv6e is the most cost-efficient chip for both prefill and decode (see \Cref{fig:motiv_energy_cost}), so we stress the system by sweeping request rates (from 60 requests/min base rate to 16$\times$ the base rate) until TPUv6e capacity becomes insufficient. \pname{} maintains near-100\% SLO satisfaction across all request rates by first using TPUv6e and then automatically failing over to TPUv5e and TPUv4 when TPUv6e chips run out. \base{} only auto-scales within the TPUv6e pool, causing significant SLO violations at high request rates.

The trends differ between prefill and decode. For prefill, throughput saturates even at small batch sizes, so increasing the request rate provides little additional batching benefit, and the per-token cost remains nearly flat. For decode, higher request rates enable larger batches, reducing per-token cost for both \pname{} and \base{}, and \pname{} improves cost efficiency by up to 11.2\% over \base{}. At 16$\times$ request rate, \pname{} incurs higher decode cost than \base{} because some requests are served on less efficient TPUv5e/TPUv4 fail-over vPods. This is a necessary tradeoff for preserving SLO under TPUv6e scarcity.
This trend is consistent with the results in our simulation studies: by leveraging NPU heterogeneity, \pname{} always maintains or improves SLO satisfaction rate over \base{} while opportunistically improving cost efficiency.

% \hl{Briefly discuss the study conclusion with real cluster match with our simulation studies. That is the critical point for this evaluaition.}

% \hlA{\mbox{\Cref{fig:real_cluster_eval}}: real tpu cluster results with llama3-8b and Azure trace with scaled request rates; since TPUv6e is always the most efficient for both prefill and decode, we sweep various request rates (starting from an average of 60 request/min) to see the effect of allocation failover onto other TPU versions. No public API for power measuring, so we focus on the monetary cost.}

% prefill throughput saturates at low batch size, so cost does not decrease significantly as request rate increases.
% decode cost decreases as batch size increases with larger request rate.
% \pname{} always maintains near-100\% SLO satisfaction rate, as it automatically fails over to TPUv5e and TPUv4 chips when TPUv6e chips run out. In contrast, \base{} only auto-scales on TPUv6e chips.
% Hence, the SLO satisfaction rate of \base{} drops significantly at higher request rates.
% For decode, the \pname{}'s cost is higher at 16$\times$ request rate since the failover TPUv5e and TPUv4 chips are less efficient. This is a necessary tradeoff for maintaining SLOs.

\subsection{Carbon Efficiency and Sustainability}
\label{sec:eval:carbon}

\begin{figure}[t]
    \centering
    \includegraphics[width=\linewidth]{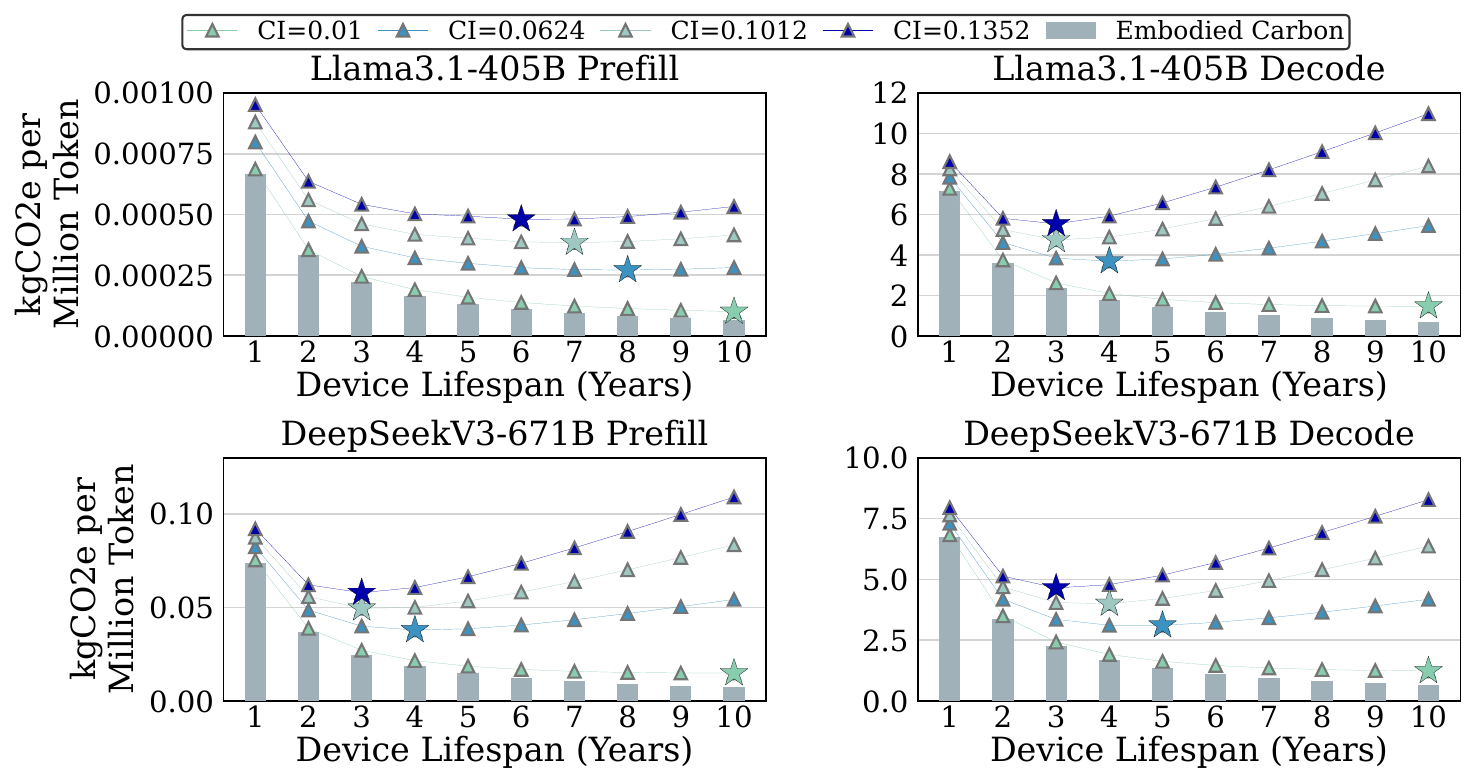}
    % \vspace{-3ex}
    \caption{Lifetime carbon emission of different device lifespans over 10 years, assuming the energy efficiency improves every year by the ratio between NPU-C and NPU-A. CI is carbon intensity. The optimal lifetime with minimal carbon emission is highlighted with stars.
    % For example, for Llama3.1-405B Prefill and \texttt{CI=0.0624}, the optimal lifespan is 8 years (i.e., we should retire old chips every 8 years).
    }
    \label{fig:carbon_efficiency_renewable}
    % \vspace{-4ex}
\end{figure}

The carbon footprint of an NPU chip consists of embodied carbon (emissions during manufacturing) and operational carbon (emissions due to runtime electricity consumption)~\cite{carbon_life_cycle}.
{{\pname{}} directly reduces the operational carbon by improving energy efficiency. It also provides an efficient way to reuse old NPU chips, thereby reducing their lifetime carbon footprint.}
% We study the carbon footprint of LLM services in \Cref{fig:carbon_efficiency_renewable}.

We quantify the embodied carbon based on Google TPU chips~\mbox{\cite{carbon_life_cycle}} and the operational carbon using the carbon intensity (CI), the carbon emission per unit energy consumed, reported by cloud providers~\mbox{\cite{google:env_report:2024}}.
For the ``datacenter taxes'' (e.g., cooling, power distribution, building maintenance), we assume a power usage effectiveness (PUE) factor of 1.09~\cite{google:env_report:2025}.
We vary the carbon intensity to study the impact of deploying more renewable energy (which reduces carbon intensity).

\Cref{fig:carbon_efficiency_renewable} quantifies the carbon footprint of LLM services, assuming various hardware update frequencies.
CI=0.01 is a hypothetical future value. Other CIs are real values from a recent Google environmental report~\cite{google:env_report:2025}.
With a 1-year lifespan (no NPU reuse), 70.0\%--98.7\% of total emissions are embodied carbon.
As we reuse NPUs, we can improve carbon efficiency by amortizing the embodied carbon.
Beyond a certain lifespan, the carbon efficiency starts to drop. This is because the operational carbon becomes relatively larger, as older chips are less energy-efficient than newer chips.
For our studied workloads, the carbon-optimal device lifespan is at least 3--6 years.
Since new NPU chips are developed and deployed at a much faster pace (e.g., every 1--2 years), reusing old NPUs to achieve their optimal lifespans is a promising way to improve datacenter carbon efficiency.

\section{Discussion}
\label{sec:discussion}

\noindent
\textbf{Support for future NPU chip design.}
\pname{} provides an efficient way to integrate new NPU generations into existing NPU clusters.
Based on our study, customizing NPU pod configurations for different workloads provides significant benefits.
This reflects the trend towards building more specialized NPU chips, such as prefill chips and decode chips~\cite{distserve_retro}.
% , as well as different chips for different request sequence lengths based on their resource demands
% to further improve the efficiency of prefill/decode disaggregation.
% \hlC{Furthermore, based on our study results, customizing NPU pod configurations for different request sequence lengths provides significant benefits. Hence, NPU chips}
Such specialization introduces more heterogeneity in the NPU cluster.
\pname{} can automatically map prefill and decode phases to the correct NPU versions.
% based on their resource demands and each chip's hardware capabilities.

\noindent
\textbf{Support for evolving ML workloads.}
\pname{} employs a generic way to determine an ML workload's best-fit NPU allocation: 
given any per-chip execution graph of an ML model, \pname{} can learn its arithmetic intensity and perform roofline-based analysis with the ML compiler.
% \revnote[3em]{\#B4}
% \hlB{While we evaluated {\pname{}} using text-only LLMs as examples, its benefits extend to multi-modal LLMs\mbox{\cite{chameleon:meta:2025,videopet:icml24,llama-imagegen,janus-pro:deepseek:2025,gpt4o-blog,gemini-google}}, as they rely on the same underlying auto-regressive transformer model and treat all modalities (text, audio, video, etc.) as tokens.}
To support a new ML model architecture, the major effort is to implement the parallelism enumeration pass. This is acceptable, as the ML framework or compiler typically needs to be updated to support new parallelism strategies~\cite{megatronlm,gshard,alpa:osdi2022}.
% \hlB{Although our evaluation in \mbox{\S\ref{sec:eval}} focuses on text-based LLMs, the benefits of \pname{} naturally extend to multi-modal LLMs~\cite{openai2023gpt4, geminiteam2023gemini, liu2023llava, alayrac2022flamingo}, as they rely on the same underlying transformer-based backbone architectures.}

\noindent
\textbf{Support for heterogeneous accelerator architectures.} As we build modern AI infrastructure, system architects should treat the heterogeneity of AI chips as a first-class citizen. The system infrastructure should be workload-aware and be able to map a workload to the best-fit chips.
\pname{} focuses on heterogeneous NPUs, but its core idea can be applied to other accelerators.
The \vpod{} abstraction already employs a generic \texttt{vChipConfig} that can represent other AI chips (e.g., GPUs), and the \texttt{ICITopology} can be generalized to express different network topologies (e.g., all-to-all NVSwitch-connected GPUs).
As cloud platforms typically deploy AI chips from different vendors, \pname{} paves the way for managing heterogeneous AI chips in a unified way.

%To provide improved LLM serving quality, we need to perform system optimizations across the full stack. For instance, our study suggests that 

% \noindent
% \hlB{\textbf{Support for other ML compilers.}
% While our implementation leverages XLA, {\pname{}} is fundamentally compiler-agnostic. Modern ML compiler stacks, such as TVM\mbox{\cite{tvm}}, AWS Neuron SDK for Trainium\mbox{\cite{aws_trainium}}, and Huawei CANN for Ascend NPUs\mbox{\cite{ascend910:huawei:hotchips31}}, share a similar compiler architecture.
% They all support different LLM parallelism strategies and operator-level cost models to estimate FLOP counts, memory traffic, and inter-chip collective latencies. \pname{}'s design space exploration can be built upon these features.}\revnoteReverse[-5em]{\#B10}

\section{Related Work}
\label{sec:related}

% These cluster-level optimizations do not consider heterogeneous accelerators.
% and are orthogonal to \pname{}.

\noindent
\textbf{{Heterogeneous cluster management.}}
Prior studies mostly focused on heterogeneous CPUs/GPUs (see \Cref{tab:related_works}).
% , introducing unique challenges.
For CPUs, they mapped heterogeneous workloads to homogeneous CPU cores by right-sizing the VM core count and memory size~\cite{SAMR,iBalloon,autoscale_webapp}. 
%They assumed the underlying CPU cores are identical. {\pname{}} maps heterogeneous LLM requests to heterogeneous NPUs.}\revnoteReverse[-2em]{\#D2}
For GPUs, they investigated optimal parallelisms for ML models across heterogeneous GPUs~\cite{yi2020optimizing,jia2022whale,um2024metis,helix:asplos25,hetis:sc25} 
% \hlF{In {\pname{}}, our {\vpod{}} abstraction explicitly avoids sharding LLM model weights across heterogeneous NPUs that have no ICI connections, as this will incur significant inter-chip communication overhead over DCN.
% As {\pname{}} targets large-scale NPU fleets that typically have more than thousands of chips for each NPU version, 
% Instead, {\pname{}} offers a practical and efficient way to exploit heterogeneity via creating \vpod{}s of different configurations.}
and heterogeneous GPU scheduling for ML training~\cite{mo2024heet,gavel:osdi20,sailor:sosp25,jabas:eurosys25}.
XSched~\cite{xsched:osdi25} developed a scheduling framework for diverse accelerator types and enabled preemptive scheduling on a single chip.
HeteCCL~\cite{heteccl:nsdi26} optimizes network communications between heterogeneous GPUs for LLM training.
{\pname{}} targets cluster scheduling and solves a completely different problem.
For resource allocation, prior works primarily relied on heuristics or profiling to build performance models and find the optimal configuration (e.g., GPU version and clock frequency)~\cite{dynamollm:hpca25,griggs2024melange}, which would incur high profiling overhead on NPUs.
Unlike CPUs/GPUs, NPUs exhibit distinct compute capabilities and performance characteristics.
%that cannot practically examine all possible heterogeneous allocations for different LLM services,  
%While profiling works well for a single GPU node with up to 8 GPUs, it cannot scale to thousands of NPU chips per {\vpod{}}.
{\pname{}} solves this challenge for NPUs using a lightweight yet accurate roofline model.
%To manage diverse accelerators, system virtualization techniques have been developed~\cite{vgpu:nvidia, mig:nvidia, preemptgpu:isca14, preemptsimt:sc16,amorphos:osdi18, 
%optimus:asplos20, vital:asplos20, synergy:asplos21, mlvital:asplos21,v10:isca23,neucloud:hotos23,neu10:micro24,heterovital:isca21}.
%They mostly focused on virtualizing a single device.
%{{\pname{}} defines the {\vpod{}} abstraction to manage NPU heterogeneity, it captures the hardware differences of heterogeneous NPUs while maintaining the best compatibility with existing ML frameworks.}
%It is the first to study how to best utilize NPU heterogeneity at scale. 
%and propose a practical solution with virtualization techniques. 

%It assumes each GPU device runs an independent data-parallel job, which does not apply to large models like LLMs. Our work focuses on the efficiency and sustainability of reusing older-generation chips. % which leads to different design challenges and design choices.

\noindent
\textbf{Auto-scaling frameworks.}
Auto-scaling for CPU workloads is well-supported by cloud platforms~\cite{autopilot:eurosys2020,autoscale_azure,aws_autoscale} and cluster orchestration frameworks~\cite{kubernetes_hpa} (see \Cref{tab:related_works}).
They primarily focused on configuring the core count and memory size of each VM~\cite{SAMR,autoscale_webapp,iBalloon}. 
%{\pname{}} configures NPU pods, which has a much larger parameter space.
% {\pname{}} addresses this ``right-sizing'' problem systematically for heterogeneous NPUs.
% State-of-the-art research leveraged
Some work leveraged reinforcement learning to make auto-scaling decisions for CPU workloads~\cite{aware:atc2023,firm:osdi20}.
None of them can directly work for NPUs. % due to the unique characteristics of ML workloads and NPU architecture.
Most prior works on GPU focused on adjusting the number of homogeneous GPU instances~\cite{gke_gpu_autoscaling,nvidia:dynamo_autoscaling,jabas:eurosys25}.
\pname{} is the first auto-scaling framework for heterogeneous NPUs.
Most recently, DynamoLLM~\cite{dynamollm:hpca25} proposed a scheduling mechanism that organizes different GPU instance types (i.e., GPU count) into pools (similar to \vpod{} groups) based on sequence lengths of incoming LLM requests. It is designed for homogeneous GPUs and relies on an expensive profiling-based allocation approach. Furthermore, it still relies on the user to configure the number of pools and the GPU instance configuration for each pool, and cannot automatically schedule across heterogeneous instance types.

% \noindent
% \textbf{Accelerator virtualization.} 
% \hl{merge to het acc management and auto-scaling frameworks}
% To facilitate the sharing of hardware accelerators in the cloud, prior studies investigated virtualization techniques for GPUs~\cite{vgpu:nvidia, mig:nvidia, preemptgpu:isca14, preemptsimt:sc16} and FPGAs~\cite{amorphos:osdi18, 
% optimus:asplos20, vital:asplos20, synergy:asplos21, mlvital:asplos21}. Most recently, researchers proposed to virtualize the internal matrix engines and generic vector engines of NPUs with hardware support~\cite{v10:isca23,neucloud:hotos23,neu10:micro24}. Different from the fine-grained virtualization of a single NPU chip, this work focused on the large-scale NPU reuse by virtualizing NPUs at the pod level and enabling NPU harvesting.

\noindent
\textbf{LLM serving.}
% Many works have focused on optimizing a single LLM serving instance (e.g., a single GPU node), including model sharding~\cite{huang2019gpipe,harlap2018pipedream,megatronlm,korthikanti2023seqparallel,fedus2022expertparallel}, KV cache management~\cite{vllm,lin2024infinite}, and request scheduling~\cite{orca:osdi22,sarathiserve:osdi24}.
Recently, many techniques have been proposed to optimize LLM serving~\cite{huang2019gpipe,harlap2018pipedream,korthikanti2023seqparallel,fedus2022expertparallel,vllm,lin2024infinite,orca:osdi22,sarathiserve:osdi24}.
%They can be deployed to optimize each \vpod{}.
% They primarily focus on a single LLM serving instance (e.g., a single GPU node).
%\pname{} manages multiple LLM serving instances (i.e., \vpod{}s).
For instance, prefill-decode disaggregation (PDD)~\cite{DistServe:osdi24,splitwise:isca24} was proposed to address the distinct resource demands between prefill and decode phases.
\pname{} maximizes the benefits of PDD by mapping the two phases to their best-fit \vpod{}s.
Load balancing~\cite{llumnix:osdi24,cheng2024slice,kossmann2024gpu} for homogeneous LLM instances can be employed to optimize request scheduling within a \vpod{} group.
{SpotServe\mbox{\cite{SpotServe}} leverages spot instances to serve LLMs. {\pname{}} can directly support ``spot \vpod{}s'' by treating {\vpod{}} preemptions as {\vpod{}} failures.}
DynamoLLM~\cite{dynamollm:hpca25} configures DVFS for different requests to optimize energy efficiency.
As future work, \pname{} can be extended to include DVFS as a configurable parameter.
% We wish to explore this in future work.
% LLM Inference: LLMCompass~\cite{LLMCompass}.
% GPU LLM Training: MAD-Max~\cite{mad:isca24}
 % or between host and accelerators~\cite{aminabadi2022deepspeed,sheng2023flexgen, song2024powerinfer}

% \noindent
% \textbf{Cloud resource harvesting.} 
% To improve cloud resource utilization, harvesting techniques have been developed  for virtual machines~\cite{smartharvest:eurosys2021, azureburst,
% awsburst, harvestvm:osdi2020,memharvest:asplos2022}. Unfortunately, all these techniques were developed for conventional server resources, including host CPU~\cite{smartharvest:eurosys2021, harvestvm:osdi2020}, memory~\cite{memharvest:asplos2022}, and storage~\cite{blockflex:osdi2022}. This work is the first to enable resource harvesting of NPU chips and address the unique challenges in NPU harvesting. 

\section{Conclusion}
\label{sec:conclusion}

% We investigate the potential of utilizing heterogeneous NPUs in cloud platforms.
We develop \pname{}, a new auto-scaling framework for LLM serving on heterogeneous NPU chips.
\pname{} improves energy/cost efficiency and SLO satisfaction for LLM services. It enables NPU reuse, facilitating its sustainable development and deployment.
% We believe that enabling efficient NPU reuse will facilitate the sustainable development of AI cloud platforms. 
% We discuss the potential solutions for supporting optimized NPU allocation, dynamic scheduling and NPU harvesting. 

\section*{Acknowledgements}

We thank the anonymous reviewers for their valuable feedback.
We thank Shaobo Li, Eric Zhou, and Yikang Yue from the Systems Platform and Intelligence Lab (Illinois PlatformX) for proofreading the paper, improving the presentation of the figures, and discussions on the LLM serving frameworks during the development of \pname{}.
This work was partially supported by Google’s TPU Research Cloud (TRC) program and NSF grants CAREER CNS-2144796 and CCF-2107470. 

%%%%%%% -- PAPER CONTENT ENDS -- %%%%%%%%

\balance
\bibliographystyle{IEEEtranS}
\bibliography{ref}
%%%%%%%%%%%%%%%%%%%%%%%%%%%%%%%%%%%%

\clearpage
\appendix
\section{Artifact Appendix}
\label{app:ae}

%%%%%%%%%%%%%%%%%%%%%%%%%%%%%%%%%%%%%%%%%%%%%%%%%%%%%%%%%%%%%%%%%%%%%
\subsection{Abstract}

This artifact contains the cluster NPU simulator used for the evaluation experiments in our paper.
The simulator is based on the open-source NeuSim codebase (\url{https://github.com/platformxlab/NeuSim}).
Our artifact provides automated scripts, sample input traces, simulator configuration files, and necessary instructions to run experiments. The artifact can be executed on any x86 machine. 
% We strongly recommend running this artifact with at least 32 GB of memory (for a smaller
% memory, disk swapping may need to be enabled).

\subsection{Artifact check-list (meta-information)}

% {\em Obligatory. Use just a few informal keywords in all fields applicable to your artifacts
% and remove the rest. This information is needed to find appropriate reviewers and gradually 
% unify artifact meta information in Digital Libraries.}

{\small
\begin{itemize}
  % \item {\bf Algorithm: }
  % \item {\bf Program: }
  % \item {\bf Compilation: }
  % \item {\bf Transformations: }
  % \item {\bf Binary: }
  % \item {\bf Model: }
  \item {\bf Data set: } \url{https://github.com/Azure/AzurePublicDataset}
  \item {\bf Run-time environment:} Ubuntu 24.04, Python 3.12.
  \item {\bf Hardware:} Any x86 machine. At least 24 CPU cores and 32 GB memory are recommended.
  % \item {\bf Run-time state: }
  % \item {\bf Execution: }
  % \item {\bf Metrics: }
  \item {\bf Output:} Simulator output files.
  % \item {\bf Experiments: }
  \item {\bf How much disk space required (approximately)?:} At least 20 GB recommended.
  \item {\bf How much time is needed to prepare workflow (approximately)?:} $\leq$10 minutes.
  \item {\bf How much time is needed to complete experiments (approximately)?: } At most one hour with the sample traces.
  \item {\bf Publicly available?: } \url{https://github.com/platformxlab/NeuSim/tree/micro26ae_neuscale}
  \item {\bf Code licenses (if publicly available)?: } Apache 2.0.
  % \item {\bf Data licenses (if publicly available)?: } 
  % \item {\bf Workflow automation framework used?: }
  \item {\bf Archived (provide DOI)?: } \url{https://doi.org/10.5281/zenodo.21544640}
\end{itemize}
}

%%%%%%%%%%%%%%%%%%%%%%%%%%%%%%%%%%%%%%%%%%%%%%%%%%%%%%%%%%%%%%%%%%%%%
\subsection{Description}

\subsubsection{How to access}
The artifact can be downloaded from
Zenodo at \url{https://doi.org/10.5281/zenodo.21544640}. For the latest
version, we recommend visiting our GitHub repository: \url{https://github.com/platformxlab/NeuSim/tree/micro26ae_neuscale}.

% {\em Obligatory}

\subsubsection{Hardware dependencies}
The artifact runs on any x86 machine with at least 8\,GB of free disk space.
We recommend a machine with high single-core performance, at least 32\,GB of
memory, and 24 physical CPU cores; 64\,GB of memory and 20\,GB of free disk
space are preferred.

\subsubsection{Software dependencies}
The artifact requires Linux, preferably Ubuntu 24.04 or newer. It uses
\texttt{conda} and \texttt{pip} to manage software dependencies.
Python 3.12 is required.

\subsubsection{Data sets}
The artifact uses a truncated, downsampled trace from the AzurePublicDataset~\cite{dynamollm:hpca25}. The sample trace is supplied with this artifact.

% \subsubsection{Models}

%%%%%%%%%%%%%%%%%%%%%%%%%%%%%%%%%%%%%%%%%%%%%%%%%%%%%%%%%%%%%%%%%%%%%
\subsection{Installation}

We strongly recommend following the \texttt{micro26ae.md} for detailed instructions on installing the artifact and running experiments. We list the key steps below. 

Clone the artifact branch and install its dependencies:
\begin{lstlisting}[style=BashStyle,escapechar=~~]
    git clone --branch micro26ae_neuscale \
      https://github.com/platformxlab/NeuSim.git
    cd NeuSim
    
    conda create -n neuscale-ae python=3.12.2
    conda activate neuscale-ae
    pip install -e ".[dev]"
\end{lstlisting}

The sampled Azure trace and its precomputed optimal-configuration cache are
distributed separately:
\begin{lstlisting}[style=BashStyle,escapechar=~~]
    git clone --depth 1 \
      https://github.com/XZman/micro26ae_supplementary_files \
      ../micro26ae_supplementary_files
    
    unzip ../micro26ae_supplementary_files/\
    AzureLLMInferenceTrace_code_3h_sampled.zip \
      -d artifact_eval/micro26/data
    
    unzip ../micro26ae_supplementary_files/\
    request_lookup_cache_deepseekv3_azure_3h_v5p_v6e.zip \
      -d artifact_eval/micro26
    
    (cd artifact_eval/micro26/data && sha256sum -c SHA256SUMS)
    python -m neusim.run_scripts.prepare_micro26ae_sample_cache \
      --validate-only
\end{lstlisting}

Archive checksums and additional installation details are provided in
\texttt{micro26ae.md}.

%%%%%%%%%%%%%%%%%%%%%%%%%%%%%%%%%%%%%%%%%%%%%%%%%%%%%%%%%%%%%%%%%%%%%
\subsection{Experiment workflow}

First, exercise the complete simulation path with a one-request smoke test:
\begin{lstlisting}[style=BashStyle,escapechar=~~]
    bash neusim/run_scripts/run_fleetsim_smoke.sh
\end{lstlisting}

Next, inspect the default simulator configurations and run a 100-request quick test:
\begin{lstlisting}[style=BashStyle,escapechar=~~]
    python -m neusim.run_scripts.run_micro26ae_sample --list
    python -m neusim.run_scripts.run_micro26ae_sample --max-requests 100
\end{lstlisting}

Finally, replay the three-hour sample trace with 3,460 requests:
\begin{lstlisting}[style=BashStyle,escapechar=~~]
    python -m neusim.run_scripts.run_micro26ae_sample
\end{lstlisting}

The launcher evaluates Base-Max, NeuScale, and Ideal for both energy and
monetary objectives with the sample trace. Energy and monetary jobs run concurrently. On the
recommended hardware, the complete post-installation workflow should take
less than 1 hour.

%%%%%%%%%%%%%%%%%%%%%%%%%%%%%%%%%%%%%%%%%%%%%%%%%%%%%%%%%%%%%%%%%%%%%
\subsection{Evaluation and expected results}

The simulation results are generated under 
\begin{center}
    \texttt{artifact\_eval/micro26/results}.
\end{center}
The top-level \texttt{ae\_run\_manifest.json} should report six completed or reused
settings. Each full setting should contain 3,460 requests in both
\texttt{request\_trace.csv} and \texttt{stats.json}; the bounded test should
contain 100 requests.

Each setting also contains \texttt{run.log} and
\texttt{run\_contract.json}. Event-driven settings additionally emit vPod
lifecycle and reconfiguration traces and a final checkpoint.
\texttt{stats.json} reports latency, throughput, and energy- and
cost-efficiency metrics. Because the artifact
uses a three-hour sampled trace, the output results may not exactly match
the full-trace results reported in the paper.

%%%%%%%%%%%%%%%%%%%%%%%%%%%%%%%%%%%%%%%%%%%%%%%%%%%%%%%%%%%%%%%%%%%%%
\subsection{Experiment customization}

\noindent
\textbf{Sensitivity analysis.}
The simulator supports alternative models, traces, objectives, NPU generations,
and sensitivity settings. For example:
\begin{lstlisting}[style=BashStyle,escapechar=~~]
    python -m neusim.run_scripts.run_micro26ae_sample \
      --trace-file /path/to/trace.csv \
      --trace-name Custom-Trace \
      --model llama3-70b \
      --request-cache-dir /path/to/cache \
      --systems Base-Max,NeuScale,Ideal \
      --goals energy
\end{lstlisting}
Here, \texttt{--trace-file} selects the request-trace CSV;
\texttt{--trace-name} names the workload in outputs and selects any
trace-specific defaults; \texttt{--model} selects a model configuration;
\texttt{--request-cache-dir} supplies its matching optimal-configuration
cache; \texttt{--systems} selects the comma-separated auto-scaling policies to
compare; and \texttt{--goals} selects one or both optimization objectives.

\noindent
\textbf{Offline optimal (best-fit) vPod allocation generation.}
A matching optimal-configuration cache must be generated for a new model,
trace, or NPU set:
\begin{lstlisting}[style=BashStyle,escapechar=~~]
    python -m neusim.run_scripts.generate_fleetsim_optimal_cache \
      --trace /path/to/trace.csv \
      --models llama3-70b \
      --versions 5p,6e \
      --top-k all \
      --output-dir /path/to/cache \
      --skip-existing
\end{lstlisting}
Here, \texttt{--trace} specifies an input request trace and may be repeated;
\texttt{--models} selects one or more comma-separated LLMs;
\texttt{--versions} selects the NPU generations to explore;
\texttt{--top-k all} retains every ranked configuration for each NPU version
and inference phase; \texttt{--output-dir} specifies the generated cache
directory; and \texttt{--skip-} \texttt{existing} reuses compatible existing results
when resuming an interrupted generation.
Note that the lookup cache generation may require many CPU-hours. Further launcher details are in
\texttt{micro26ae.md}; the general simulator interface is documented in
\texttt{neusim/fleetsim/README.md}.

\noindent
\textbf{Custom NPU cluster simulations.}
For maximum flexibility, the \texttt{neusim.fleetsim} module can also be invoked directly:
\begin{lstlisting}[style=BashStyle,escapechar=~~,
                   label=code:fleetsim-main]
    python -m neusim.run_scripts.fleetsim_main \
      --configs_path="$PWD/configs" \
      --model=llama3-70b \
      --system=NeuScale \
      --opt_goal=energy \
      --request_pattern=trace \
      --trace=Custom-Trace \
      --trace_file=/path/to/trace.csv \
      --request_rate=2 \
      --max_timestamp_hours=3 \
      --max_num_requests=1000 \
      --chip_versions=5p,6e \
      --prefill_chip_versions=6e \
      --decode_chip_versions=5p,6e \
      --max_chips_per_version=5p=256,6e=512 \
      --allocation_success_rate=0.95 \
      --hs_interval_minutes=15 \
      --vs_interval_minutes=5 \
      --hs_window_minutes=30 \
      --vs_window_minutes=15 \
      --ewma_alpha=0.5 \
      --scaling_headroom_factor=1.2 \
      --queue_drain_target_seconds=30 \
      --output_prediction_accuracy=0.8 \
      --request_results_cache_dir=/path/to/cache \
      --npusim_backend_cache_dir=/path/to/backend-cache \
      --output_dir="$PWD/results/fleetsim/custom-neuscale"
\end{lstlisting}
This interface exposes the model, policy, objective, trace request rate, NPU versions, allocation failure rate, auto-scaling configurations, output-length prediction accuracy, etc. Further launcher details are in
\texttt{micro26ae.md}; all FleetSim flags are documented in
\texttt{neusim/fleetsim/README.md}.

%%%%%%%%%%%%%%%%%%%%%%%%%%%%%%%%%%%%%%%%%%%%%%%%%%%%%%%%%%%%%%%%%%%%%
% \subsection{Notes}

%%%%%%%%%%%%%%%%%%%%%%%%%%%%%%%%%%%%%%%%%%%%%%%%%%%%%%%%%%%%%%%%%%%%%
\subsection{Methodology}

Submission, reviewing and badging methodology:

\begin{itemize}
  \item \url{https://www.acm.org/publications/policies/artifact-review-and-badging-current}
  \item \url{https://cTuning.org/ae}
\end{itemize}

%%%%%%%%%%%%%%%%%%%%%%%%%%%%%%%%%%%%%%%%%%%%%%%%%%%%
% When adding this appendix to your paper, 
% please remove below part
%%%%%%%%%%%%%%%%%%%%%%%%%%%%%%%%%%%%%%%%%%%%%%%%%%%%

% \end{document}

\end{document}